\documentclass[twocolumn]{aa}

\usepackage[varg]{txfonts}
\usepackage{natbib}
\usepackage{graphicx}
\usepackage{times}
\usepackage{longtable}
\usepackage{colortbl}
\usepackage{enumitem}
\usepackage{color}
\usepackage{epstopdf}

\newcommand{\ble}{$\beta$~Lyr~A}

\newcommand{\spefo}{{\tt SPEFO} }

\newcommand{\ks}{km~s$^{-1}$}

\newcommand{\ms}{M$_{\odot}$}
\newcommand{\rs}{R$_{\odot}$}

\newcommand{\ha}{H$\alpha$ }

\begin{document}

\title{Optically thin circumstellar medium in $\beta$~Lyr~A system}

\author{
M.~Bro\v z\inst{1}\and
D.~Mourard\inst{2}\and
J.~Budaj\inst{3}\and                   
P.~Harmanec\inst{1}\and
H.~Schmitt\inst{4}\and                 
I.~Tallon-Bosc\inst{5}\and             
D.~Bonneau\inst{2}\and                 
H.~Bo\v{z}i\'c\inst{6}\and             
D.~Gies\inst{7}\and                    
M.~\v{S}lechta\inst{8}                 
}

\offprints{\email mira@sirrah.troja.mff.cuni.cz}

\institute{
  Astronomical Institute of the Charles University, Faculty of Mathematics and
  Physics,\\ V~Hole\v{s}ovi\v{c}k\'ach~2, 180~00~Praha~8, Czech~Republic \and
  Universit\'{e} C\^{o}te d'Azur, OCA, CNRS, Lagrange, Parc~Valrose,
  B\^{a}t. Fizeau, 06108~Nice, France \and
  Astronomical Institute, Slovak Academy of Sciences,
  059~60 Tatransk\'a Lomnica, Slovak~Republic \and
  Naval Research Laboratory, Remote Sensing Division, Code~7215,
  4555~Overlook~Ave.~SW, Washington, DC~20375, USA \and
  Univ Lyon, Univ Lyon1, Ens de Lyon, CNRS, Centre de Recherche Astrophysique de Lyon UMR5574, F-69230, Saint-Genis-Laval, France \and
  Hvar Observatory, Faculty of Geodesy, University of Zagreb,
  Ka\v{c}i\'ceva~26, 10000~Zagreb, Croatia \and
  The CHARA Array of Georgia State University, Mount Wilson Observatory,
  Mount Wilson, California~91023, USA \and
  Astronomical Institute, Czech Academy of Sciences,
  251~65~Ond\v{r}ejov, Czech~Republic
}

\date{Received \today}


\abstract{%
$\beta$~Lyr~A is a complex binary system with an extensive
observational dataset:
light curves (from FUV to FIR),
interferometric squared visibility,
closure phase,
triple product measurements,
spectral-energy distribution (SED),
high-resolution spectroscopy,
differential visibility amplitude, and also
differential phase.
In particular, we use spectra from Ond\v rejov 2m telescope from 2013 to 2015,
to measure the emission in H$\alpha$, \ion{He}{i}, \ion{Si}{ii}, \ion{Ne}{i}, or \ion{C}{ii} lines,
and differential interferometry by CHARA/VEGA from the 2013 campaign
to measure wavelength-dependent sizes across H$\alpha$ and \ion{He}{i} 6678.
This allows us to constrain not only
optically thick objects
(primary, secondary, accretion disk),
but also optically thin objects
(disk atmosphere, jets, shell).
We extended our modelling tool Pyshellspec
(based on Shellspec; a 1D LTE radiative transfer code)
to include all new observables,
to compute differential visibilities/phases, 
to perform a Doppler tomography, and
to determine a joint $\chi^2$ metric.
After an optimisation of 38 free parameters,
we derive a robust model of the $\beta$~Lyr~A system.
According to the model, the emission is formed in
an extended atmosphere of the disk,
two perpendicular jets expanding at ${\sim}\,700\,{\rm km}\,{\rm s}^{-1}$,
and a symmetric shell with the radius ${\sim}\,70\,R_\odot$.
The spectroscopy indicates a low abundance of carbon,
$10^{-2}$ of the solar value.
We also quantify systematic differences between datasets
and discuss alternative models, with
higher resolution,
additional asymmetries, or
He-rich abundance.
}



\authorrunning{Bro\v z et al.}
\titlerunning{Optically thin medium in $\beta$~Lyr~A}
\maketitle


\section{Introduction}
$\beta$~Lyr~A (HR 7106, HD 174638) is an archetype of a semidetached binary in a rather rapid phase of mass transfer (of the order of 
$2\cdot 10^{-5}\,M_\odot\,{\rm yr}^{-1}$) between binary components.
Its orbital period has been increasing by the high rate
of 19~sec per year. While during the interferometric campaign in 2013
the value of the period was 12\fd9427, in 2020 it is already 12\fd9440.
The gainer (primary) is an early B star hidden in an optically thick accretion disk. The donor (secondary) is a late B star filling its Roche lobe.
For a summary of numerous investigations of this object,
we refer the readers to \citet{sahade66}, \citet{hec2002}, \citet{skul2020}.

In \citet{Mourard_etal_2018A&A...618A.112M}, we summarized
more recent studies and carried out an attempt to model optically thick matter within the system.
The model of $\beta$~Lyr~A was constrained by wide-band light curves (FUV to FIR) and continuum interferometric measurements.
It was thus sensitive to the properties of the Roche-filling secondary, the primary and its opaque accretion disk. 
The following values were adopted from the previous studies:
semi-amplitudes of the radial-velocity (RV) curves
$K_1=41.4$\,\ks, $K_2=186.3$\,\ks, implying
the inverse mass ratio $q=0.223$,
the projected semimajor axis $a\sin{i}=58.19$\,\rs\
and the masses
$M_2=13.048$\,\ms,
$M_1=2.910$\,\ms.
The model led to an estimated distance to the system of $\sim\,320\,{\rm pc}$
and to the finding that the accretion disk fills the whole available
space of the Roche lobe in the orbital plane.

This study represents an extension of that work to optically thin parts 
of circumstellar matter within the system. To this goal,
we shall use additional observational data, in particular
spectral-energy distribution (SED),
high-resolution spectroscopy,
and differential interferometry
to measure
absolute fluxes,
emission line profiles,
wavelength-dependent brightness distribution
at the same time.
This allows us to model the properties of
the disk atmosphere,
jets, or possible
shell-like structures.
To this point, we use a {\em geometrically constrained\/} model,
described by a limited set of geometrical objects
and a limited number of parameters.
This method has been preferred because an image reconstruction
from limited spatial frequencies of interferometric measurements
was not possible.
On the other hand,
any geometrical model uses numerous assumptions,
e.g., the Roche geometry for stellar surfaces,
some symmetries,
or an a-priori knowledge.


\section{Observational data}

All observational data used in our previous study \citep{Mourard_etal_2018A&A...618A.112M}
remain the same, i.e., the light curves and optical interferometric data.
We thus refer to this work for their detailed description.
We just recall that the interferometric measurements give access to
information on the brightness spatial distribution of the source. The squared
visibilities sample the Fourier transform of the distribution at a spatial
frequency defined by the baseline vector projected on the plane of sky divided
by the central wavelength of the observed band, $\vec B/\lambda$. Closure phase
and triple product amplitudes are self-calibrated estimators based on the
interferometric data considered on a triplet of telescopes.
Hereinafter, we describe only the additionally used data.


\subsection{Spectral-energy distribution (SED)}

To constrain the absolute flux of $\beta$~Lyr~A, we use data from \cite{burnashev78}.
This low-resolution absolute spectrophotometry covers the wavelength range
from 3300 to 7400\,\AA, i.e., including the H$\alpha$, H$\beta$, H$\gamma$,
H$\delta$, the Balmer jump, as well as the \ion{He}{i} 5876 and 6678 lines.
The effective bands are 25\,\AA\ wide which is not enough to resolve
the spectral lines. These are well represented by our high-resolution spectra
described below.
The fluxes were calibrated on Vega ($\alpha$~Lyr), based on its absolute calibration by \citet{teres72}.
When interpreting these fluxes, one should be more careful
in the NUV region, where the calibration is generally more difficult.

We performed a dereddening of the absolute fluxes to account
for interstellar extinction. For the galactic coordinates
$l = 63.1876^\circ$, $b = 14.7835^\circ$
and the distance moduli
$\mu = 5\log_{10}[d]_{\rm pc}-5 \doteq 7.3$ to $7.6$
we would expect a value at most
$E(B-V) = 0.020$
\citep{Green_etal_2015ApJ...810...25G}.
Using standard relations for
$A_V = 3.1\,E(B-V)$
and
$A_\lambda/A_V$
\citep{Schlafly_Finkbeiner_2011ApJ...737..103S},
we increased the observed absolute fluxes~$F_\lambda$ accordingly.


\subsection{High-resolution spectroscopy (SPE)}

We have at our disposal 72 Ond\v{r}ejov CCD spectra from 2013 to 2015
(JDs~2456450.37 to 2457294.30). The spectra have a linear dispersion of
$17.2\,\mbox{\AA}\,{\rm mm}^{-1}$ and a two-pixel resolution of 12700.
They cover the wavelength region of approximately 6300 to 6730\,\AA.
Their initial reductions (flatfielding, wavelength calibrations and creation
of 1-D spectra) were carried out by M\v{S} in IRAF. Normalization and measurement
of a selection of telluric lines to be used for a fine correction of radial-velocity
(RV) zero point were carried out by PH in \spefo \citep{sef0,spefo3}. 
For modelling, we used a selection of 11 spectra, well covering different phases
of the orbital period.

Given the nature of the $\beta$~Lyr~A system,
we shall perform a Doppler-tomography analysis.
This should enable us to resolve a 3-D structure and velocity fields
of the circumstellar matter.


\begin{figure}
\centering
\includegraphics[width=6cm]{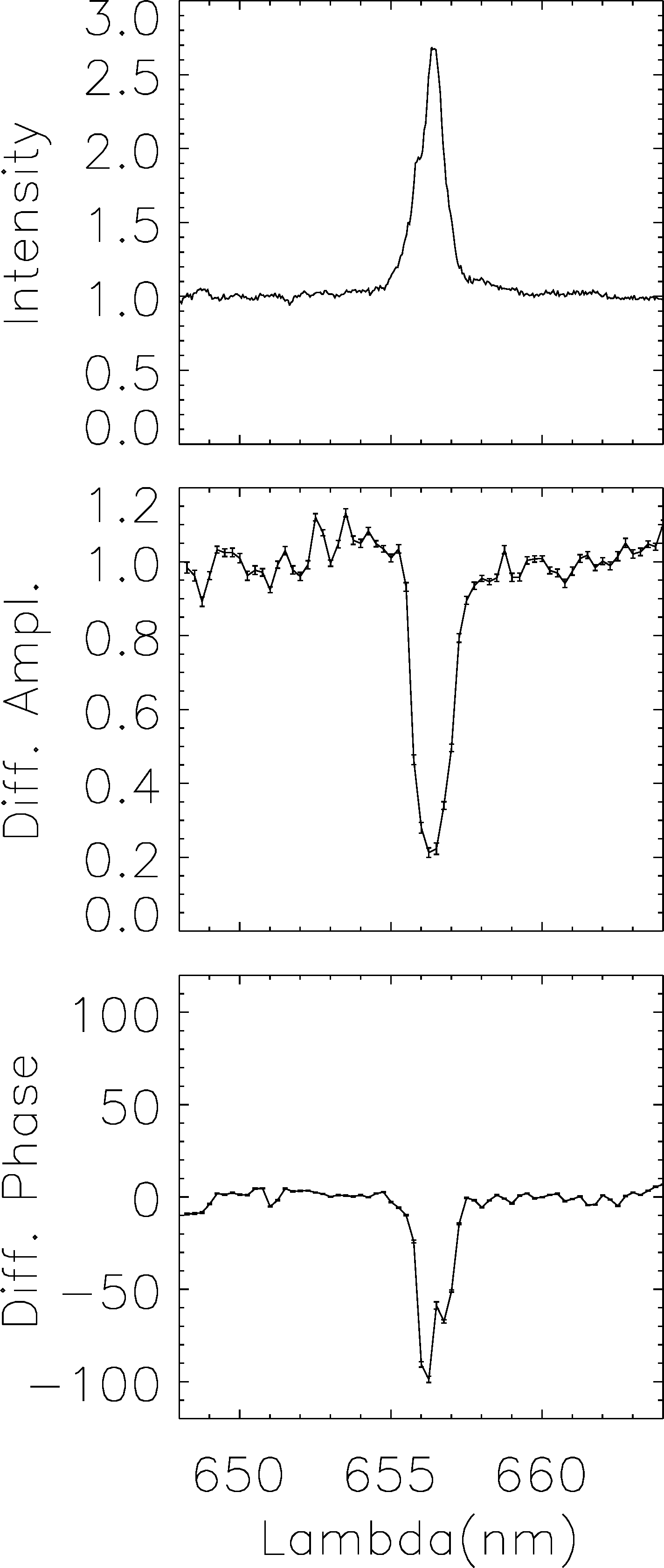}
\caption{
An example of the VEGA differential visibility measurement during the night of 27 Jun 2013 with the 140\,m E2W2 baseline. Top: the \ha line (continuum normalized to 1); middle: Amplitude of the differential visibility (normalized to 1 in the continuum); bottom: phase of the differential visibility in degrees.}
\label{fig:visdiff}
\end{figure}

\begin{figure}
\centering
\includegraphics[width=8cm]{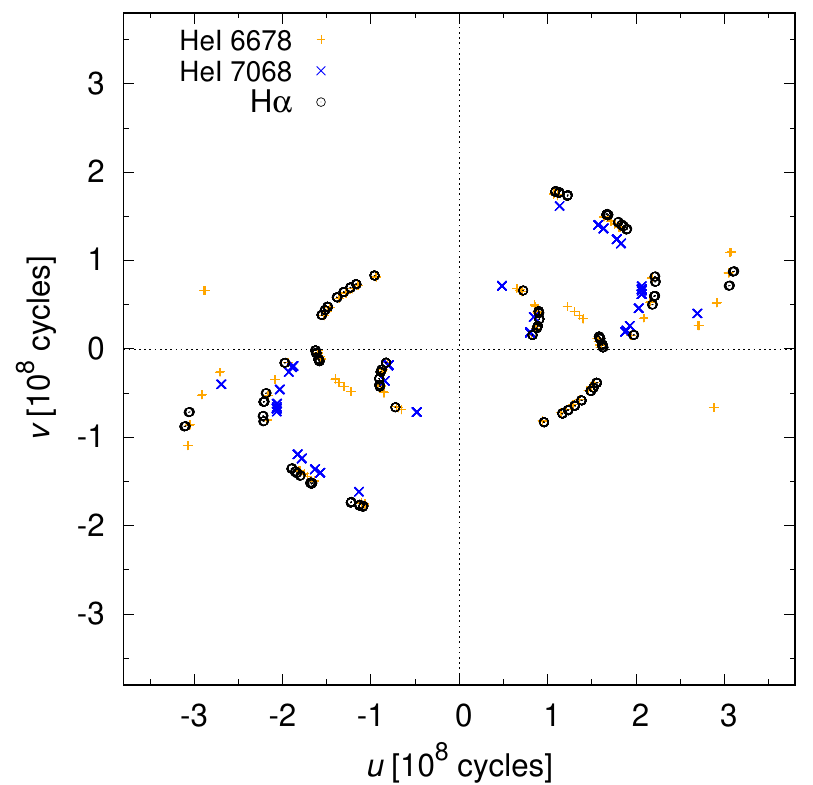}
\caption{
The $(u,v)$ coverage of differential visibility and phase
measurements from the CHARA/VEGA instrument.
A subset of H$\alpha$ data used for modelling is shown (black),
together with additional \ion{He}{i} 6678 (green) and 7065 (blue) data.
The orientation is $v>0$ north and $u>0$ east.}
\label{fitting_shell8_20200415__20_uv2}
\end{figure}

\subsection{Differential interferometry (VAMP, VPHI)}
\label{sec:diffinterf}
In \citet{Mourard_etal_2018A&A...618A.112M} we have presented an extensive interferometric data set recorded during a coordinated campaign in 2013. Data were obtained on the NPOI array \citep{npoi}, and on the CHARA Array \citep{chara} with the MIRC \citep{mirc} and VEGA \citep{vega,vega2} instruments. This first paper was dedicated to the study of the opaque accretion disk and has made use of all the interferometric measurements in the continuum bands from 525 to 861\,nm and in the H~band. 

For the work presented here, the VEGA data are used to measure differential
complex visibilities in the H$\alpha$ (6562\,\AA) and \ion{He}{i} (6678 and
7065\,\AA) lines.
Differential complex visibilities are estimated through the cross-spectrum
of the interferometric data between a first wide reference spectral band and a
second narrow analysis band crossing the first one. Knowing the shape of the
object from the squared visibilities in the reference band, the differential
data permit to extract information on the variation of the shape of the object
at high spectral and spatial resolution over a small band. The amplitude gives
information on the chromatic dependence of the size of the object whereas the
phase provides valuable information on the position on the sky.

202 measurements (87 for H$\alpha$, 87 for \ion{He}{i} 6678, and 28 for
\ion{He}{i} 7065) are available with both differential amplitude and phase,
with a standard deviation of phase in the continuum smaller than $15^\circ$. In
the case of a larger standard deviation of the differential phase, we got 498
additional measurements of the amplitude of the differential visibility (195
for H$\alpha$, 204 for \ion{He}{i} 6678, and 99 for \ion{He}{i} 7065) with a
signal to noise ratio better than~5 in the continuum. The details of the
observations are presented in \cite{Mourard_etal_2018A&A...618A.112M}.

The standard differential processing \citep{vega} of the VEGA data has been
modified to avoid the underestimation of the uncertainties of the differential
quantities. For this work, we have replace them by the standard deviation of
the measurements (both amplitude and phase) computed in the continuum part and
multiplied by a factor equal to the square root of the flux of each narrow-band
channel in order to correctly match the behavior of the photon noise.


One example of an individual measurement is presented in
Figure~\ref{fig:visdiff}. It should be noted that the amplitude of the
differential visibility is normalized to 1 in the continuum and that the phase
is arbitrarily set to a mean value of 0 in the continuum. It is also important
to note that, in some cases, phase jumps may occur as the differential phase is
defined only modulo $2\pi$. The $(u,v)$ coverage is shown in
Figure~\ref{fitting_shell8_20200415__20_uv2}.


\section{Pyshellspec model}

To account for all types of observational data,
we had to significantly extend and improve our modelling tool called Pyshellspec\footnote{\url{http://sirrah.troja.mff.cuni.cz/~mira/betalyr/}}.
Its purpose is to calculate radiative transfer through the volume
surrounding the binary.
We now use a joint $\chi^2$ metric as follows:
\begin{equation}
\chi^2 = \chi^2_{\rm lc} + \chi^2_{\rm vis} + \chi^2_{\rm clo} + \chi^2_{\rm t3} + \chi^2_{\rm sed} + \chi^2_{\rm spe} + \chi^2_{\rm vamp} + \chi^2_{\rm vphi}\,,
\end{equation}
with individual contributions:
\begin{equation}
\chi^2_{\rm lc} = \sum_{k=1}^{N_{\rm band}} \sum_{i=1}^{N_{{\rm lc}\,k}} \left(\frac{m_{ki}^{\rm obs}-m_{ki}^{\rm syn}}{\sigma_{ki}}\right)^2\,,
\end{equation}
\begin{equation}
\chi^2_{\rm vis} = \sum_{i=1}^{N_{\rm vis}} \left(\frac{|V_i^{\rm obs}|^2-|V_i^{\rm syn}|^2}{\sigma_i}\right)^2\,,
\end{equation}
\begin{equation}
\chi^2_{\rm clo} = \sum_{i=1}^{N_{\rm clo}} \left(\frac{\arg T_{3i}^{\rm obs}-\arg T_{3i}^{\rm syn}}{\sigma_i}\right)^2\,,
\end{equation}
\begin{equation}
\chi^2_{\rm t3} = \sum_{i=1}^{N_{\rm clo}} \left(\frac{|T_{3i}|^{\rm obs}-|T_{3i}|^{\rm syn}}{\sigma_i}\right)^2\,,
\end{equation}
\begin{equation}
\chi^2_{\rm sed} = \sum_{i=1}^{N_{\rm sed}} \left(\frac{F_{\lambda i}^{\rm obs}-F_{\lambda i}^{\rm syn}}{\sigma_i}\right)^2\,,
\end{equation}
\begin{equation}
\chi^2_{\rm spe} = \sum_{i=1}^{N_{\rm spe}} \left(\frac{I_{\lambda i}^{\rm obs}-I_{\lambda i}^{\rm\,syn}}{\sigma_i}\right)^2\,,
\end{equation}
\begin{equation}
\chi^2_{\rm vamp} = \sum_{k=1}^{N_{\rm set}} \sum_{i=1}^{N_{{\rm vamp}\,k}} \left(\frac{V_i^{\rm obs}-V_i^{\rm syn} f_k}{\sigma_i}\right)^2\,,
\end{equation}
\begin{equation}
\chi^2_{\rm vphi} = \sum_{k=1}^{N_{\rm set}} \sum_{i=1}^{N_{{\rm vphi}\,k}} \left(\frac{\arg V_i^{\rm obs}-\arg V_i^{\rm syn} + g_k + h_k}{\sigma_i}\right)^2\,,
\end{equation}
where
$m$~denotes magnitudes in given passbands,
$|V|^2$ squared visibility,
$\arg T_3$ closure phase,
$|T_3|$ triple product amplitude,
$F_\lambda$ absolute monochromatic flux,
$I_\lambda$ normalized monochromatic flux,
$|V|$ differential visibility amplitude,
$\arg V$ differential visibility phase.
The latter two interferometric quantities are modified by
a~multiplicative factor $f_k$,
an~additive offset $g_k$, and
a~phase slip $h_k$ ($\pm360^\circ$)
to correctly match the way these quantities are estimated
as explained in Sec.~\ref{sec:diffinterf}.

Apart from new observables,
more objects (jet, flow, shell) were implemented in Python,
and correspondingly more free parameters.
We performed some corrections necessary
for the high-resolution spectroscopy, in particular
velocity fields of all object are propagated to Shellspec;
Phoenix absolute spectra \citep{husser2013},
used as boundary conditions at the stellar surfaces of our 3D model,
were converted from vacuum to air wavelengths
and a higher resolution $0.1\,\mbox{\AA}$ was used.
Fitting of factors, offsets and slips per each interferometric dataset was included,
to minimize the difference between observed and synthetic differential visibilities and phases.
Optionally, we use the subspace-searching simplex algorithm
(or subplex; \citealt{Rowan_1990}) for the $\chi^2$ minimisation,
which is sometimes very useful.
The Openmp (threads) parallelisation is done
per wavelengths (for LC, VIS, VAMP, VPHI, etc. datasets)
or per phases (for SED, SPE).
To model extended optically thin structures,
we had to extend the grid
(usually $80\times 80\,R_\odot$),
and optionally use a lower resolution
($2\,R_\odot$ instead of $1\,R_\odot$).
A majority of rays is in a non-empty space
even with these approximations,
and the task is thus computationally substantially more demanding than before.
For our extensive dataset, we need
$3564$ synthetic images per iteration,
and the number of iterations is about $10^3$ to achieve a convergence.

Some improvements of the original Shellspec 
\citep{budaj2004,budaj2005,budaj2011b}
were also implemented in Fortran.
This includes a radial velocity field in disk
(added on top of the Keplerian field),
simple shadowing with prescribed scale height $H$,
which allows to switch on scattering in the disk atmosphere,
a variable step in the optical depth to prevent integration artefacts.
We modified priorities of overlapping objects
(jets priority is higher than nebulas, and envelopes is lower than nebulas).
There is a possibility to use two embedded grids,
with a lower resolution for extended structures
and a higher resolution in the centre.
We tested also asymmetric jets, or temperature gradients in shells.

Nevertheless, we shall recall all physical properties of our model. We assume
LTE level populations,
LTE ionisation equilibrium,
the line profile is determined by
thermal,
microturbulent,
natural,
Stark,
Van der Waals broadenings, and
the Doppler shift.
The continuum opacity is caused by
\ion{H}{i}~bound-free,
\ion{H}{i}~free-free,
${\rm H}^-$~bound-free,
${\rm H}^-$~free-free transitions,
the Thomson scattering on free electrons,
and the Rayleigh scattering on neutral hydrogen.
The scattering processes are implemented only for optically thin
environment (single scattering process),
with the shadowing mentioned above.
The scattering is non-isotropic and is described by the dipole phase function.
We also account for the line opacity of H$\alpha$, \ion{He}{i}, \ion{Si}{ii}, \ion{Ne}{i}, and \ion{C}{ii}.
Abundances are assumed to be either solar,
increased up to 3~times (0.5\,dex),
sub-solar (in C; Section~\ref{LOWC}),
or He-rich (Section~\ref{H0.4_He0.6}).
We use a small grid of synthetic spectra for the stars,
generated by Pyterpol \citep{jn2016} from Phoenix, BSTAR, and OSTAR grids \citep{husser2013,lanz2007,lanz2003}.
The stars are subject to the Roche geometry,
limb darkening and
gravity darkening (in particular the Roche-filling donor).

On the other hand, we do {\em not\/} include
optical irradiation of stars,
reflection (because the hot primary is mostly hidden in the disk),
Mie absorption on dust,
Mie scattering, or
dust thermal emission.
We consider these missing opacity sources negligible,
because temperatures in the system are too high for dust condensation.

As in the previous study, we use the quadratic ephemeris by \citet{ak2007}:
\begin{eqnarray}
T_{\rm min.I}({\rm HJD}) &=& \,2\,408\,247.968(15) + 12.913779(16) \cdot E \,+\nonumber \\
& + & 3.87265(369)\times10^{-6} \cdot E^2\,,\label{eq:ephemeris}
\end{eqnarray}
which corresponds to the primary minimum of the optical light curve
and in our particular case of $\beta$~Lyr~A the donor (secondary)
is {\em behind\/} the gainer (primary; hidden in its opaque disk).
Initial conditions for further convergence generally correspond
to our previous model based on the optically thick medium
\citep{Mourard_etal_2018A&A...618A.112M},
although this model did not produce sufficient emission in lines.

\begin{figure*}
\centering
\includegraphics[width=18cm]{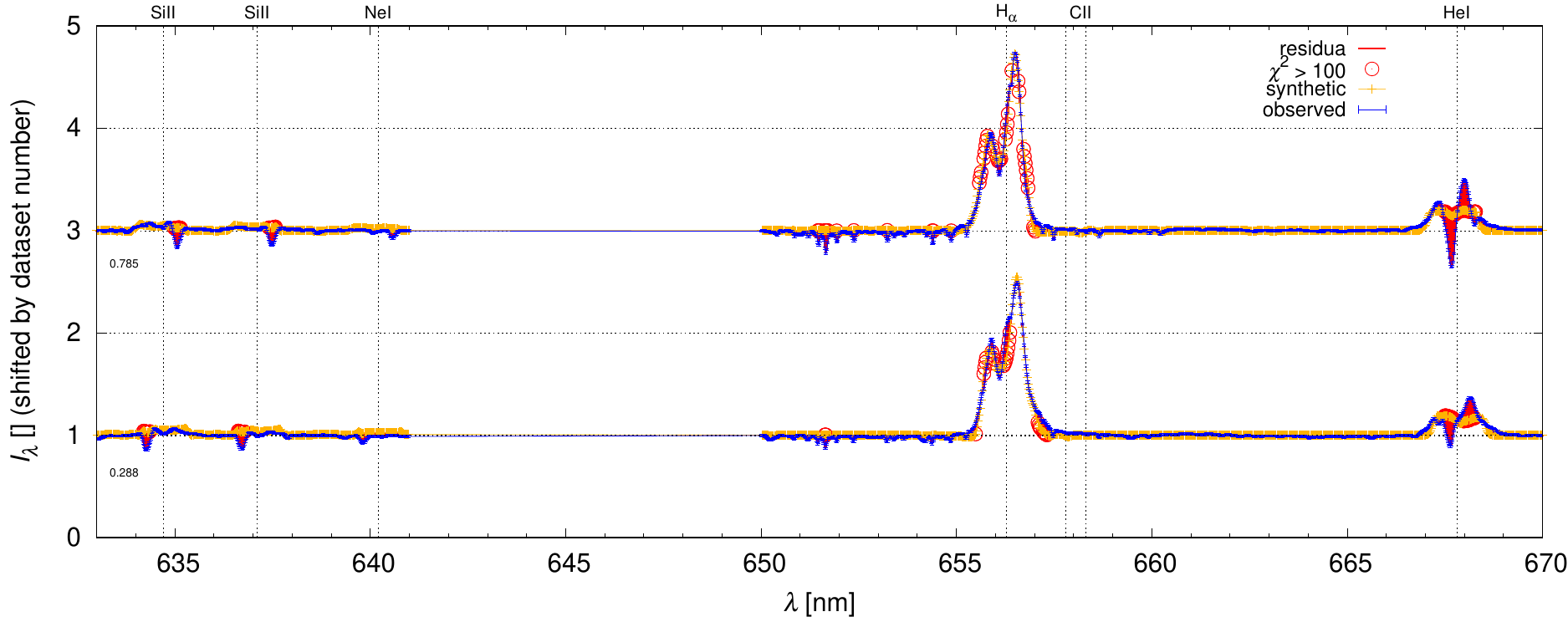}
\caption{
Normalized spectra of $\beta$~Lyr~A
for two out-of-eclipse phases (0.288 and 0.785).
There are observed data with uncertainties indicated (\color{blue}blue\color{black}),
as well as synthetic data (\color{yellow}yellow\color{black}),
residua (\color{red}red\color{black}),
and large $\chi^2$ contributions (red circles).
This 1st model was only fitted to these 2~spectra.
The H$\alpha$ emission profile and its EW is well described,
but there are systematic differences for the \ion{He}{i} 6678 line;
synthetic \ion{Si}{ii} 6347, 6371 and \ion{Ne}{i} 6402 lines have low EW.
}
\label{fitting_shell2_vsh50_SBPLX__28_chi2_SPE}
\end{figure*}

\paragraph{Parameter relations.}

For an easier interpretation of results,
we review some of the parameter relations
(i.e., the geometrical constraints),
as they are implemented in the current version of Shellspec.

The disk (a.k.a. nebula) object is described in cylindrical coordinates $(R,z)$ (see also Tab.~\ref{tab1b} for basic lenght scales):
\begin{equation}
H(R) = h_{\rm cnb} \sqrt{\frac{\gamma k_{\rm B}T}{\mu m_{\rm u}}} \frac{1}{\Omega_{\rm k}}\,,
\end{equation}
\begin{equation}
\Sigma(R) = \Sigma_{\rm nb}\left(\frac{R}{R_{\rm innb}}\right)^{e_{\rm densnb}}\!,
\end{equation}
\begin{equation}
\rho(R,0) = \frac{\Sigma}{\sqrt{2\pi}H}\,,
\end{equation}
\begin{equation}
\rho(R,z) = \rho(R,0)\exp\left[-\min\left(\frac{z^2}{2H^2}; \frac{h_{\rm windnb}^2}{2}\right)\right]\,,
\end{equation}
\begin{equation}
T(R,0) = T_{\rm nb}\left(\frac{R}{R_{\rm innb}}\right)^{e_{\rm tmpnb}}\!,
\end{equation}
\begin{equation}
T(R,z) = T(R,0)\max\left(1; 1+(t_{\rm invnb}\!-\!1)\frac{|z|-h_{\rm invnb}H}{a_{\rm neb}H-h_{\rm invnb}H}\right)\,,
\end{equation}
\begin{equation}
v_r(R) = {\cal H}(|z|-h_{\rm velnb} H)\, v_{\rm nb}\left(1-\frac{R_{\rm innb}}{R}\right)^{e_{\rm velnb}}\!,
\end{equation}
\begin{equation}
v_\phi(R) = \sqrt{\frac{GM_\star}{R}}\,,
\end{equation}
where
$H$~is the scale height,
$\gamma$~the adiabatic exponent,
$k_{\rm B}$~the Boltzmann constant,
$\mu$~the mean molecular weight,
$m_{\rm u}$~the atomic mass unit,
$G$~the gravitational constant,
$\Omega_{\rm k} = v_\phi /R$ the Keplerian angular velocity,
$\Sigma$~surface density,
$\rho$~volumetric density,
$T$~temperature,
$v_r$~radial velocity, 
$v_\phi$~azimuthal velocity;
${\cal H}(x)$ denotes the Heaviside step function.

The jet has a shape of a double cone with the opening angle~$a_{\rm jet}$
and is described in spherical coordinates $(R,\theta)$:
\begin{equation}
\rho(R) = \rho_{\rm jt} \left(\frac{R_{\rm injt}}{R}\right)^2 \frac{v_r(R_{\rm injt})}{v_r(R)}\,(1\pm a_{\rm symjt})\,,\label{rhojt}
\end{equation}
\begin{equation}
T(R) = T_{\rm jt}\left(\frac{R}{R_{\rm injt}}\right)^{e_{\rm tmpjt}}\!,
\end{equation}
\begin{equation}
v_r(R) = v_{\rm jt}\left(1-\frac{R_{\rm cjt}}{R}\right)^{e_{\rm veljt}}\!.
\end{equation}
In our case, the base plane corresponds to the orbital plane
and the cone position is determined by the radial offset~$R_{\rm poljt}$
and the polar angle~$\alpha_{\rm jet}$.

Similarly, the shell is also described in spherical coordinates:
\begin{equation}
\rho(R) = \rho_{\rm sh} \left(\frac{R_{\rm insh}}{R}\right)^2 \frac{v_r(R_{\rm insh})}{v_r(R)}\,,
\end{equation}
\begin{equation}
T(R) = T_{\rm sh} \left({R\over R_{\rm insh}}\right)^{e_{tmpsh}}\!,
\end{equation}
\begin{equation}
v_r(R) = v_{\rm sh}\left(1-\frac{R_{\rm csh}}{R}\right)^{e_{\rm velsh}}\!,\label{vrsh}
\end{equation}
We assume the spherical shell is centered on the primary
and encompasses also other objects. It represents circumstellar
matter which escaped farther away from the binary.
All remaining parameters are explained in the caption of Tab.~\ref{tab1}.


\subsection{Doppler tomography}

In order to create model spectra with enough emission, we started with two Ond\v{r}ejov spectra
taken at 0.288 and 0.785 phases, i.e., out of eclipses, and converged our new
model with additional optically-thin objects. This simplified Doppler tomography was carried out to verify that the model is indeed capable 
to fit the \ha profile.
There is always a question which objects should be included in the model
and which shouldn't. If the model were too simplistic, the objects would be
distorted; if it were too complex, the objects could be unconstrained.
After some preliminary tests we used 5~objects
(primary, secondary, disk, jet, shell)
out of~8
(spot, envelope, flow).%
\footnote{A flow is presumably a relatively small structure which can
overlap with a jet or a spot.
An envelope is co-rotating with the binary and does not have
a radially-expanding velocity field;
we verified that even a Roche-filling (L2) envelope does not
create enough emission.
A spot is tested later as an alternative model
(in Section~\ref{SPOT}).}

Given the observed \ha profiles and their overall width,
our model should include a large positive velocity with respect
to the line of sight, or gradient of~$v$.
Because \ion{H}{i}, \ion{Si}{ii}, as well as \ion{He}{i} and \ion{Ne}{i} are excited,
we expect both low and high temperatures, or gradient of~$T$
in the circumstellar medium.
Apart from absorption lines arising in stellar atmospheres,
the model is capable of creating a P~Cygni profile due to winds,
either in a disk atmosphere or in a surrounding shell.
Alternatively, line profiles may be formed by overlapping
velocity fields, in accord with the priorities of objects.
We have to look for suitable net velocities of whole objects,
but also for turbulent velocities,
which significantly affect the optical depth along the line of sight;
one should converge both at the same time.

Results of the first two-spectra model are shown in Figure~\ref{fitting_shell2_vsh50_SBPLX__28_chi2_SPE}.
The model easily created enough emission
and the EW of H$\alpha$ is fitted very well.
There are relatively minor systematics in the H$\alpha$ profile,
but major systematics can be seen for other spectral lines, especially \ion{He}{i} 6678.
The \ion{Si}{ii} 6347, 6371 and \ion{Ne}{i} 6402 model lines have lower EW and depth than the observed ones,
because there is a tension between the overall emission and the respective absorption.
The region between 6500 and 6550\,\AA\  contains telluric lines
which are {\em not\/} included in our model,
but they should not affect the convergence in a negative way.
Although we varied the chemical composition,
there might be non-LTE effects (for \ion{He}{i})
or some unaccounted temperature gradients.

\begin{figure}
\includegraphics[width=9cm]{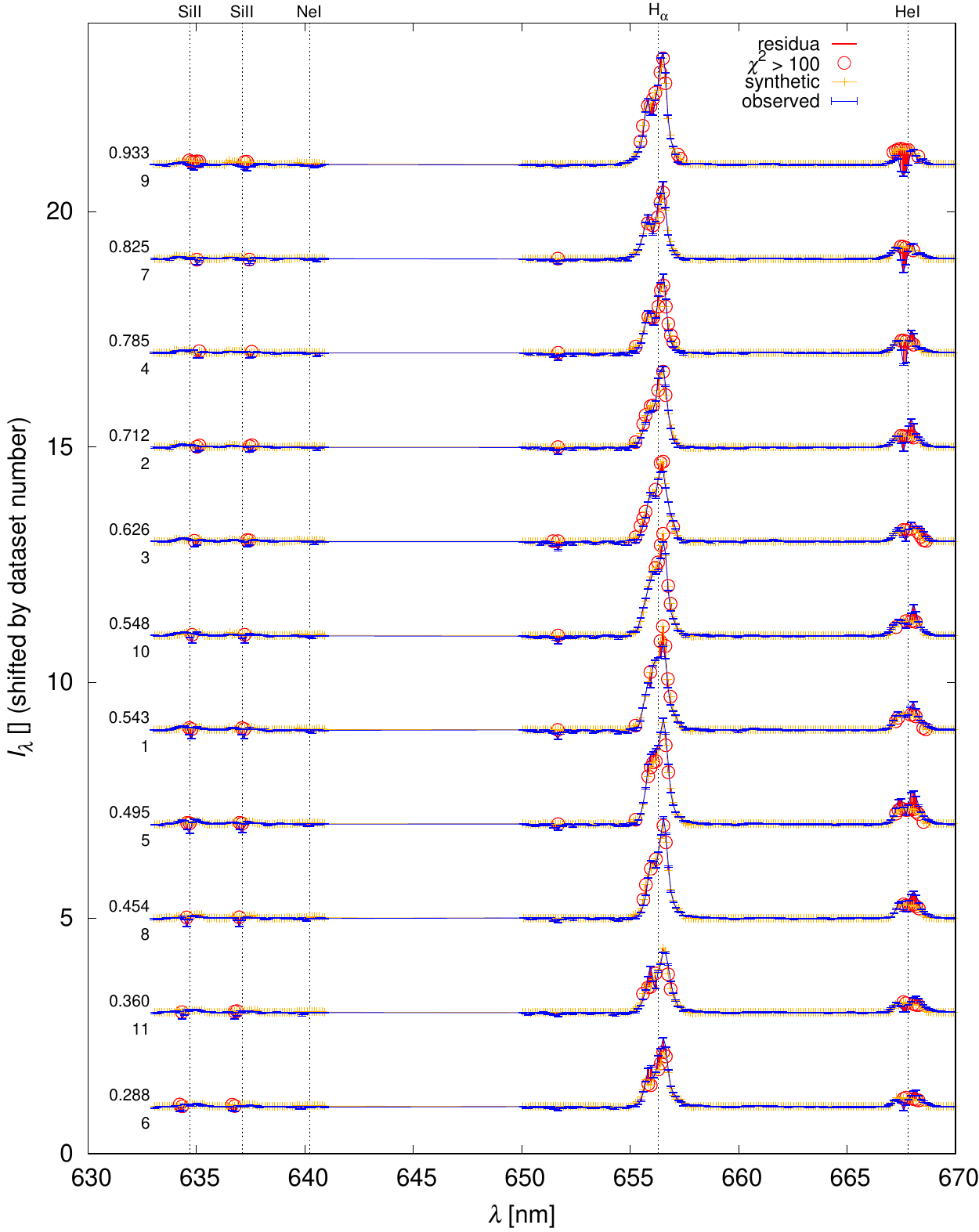}
\caption{
Normalized spectra
for 11 phases covering the whole light curve,
including primary and secondary eclipses.
Colours in this Fig. as well as in others have the same meaning
as in Fig.~\ref{fitting_shell2_vsh50_SBPLX__28_chi2_SPE}.
The synthetic H$\alpha$ profiles for the 2nd model exhibit
a variability similar to the observed ones, although at several
phases there are systematic differences (both $+$ and~$-$).
}
\label{fitting_shell2_vsh50_ONDREJOV6670__41_chi2_SPE}
\end{figure}

A more representative set of 11 spectra covering a~representative range of orbital phases was fitted in the second step. The results are shown in Figure~\ref{fitting_shell2_vsh50_ONDREJOV6670__41_chi2_SPE}.
All optically-thin objects (disk atmosphere, jet, shell)
contribute substantially to the H$\alpha$ emission flux.
Because the spectra are normalized to the continuum flux,
which is larger outside eclipses,
the emission appears to vary in strength twice each orbit.
The synthetic \ha profiles for this 2nd model exhibit
a variability similar to the observed ones, although at several
phases there are systematic differences (both positive and negative).
Comments related to \ion{He}{i}, \ion{Si}{ii}, \ion{Ne}{i} lines remain essentially the same.

It might seem easy to improve the fit further,
but it is {\em not\/} the case for a geometrically constrained model,
where all parameters have either geometrical or physical limits.
We are practically sure the convergence works all right and
it is not a matter of one local minimum of~$\chi^2$.
In order to improve the fit, it may be inevitable to relax some
of our assumptions, e.g.,
the axial symmetry of the disk,
the vertical symmetry of the jets, or
the radial symmetry of the shell.
We also do not account for any intrinsic variability of the source.
However, we prefer to keep our model as simple as possible,
at least at this stage.

\subsection{Differential interferometry}

\begin{figure}
\centering
\includegraphics[width=9cm]{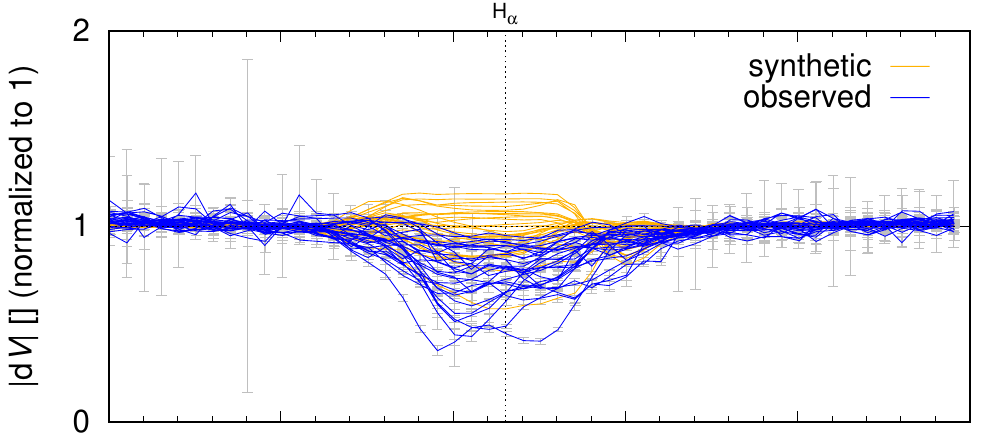}
\includegraphics[width=9cm]{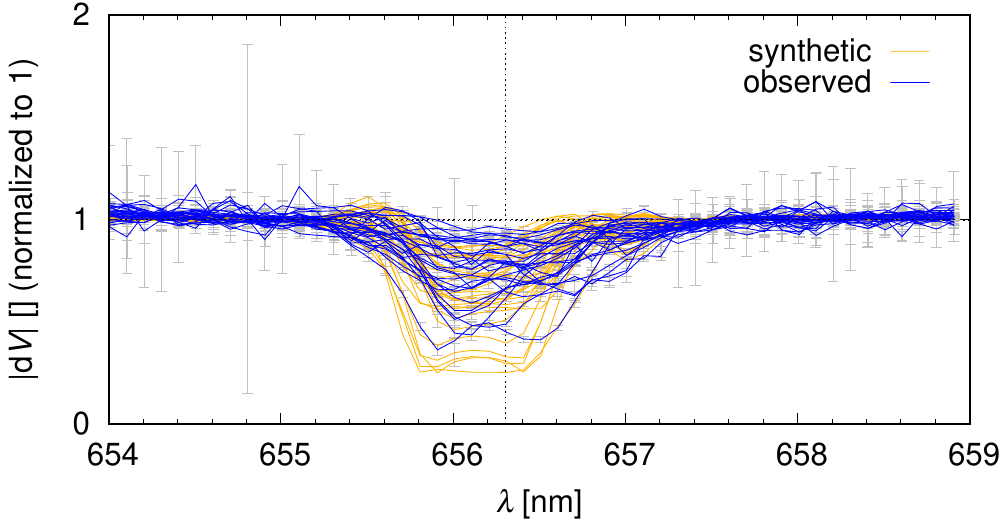}
\caption{
Observed differential visibility amplitude $|{\rm d}V|$ versus wavelength $\lambda$,
normalized to 1 in the continuum (blue),
and its decrease across the H$\alpha$ profile.
Uncertainties of $|{\rm d}V|$ are also plotted (gray).
Synthetic visibilities (yellow) are shown for the two 'extreme'
values of the shell outer radius $R_{\rm outsh} = 40\,R_\odot$ (top)
and $120\,R_\odot$ (bottom).
}
\label{fitting_shell5_ROUTSH040}
\end{figure}

Interestingly, all interferometric data indicate a decrease
of the visibility amplitude $|{\rm d}V|$
when scanning across the H$\alpha$ profile
(see Figure~\ref{fitting_shell5_ROUTSH040})
and it seems to be almost independent on baseline length and orientation
(Figure~\ref{fitting_shell8_20200415__20_uv2}). Such a general finding means that the core of the H$\alpha$ emitting region is clearly resolved for all baseline lengths between 50 and 200\,m.
Moreover, the respective velocities must be large enough
to occur in the wings of H$\alpha$.

Only an extended symmetric shell may cause $|{\rm d}V|$ to decrease. 
On contrary, a disk (nebula) or jets emit usually from small (hot) areas,
and they are both asymmetric, which would force $|{\rm d}V|$ to increase,
at least for the shortest baselines. This is not observed.
Our preliminary tests thus demonstrate the need to include the shell
in our model.

Regarding the differential phase $\arg{\rm d}V$ measurements,
it should be noted that some phase wrappings ($\pm360^\circ$) are present in the data
(cf., e.g., datasets 1, 11, 12),
which should not be a problem as we account for them in the model.
The differential phases are obtained on three baselines, two of them (E1E2 and E2W2) being oriented almost perpendicular to the orbital plane, whereas the third one (W1W2) is very close in orientation to the orbital plane. Interestingly the worst fits to the model are obtained systematically for this last orientation. This general finding is in agreement with the fact that the differential phases in H$\alpha$ are dominated by the jets but we should also conclude that our geometrically\discretionary{-}{-}{-}constrained model is not flexible enough to explain all the observed features.


\subsection{A joint 'compromise' model}

Our complete dataset is very heterogeneous.
On one hand, this is an advantage which allows us to construct
a very robust model of $\beta$~Lyr~A when everything is fitted together.
On the other hand, when some types of measurements exhibit systematics,
as mentioned above, the $\chi^2$ contributions of the joint model will be worse
while observation-specific models would be better (see Sec.~\ref{sec:obsspe}).

\begin{figure}
\includegraphics[width=9cm]{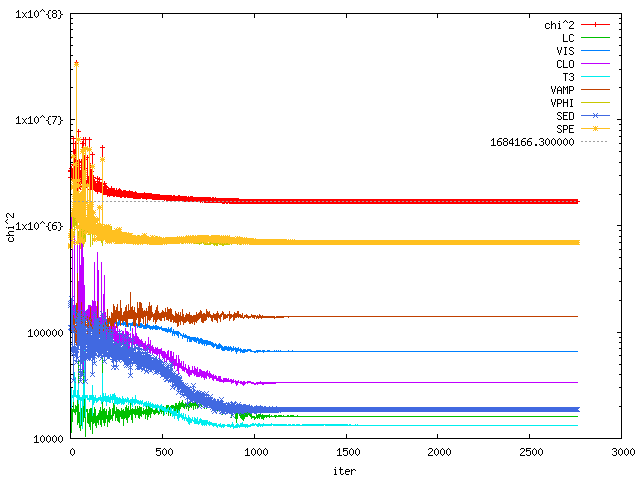}
\caption{
The $\chi^2$ convergence (red) for the joint model;
individual contributions (LC, VIS, CLO, T3, SED, SPE, VAMP, VPHI) are also indicated.
The model successfully converges to a local minimum.
Some datasets have a substantially larger number of observations,
i.e., effectively a larger weight.
The $\chi^2$ values are different from Tab.~\ref{tab1},
because the model was re-converged several times
and uncertainties of some datasets were modified.
}
\label{fitting_shell2_vsh50_20191216__37_chi2_iter}
\end{figure}

After 2761 iterations (Figure~\ref{fitting_shell2_vsh50_20191216__37_chi2_iter}),
we obtained a model with reduced $\chi^2_{\rm R} = \chi^2/N$ values
which are summarized in Table~\ref{tab1} (1st column).
Previously used datasets are fitted only slightly worse
than in \cite{Mourard_etal_2018A&A...618A.112M}, with
$\chi^2_{\rm lc}   = 3.1$,
$\chi^2_{\rm vis}  = 4.0$,
$\chi^2_{\rm clo}  = 3.4$,
$\chi^2_{\rm t3}   = 5.6$;
the new datasets resulted in
$\chi^2_{\rm sed}  = 4.7$,
$\chi^2_{\rm spe}  = 44.1$,
$\chi^2_{\rm vamp} = 4.5$,
$\chi^2_{\rm vphi} = 44.0$,
and the total
$\chi^2_{\rm R}    = 17.0$.


We have to explain why some contributions are large.
For $\chi^2_{\rm sed}$,
there is a difference in NUV below the Balmer jump,
which is caused by a relatively low number of measurements
compared to other datasets and thus a low weight.
Lines are not fitted, because the flux is computed at monochromatic wavelengths;
we would have to use about 5 times higher wavelength resolution
and perform a convolution with the instrumental profile of \cite{burnashev78}.
Nevertheless, the line-to-continuum flux ratio is fully described by our SPE dataset.
In case of $\chi^2_{\rm spe}$,
there are systematic differences at certain phases (0.288, 0.825),
although others are good fits (0.454, 0.548, 0.712).
A substantial contribution arises from high-temperature \ion{He}{i} and \ion{Ne}{i} lines
and also from telluric lines which are not fitted by our model.
For $\chi^2_{\rm vamp}$,
synthetic $|{\rm d}V|$ sometimes exhibit a narrower decrease (cf. dataset 1),
or a peak in the middle of H$\alpha$ (2, 3),
although others are almost perfect fits (4, 5, 6, \dots).
Finally, $\chi^2_{\rm vphi}$ is substantially increased because
synthetic $\arg{\rm d}V$ are sometimes smoother (4, 6),
there are possibly remaining phase slips (1, 11, 12),
or mirroring of phases (2, 3);
these numerous measurements have relatively high weight.

\begin{figure}
\includegraphics[width=9cm]{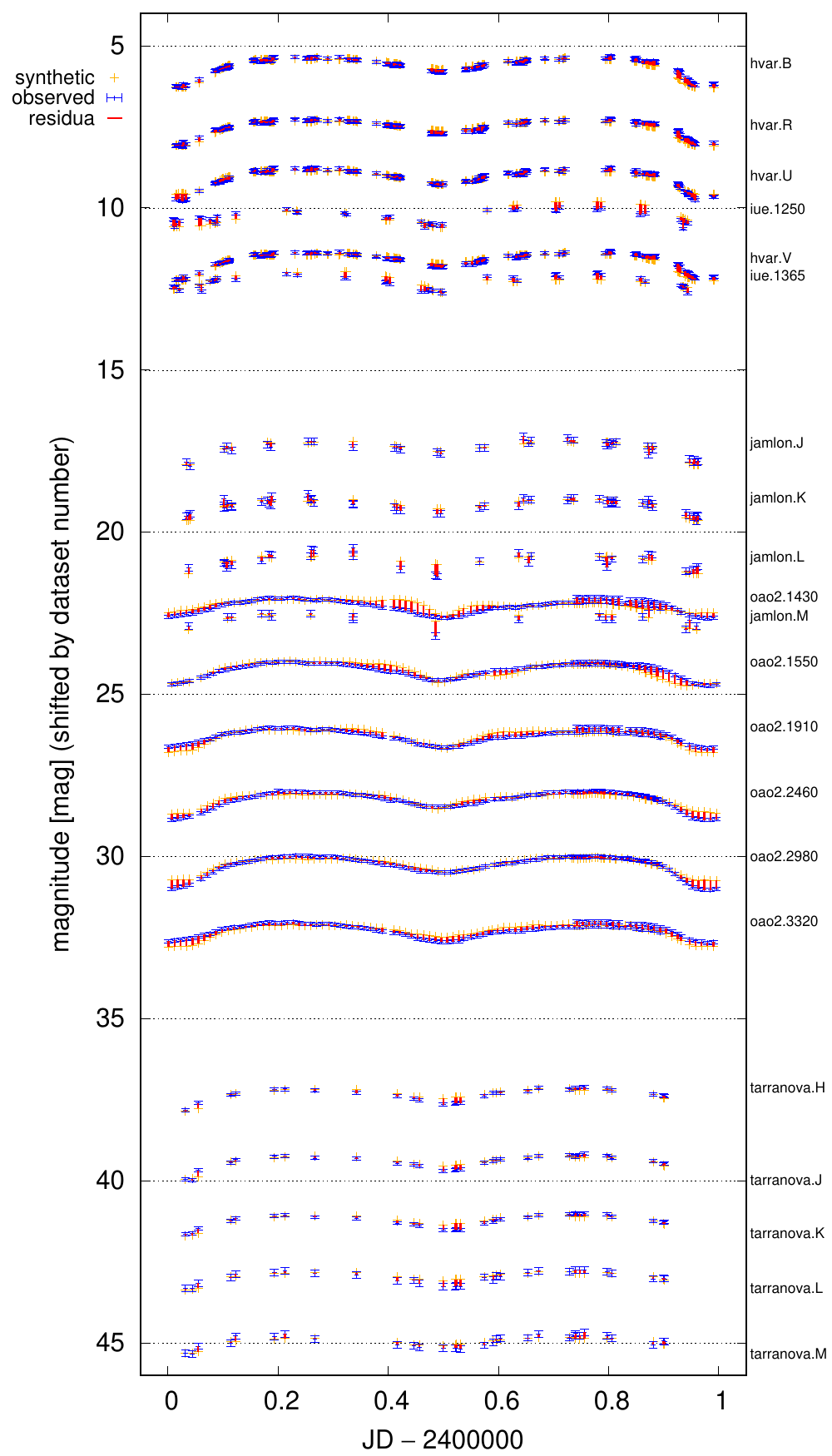}
\caption{
Comparison of observed and synthetic phased light curves from FIR to FUV,
computed for the joint ('compromise') model.
The names of datasets are show in the right column.
The light curves were arbitrarily shifted in the vertical direction.
}
\label{fitting_shell8_20200415__20_chi2_LC_PHASE}
\end{figure}

\begin{figure}
\includegraphics[width=9cm]{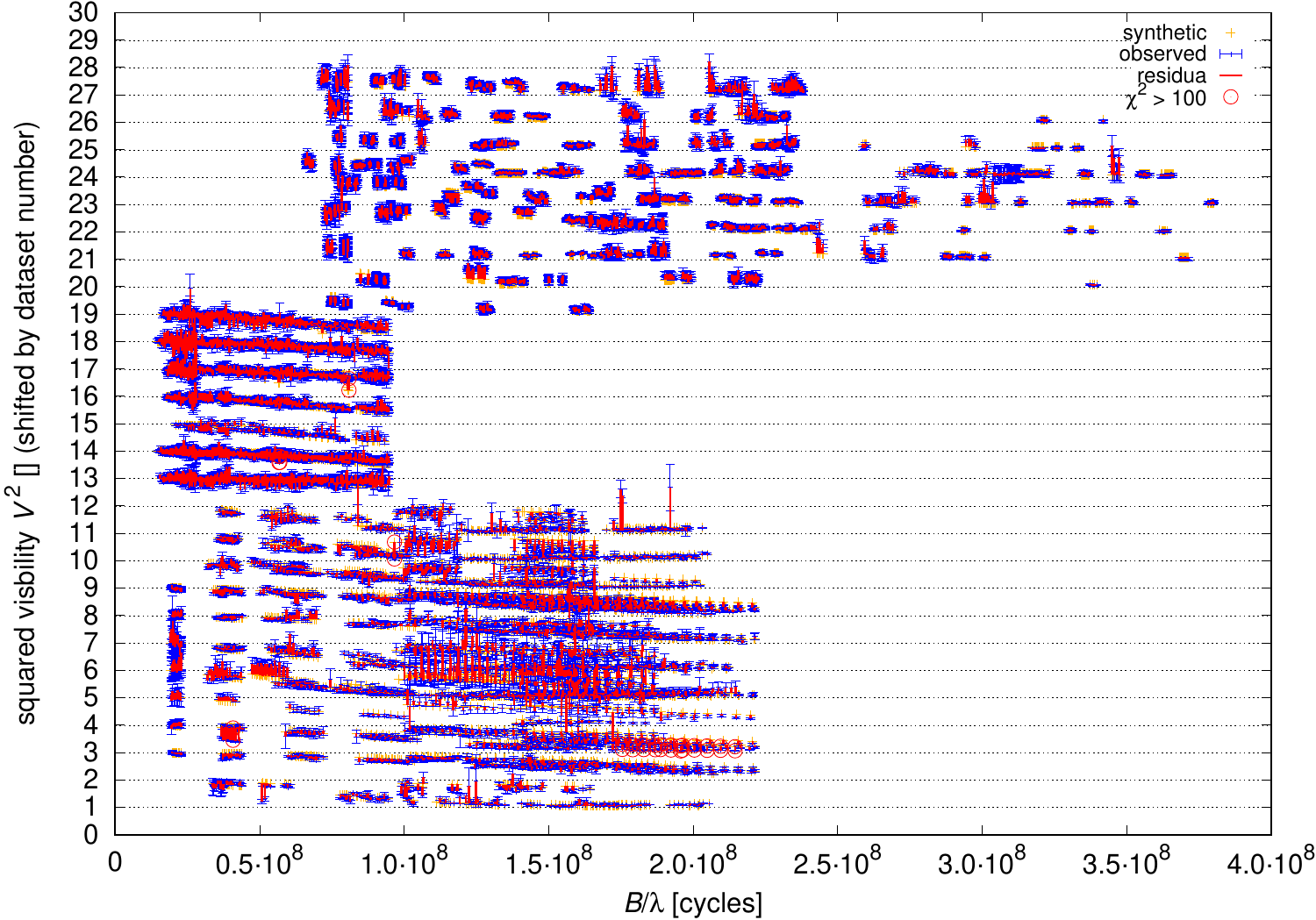}
\caption{
Squared visibility $|V^2|$ (shifted by dataset number)
for different baselines $B/\lambda$ (in cycles),
The are CHARA/MIRC observations at the bottom, NPOI in the middle, and CHARA/VEGA at the top (blue);
synthetic data for the joint model are plotted for comparison (yellow).
Overall trends ($|V^2|(B/\lambda)$) seem to be correctly described,
although there are some systematics for small groups of data.
}
\label{fitting_shell8_20200415__20_chi2_VIS}
\end{figure}

\begin{figure}
\includegraphics[width=9cm]{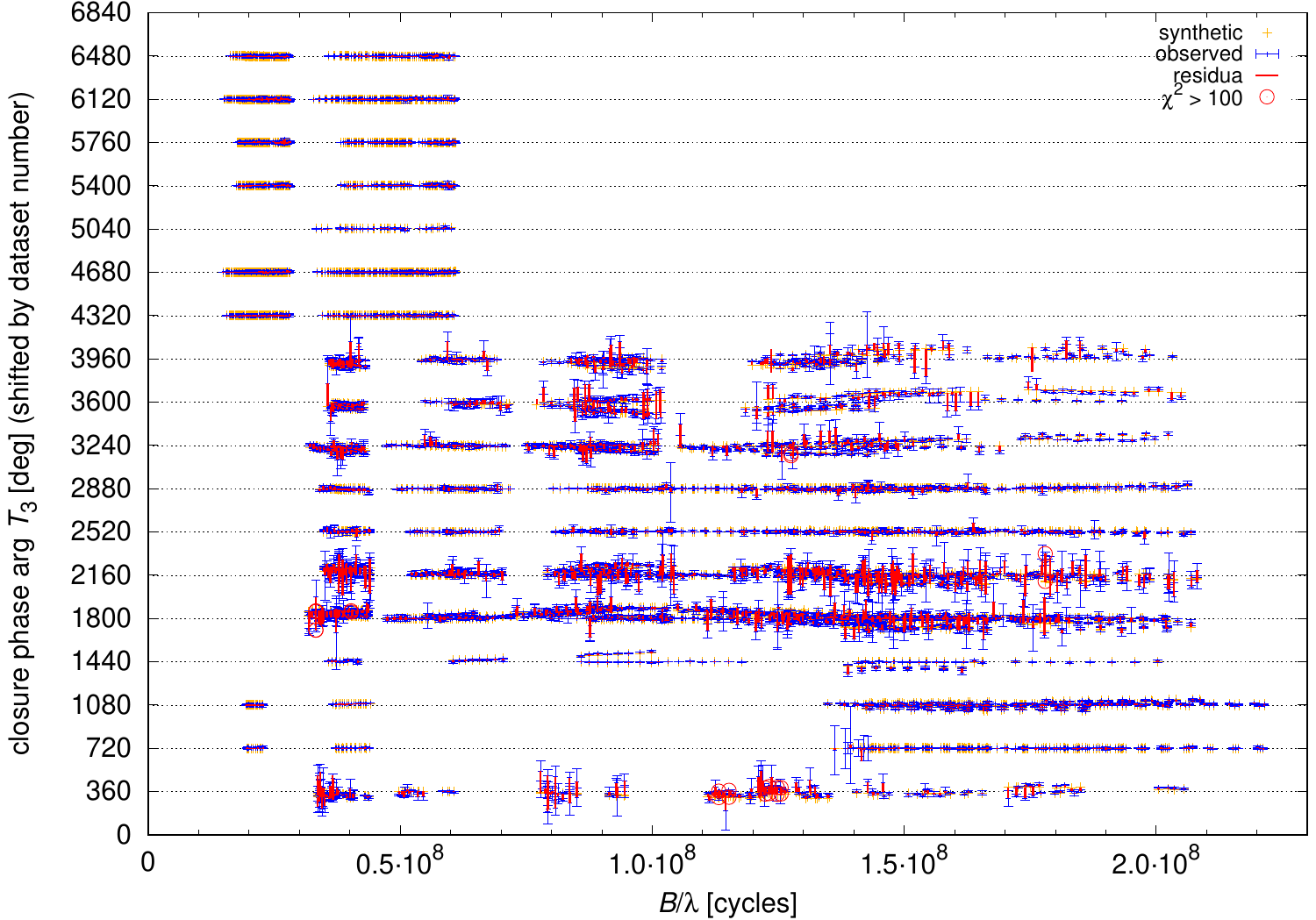}
\caption{
Similar as Fig.~\ref{fitting_shell8_20200415__20_chi2_VIS},
for the closure phase $\arg T_3$ versus $B/\lambda$.
}
\label{fitting_shell8_20200415__20_chi2_CLO}
\end{figure}

\begin{figure}
\includegraphics[width=9cm]{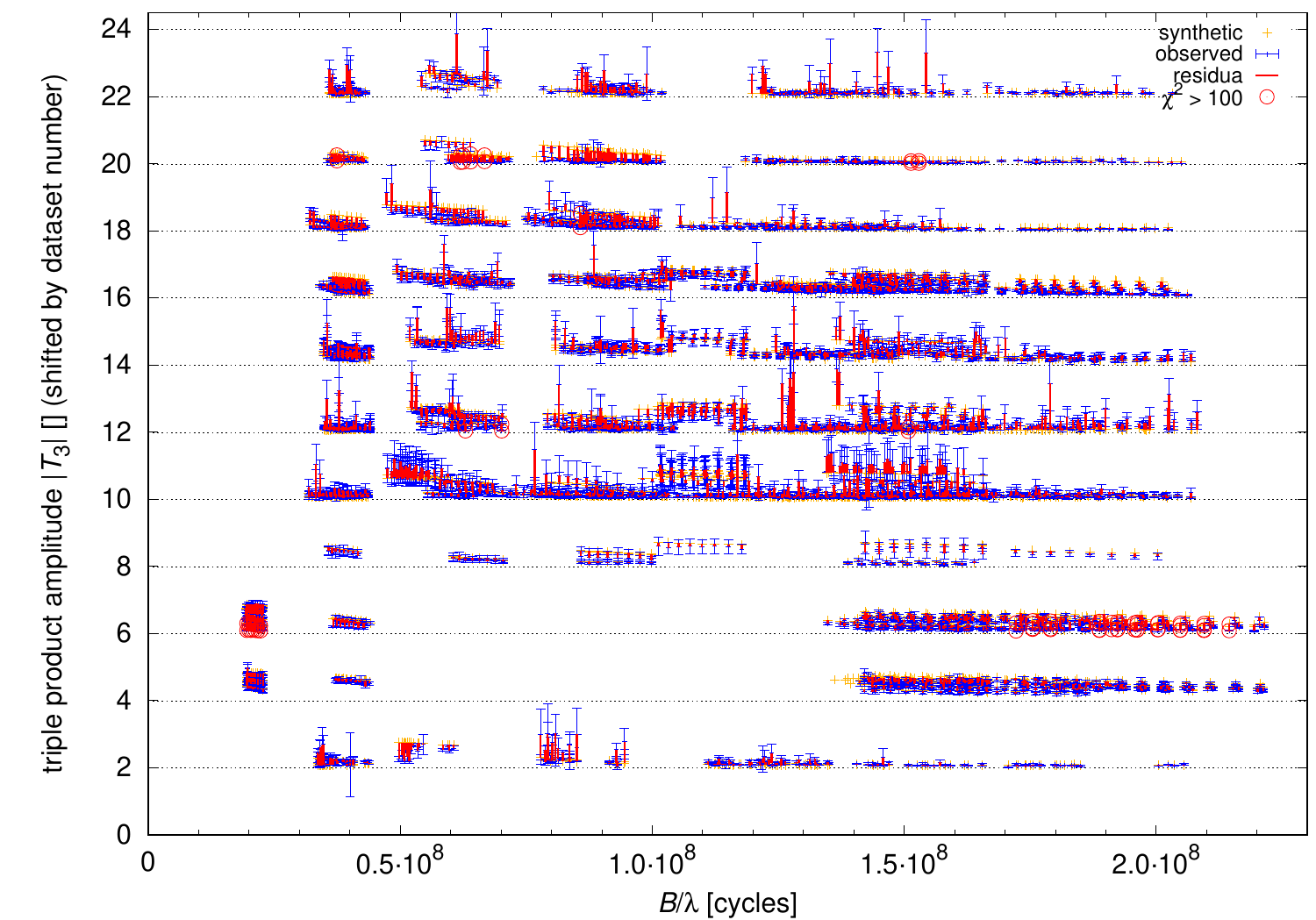}
\caption{
Similar as Fig.~\ref{fitting_shell8_20200415__20_chi2_VIS},
for the triple product $|T_3|$ versus $B/\lambda$.
}
\label{fitting_shell8_20200415__20_chi2_T3}
\end{figure}

\begin{figure}
\includegraphics[width=9cm]{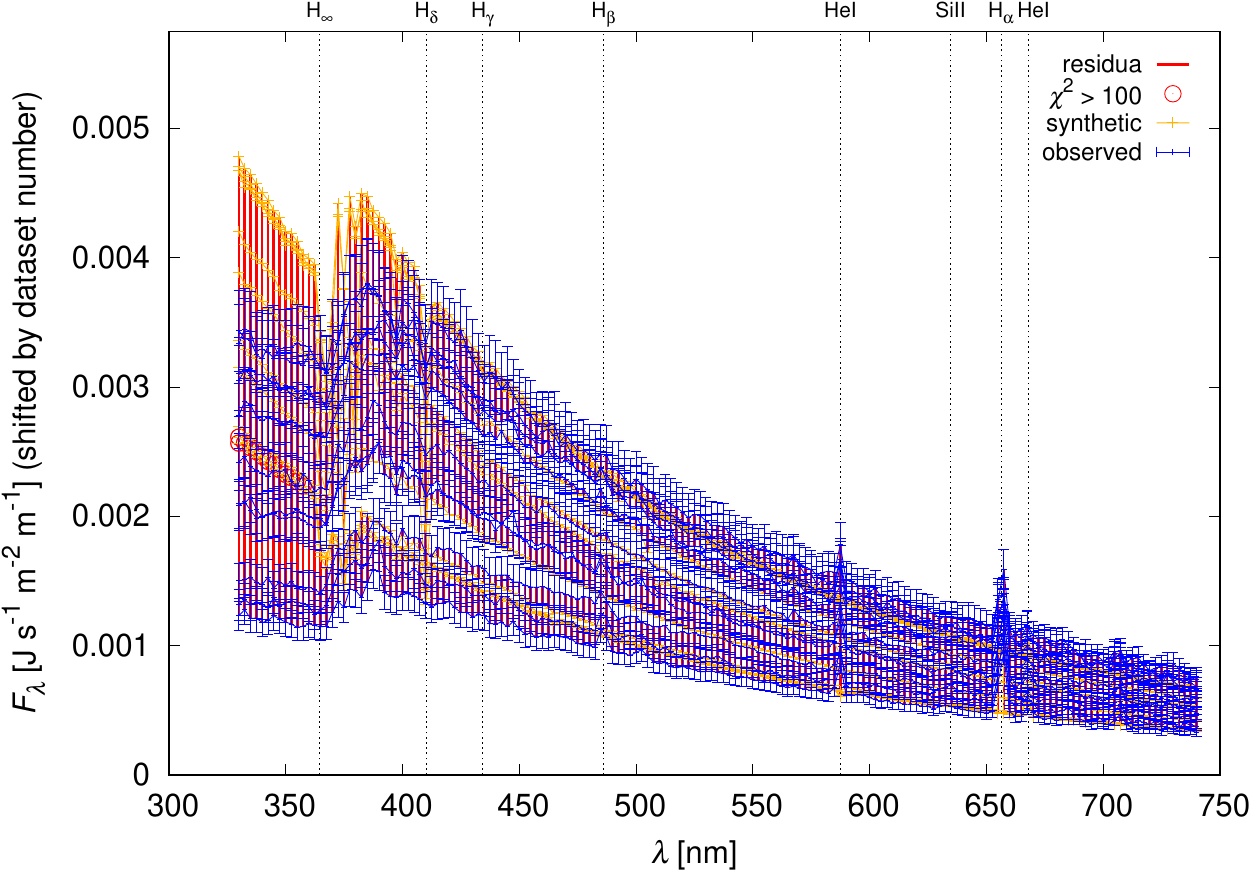}
\caption{
Spectral-energy distribution (SED) overplotted for 10 phases.
Synthetic SEDs systematically differ from observations,
especially in NUV and in the vicinity of the Balmer jump.
This is not the best model in terms of $\chi^2_{\rm sed}$,
but cf.~Fig.~\ref{fitting_shell9_20200415_SED_chi2_SED}.
}
\label{fitting_shell8_20200415__20_chi2_SED}
\end{figure}

\begin{figure}
\includegraphics[width=9cm]{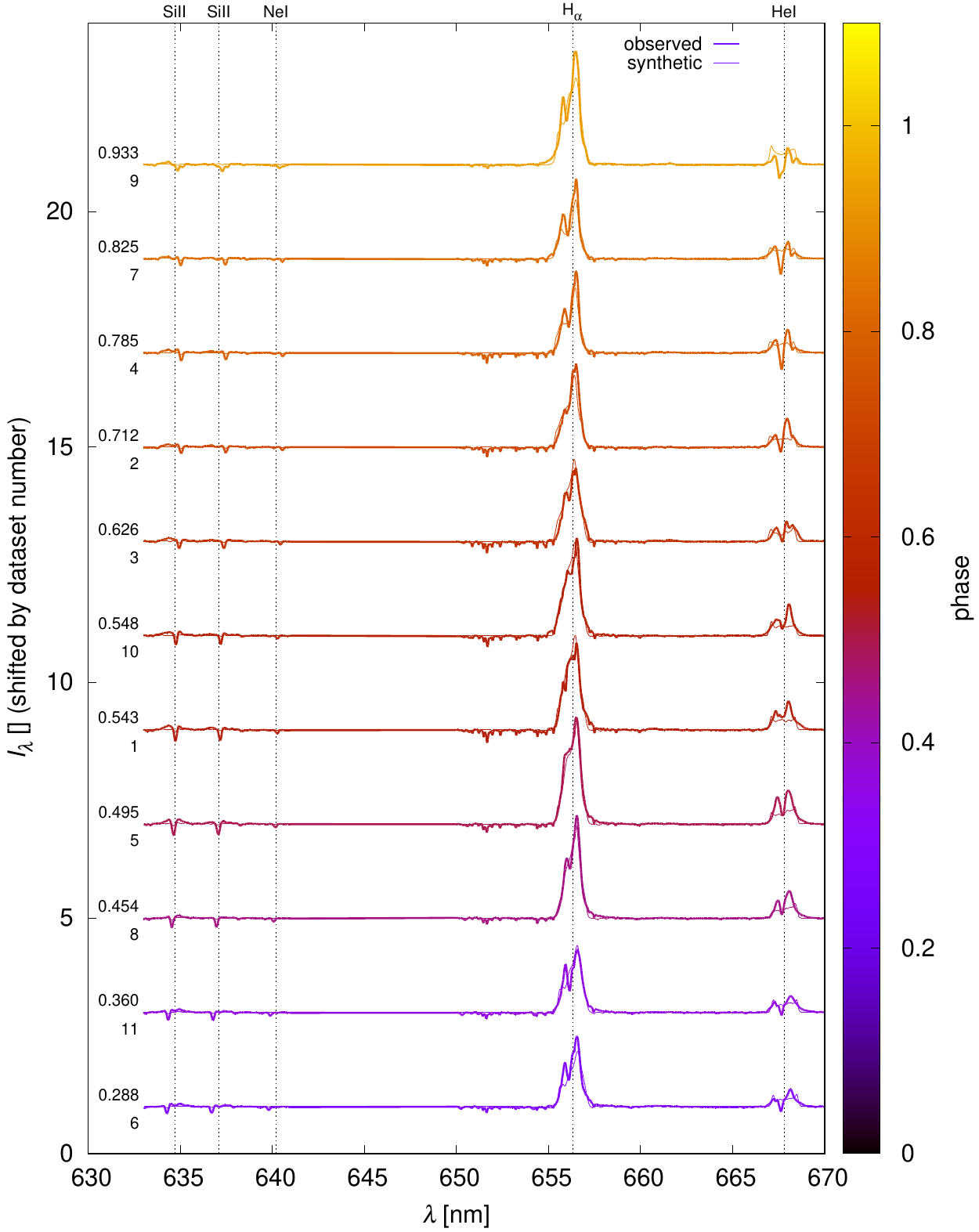}
\caption{
Normalized spectra $I(\lambda)$ for 11 phases.
Both observed (thick) and synthetic (thin) spectra are plotted for comparison.
Colours correspond to the phases (see also the values on the left).
Uncertainties are not plotted, but they are of the order of 0.01.
The vertical lines correspond to (from left to right):
\ion{Si}{ii} 6347, \ion{Si}{ii} 6371, \ion{Ne}{i} 6402, H$\alpha$, \ion{He}{i} 6678 wavelengths.
}
\label{fitting_shell8_20200415__20_chi2_SPE_PHASE}
\end{figure}

\begin{figure}
\includegraphics[width=9cm]{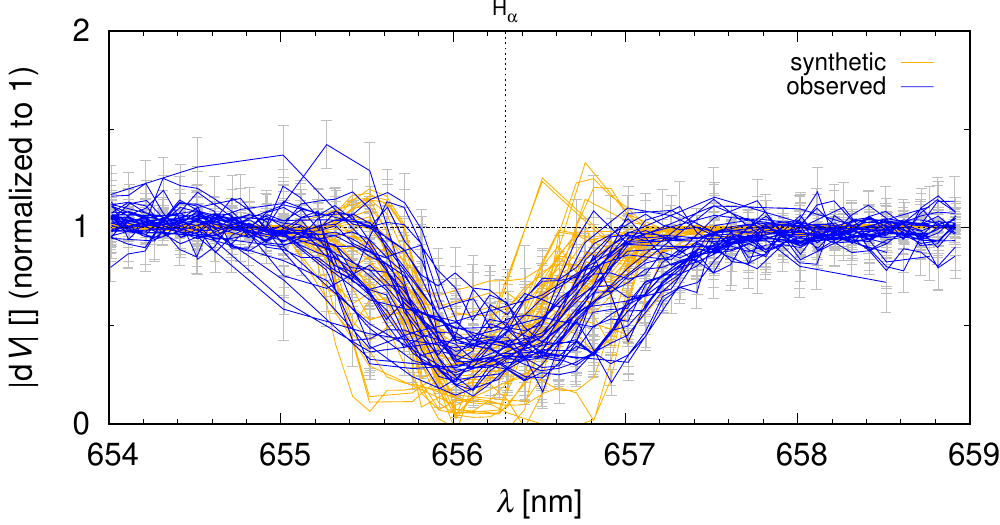}
\caption{
Differential visibilities $|{\rm d}V|$ vs $\lambda$
normalized in continuum.
Synthetic (yellow) visibilities for the joint model
exhibit a similar decrease across the H$\alpha$ profile
as the observed ones (blue). However, some of
the synthetic $|{\rm d}V|$'s are systematically
lower in the blue wing and higher in the red wing.
}
\label{fitting_shell8_20200415__20_chi2_VAMP_null}
\end{figure}

\begin{figure}
\includegraphics[width=9cm]{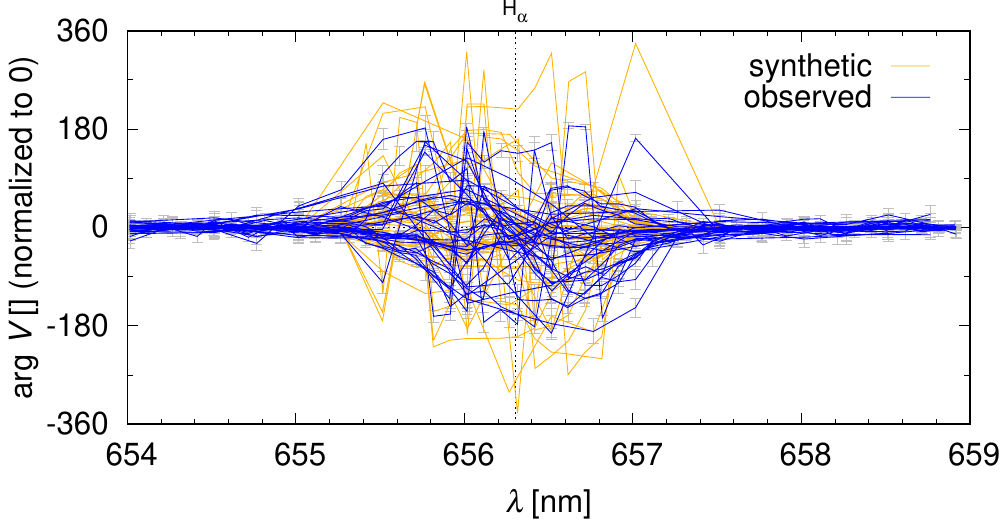}
\caption{
Similar as Fig.~\ref{fitting_shell8_20200415__20_chi2_VAMP_null},
for the differential phases $\arg{\rm d}V$ versus $\lambda$.
On average, phase changes are comparable in both observed
and synthetic data.
There are some remaining phase slips which are also fitted for.
}
\label{fitting_shell8_20200415__20_chi2_VPHI_null}
\end{figure}

\begin{figure}[h]
\centering
\includegraphics[width=7cm]{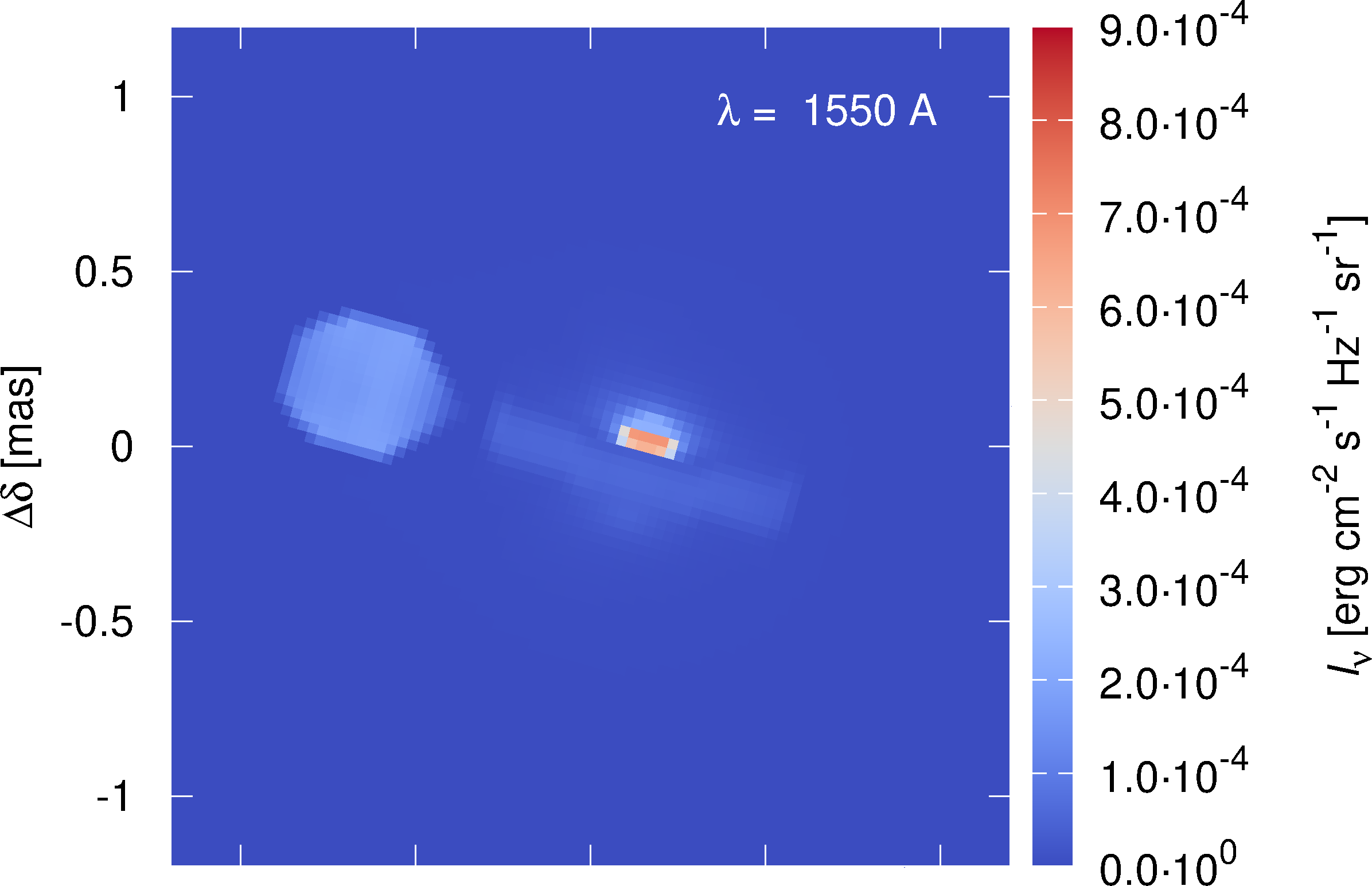}
\includegraphics[width=7cm]{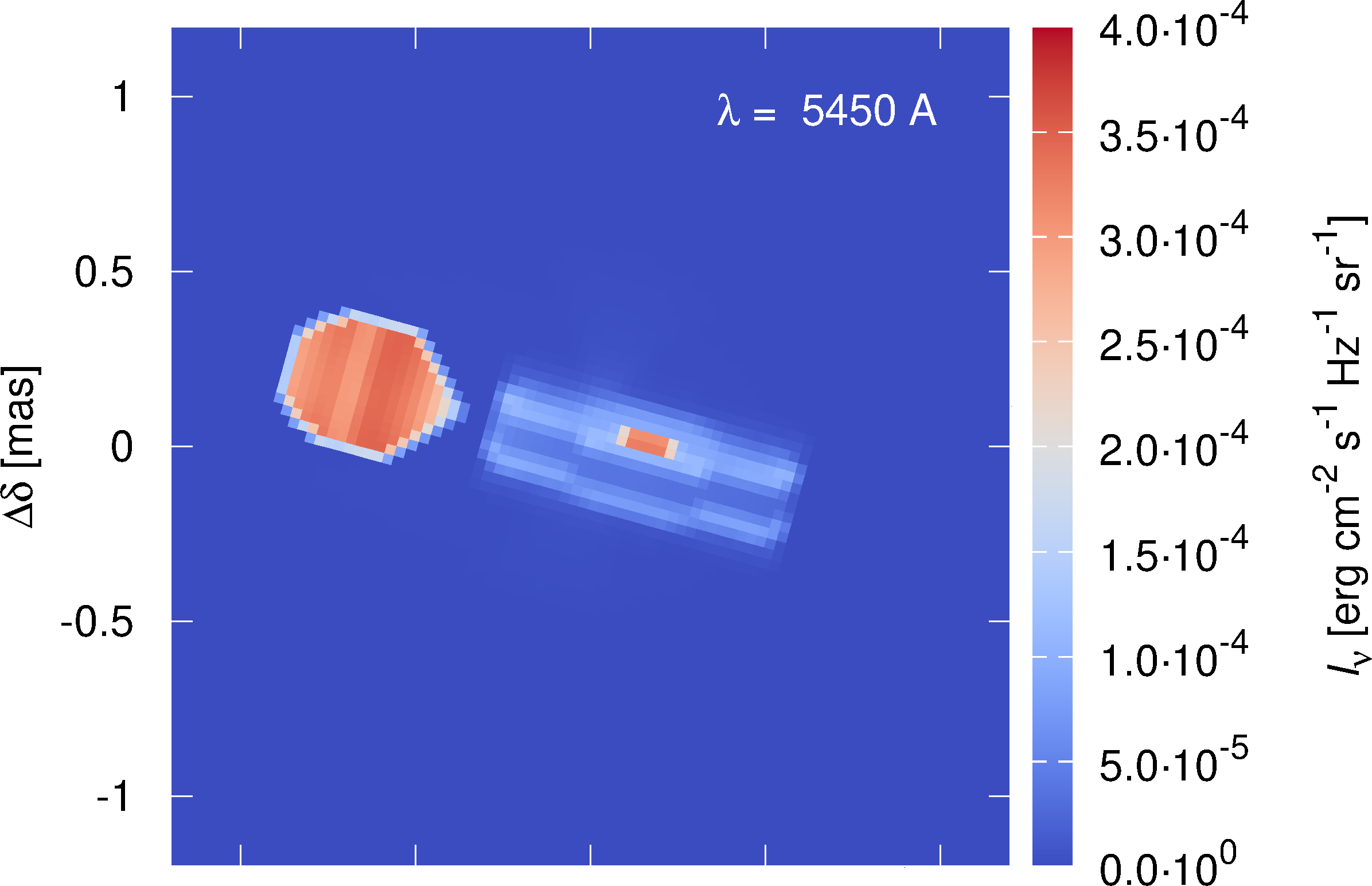}
\includegraphics[width=7cm]{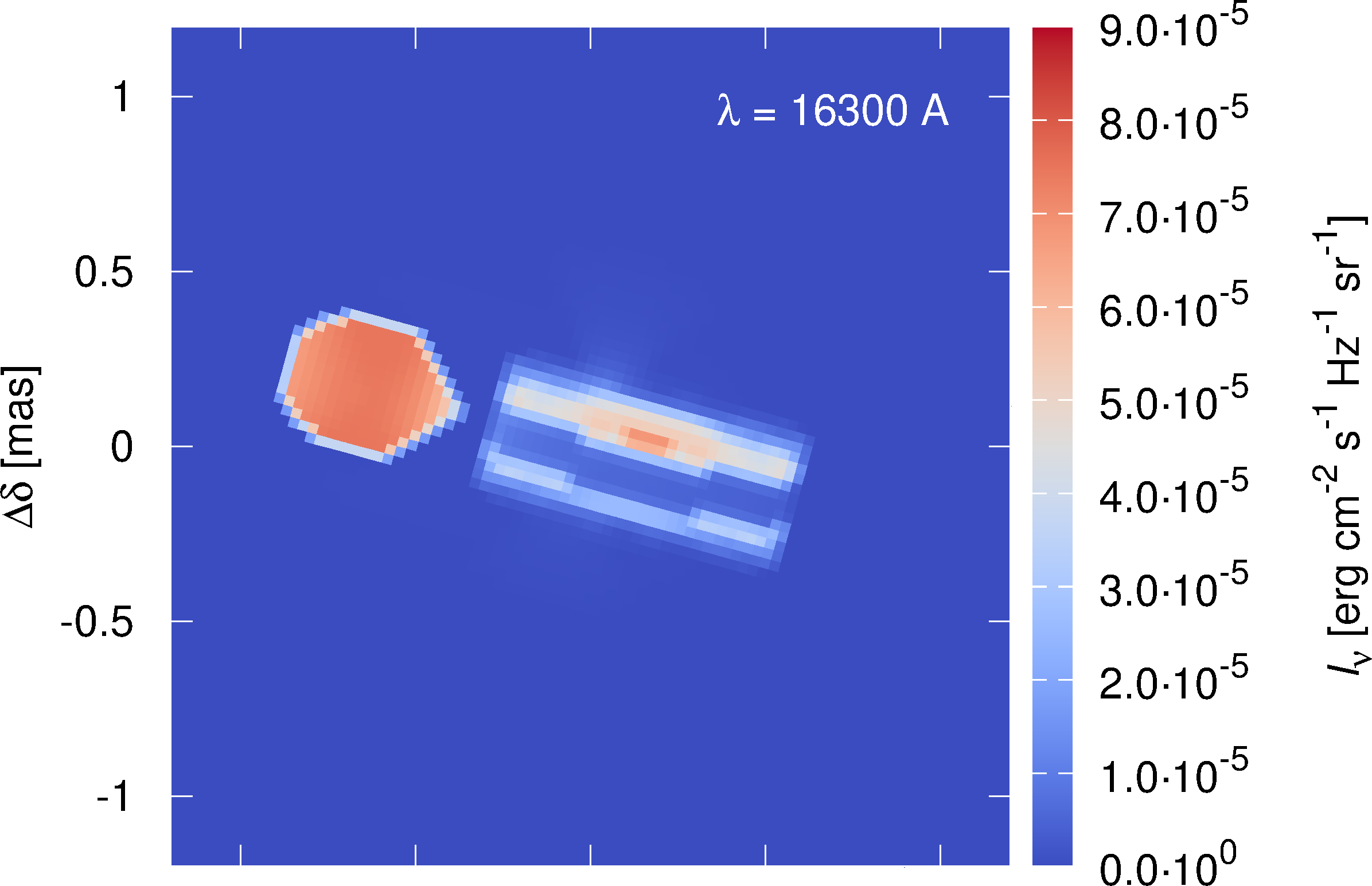}
\includegraphics[width=7cm]{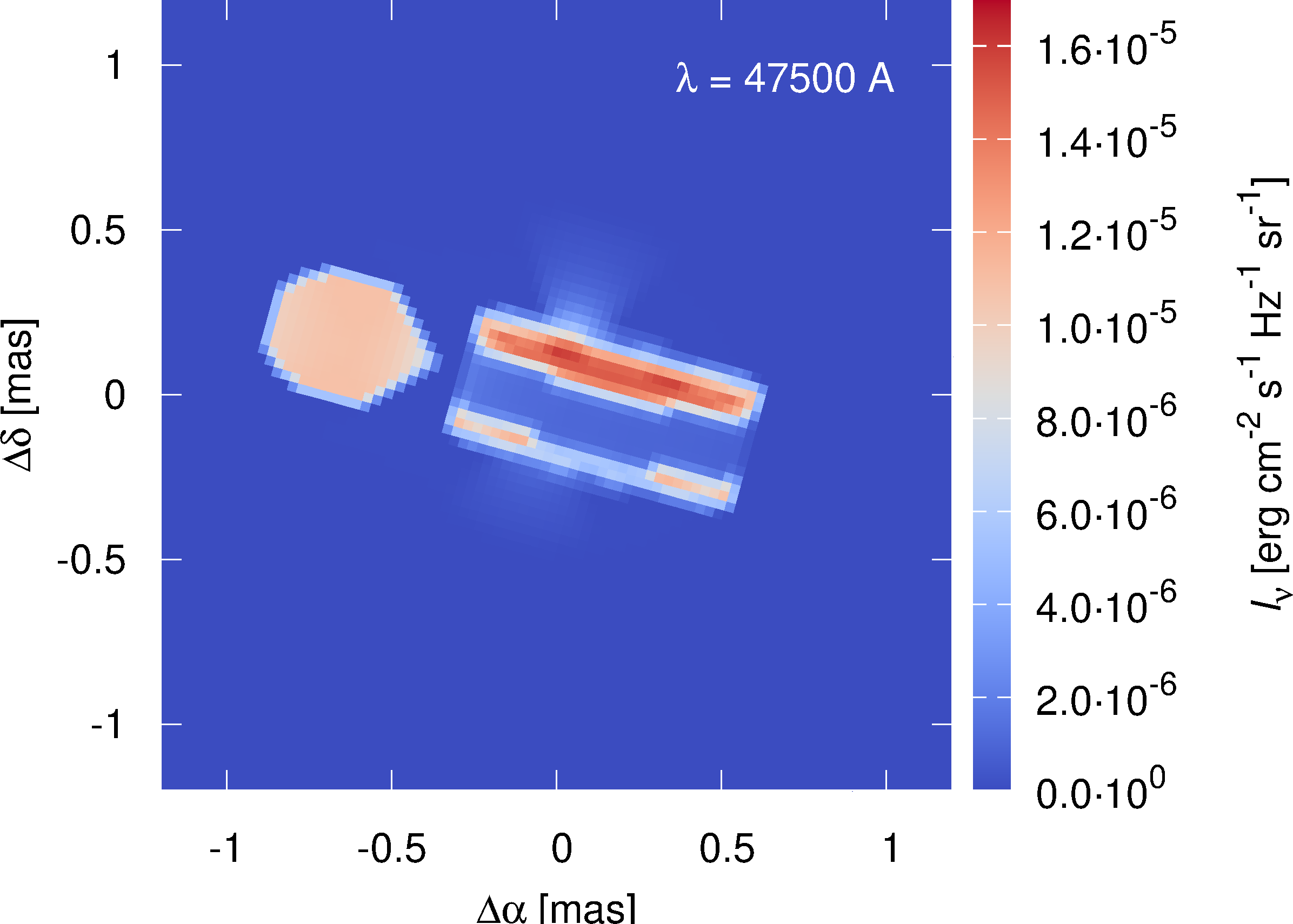}
\caption{
Continuum synthetic images of $\beta$~Lyr~A
for the joint model,
computed for four monochromatic wavelengths
(from top to bottom):
155\,nm (FUV),
545\,nm (V),
1630\,nm (H),
4750\,nm (FIR).
The orbital phase is always 0.25.
The appearance of the system changes substantially.
Optically-thin components (disk atmosphere, jet, shell)
are not seen very well in continuum radiation.
}
\label{img_1550.0}
\end{figure}

\begin{figure*}
\centering
\newdimen\tmpdim\tmpdim=3.0cm
\begin{tabular}{l@{\kern1mm}l@{\kern1mm}l@{\kern1mm}l@{\kern1mm}l}
\includegraphics[width=\tmpdim]{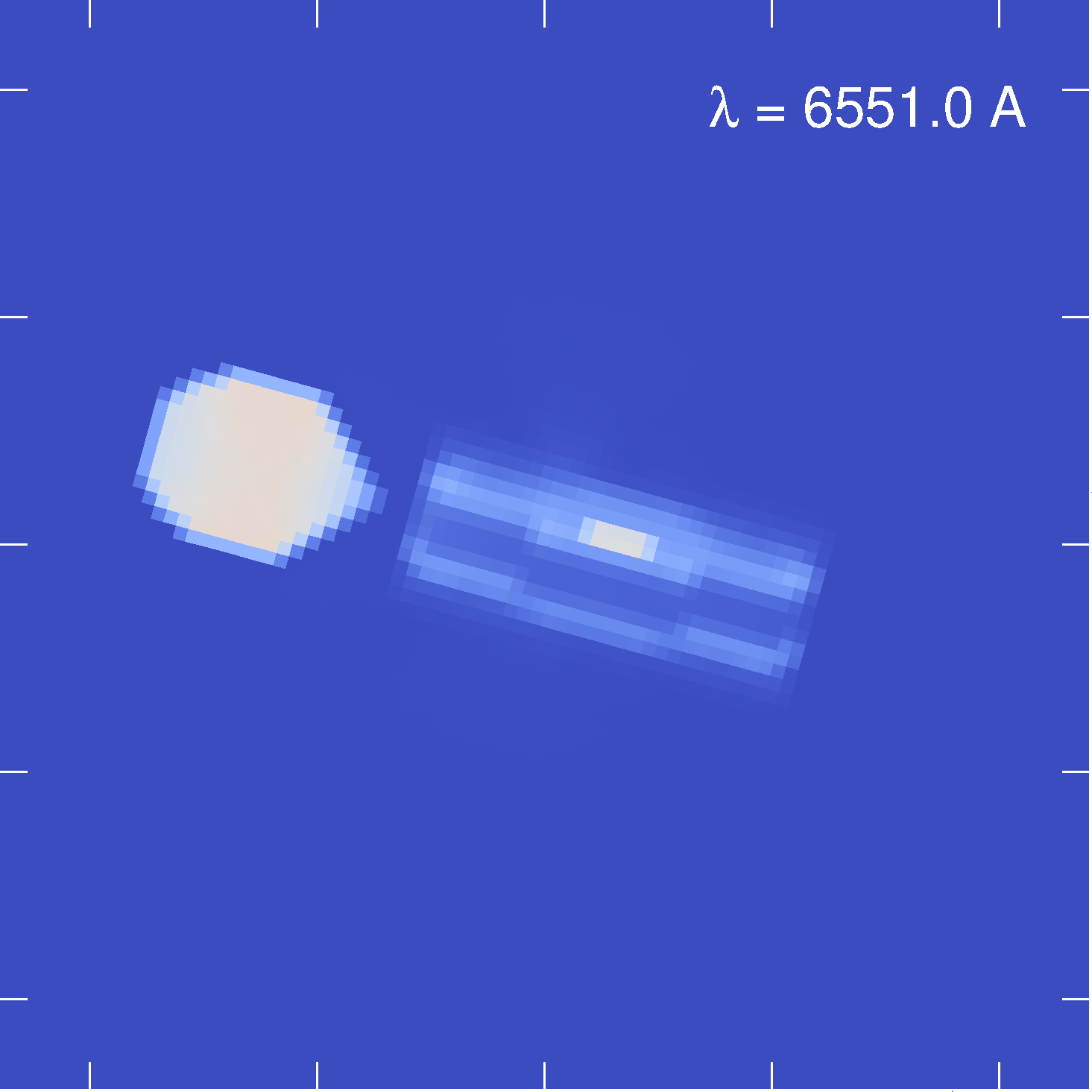} &
\includegraphics[width=\tmpdim]{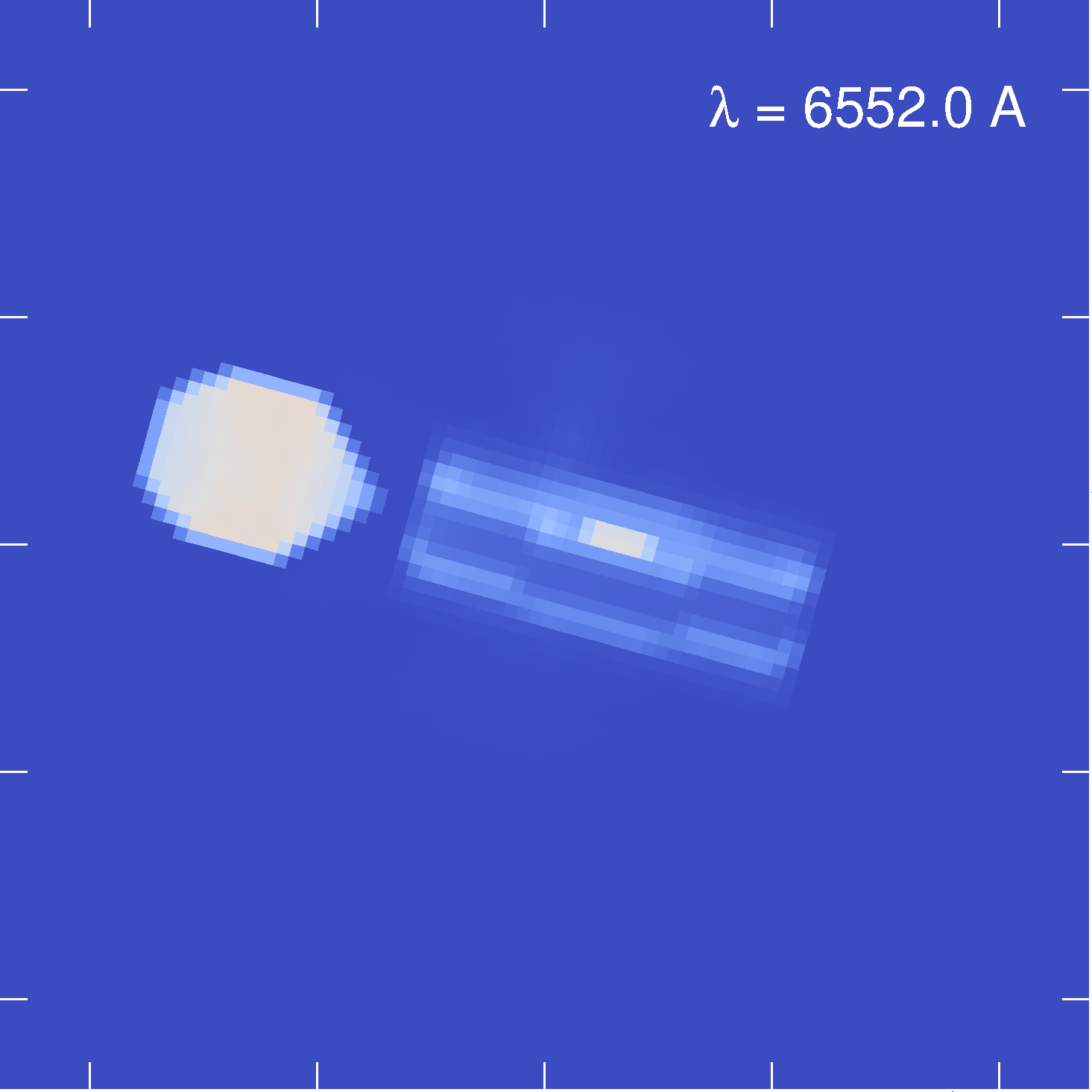} &
\includegraphics[width=\tmpdim]{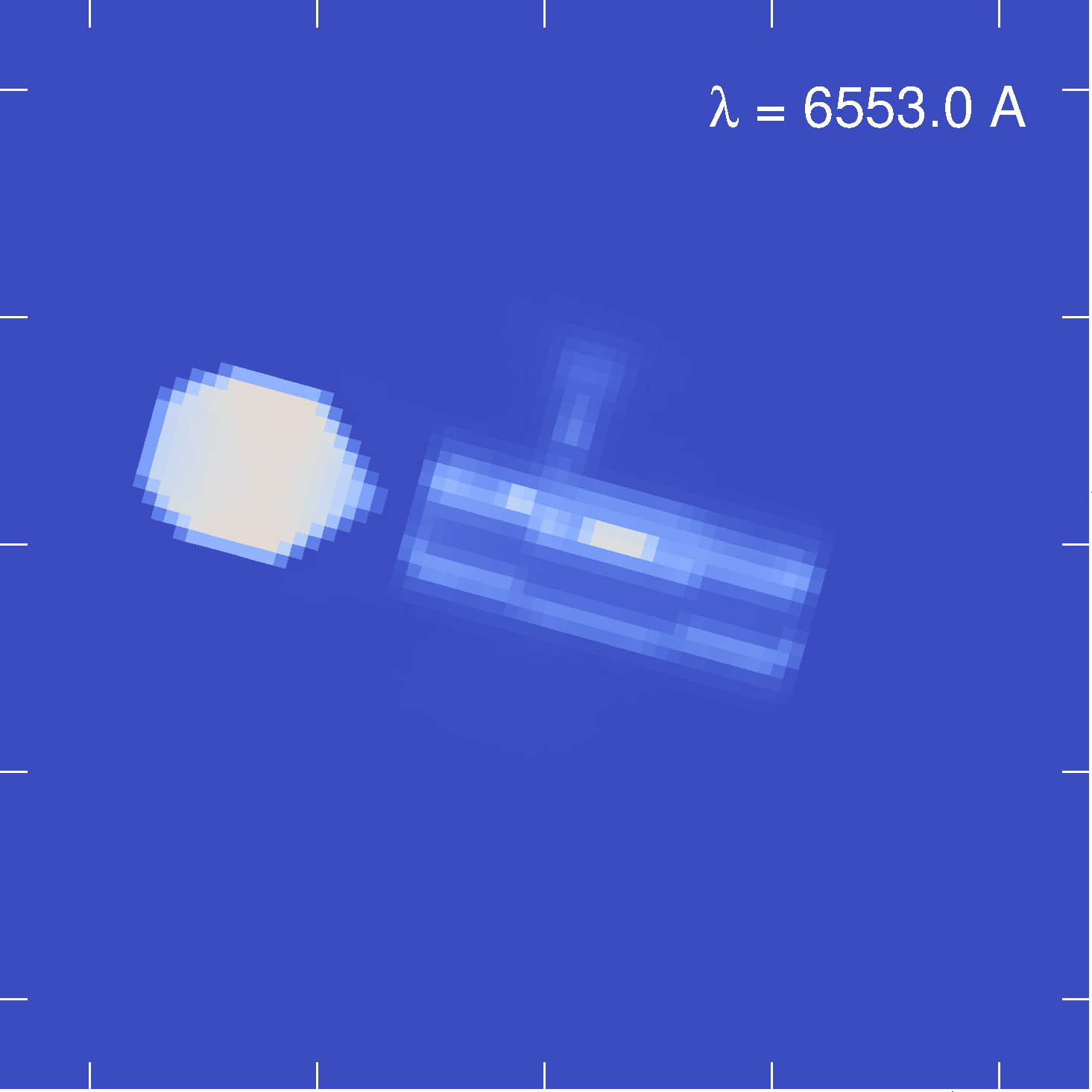} &
\includegraphics[width=\tmpdim]{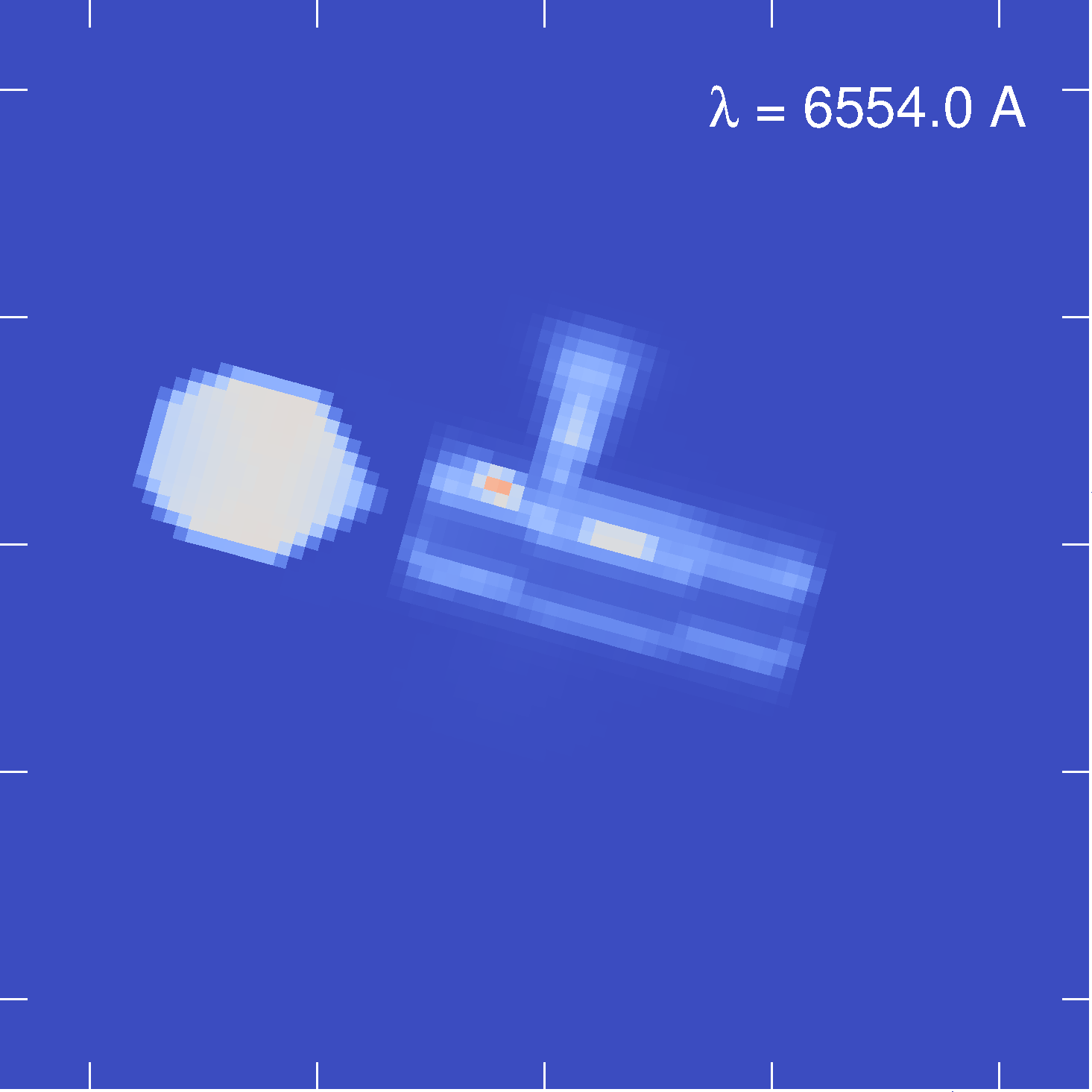} \\
\includegraphics[width=\tmpdim]{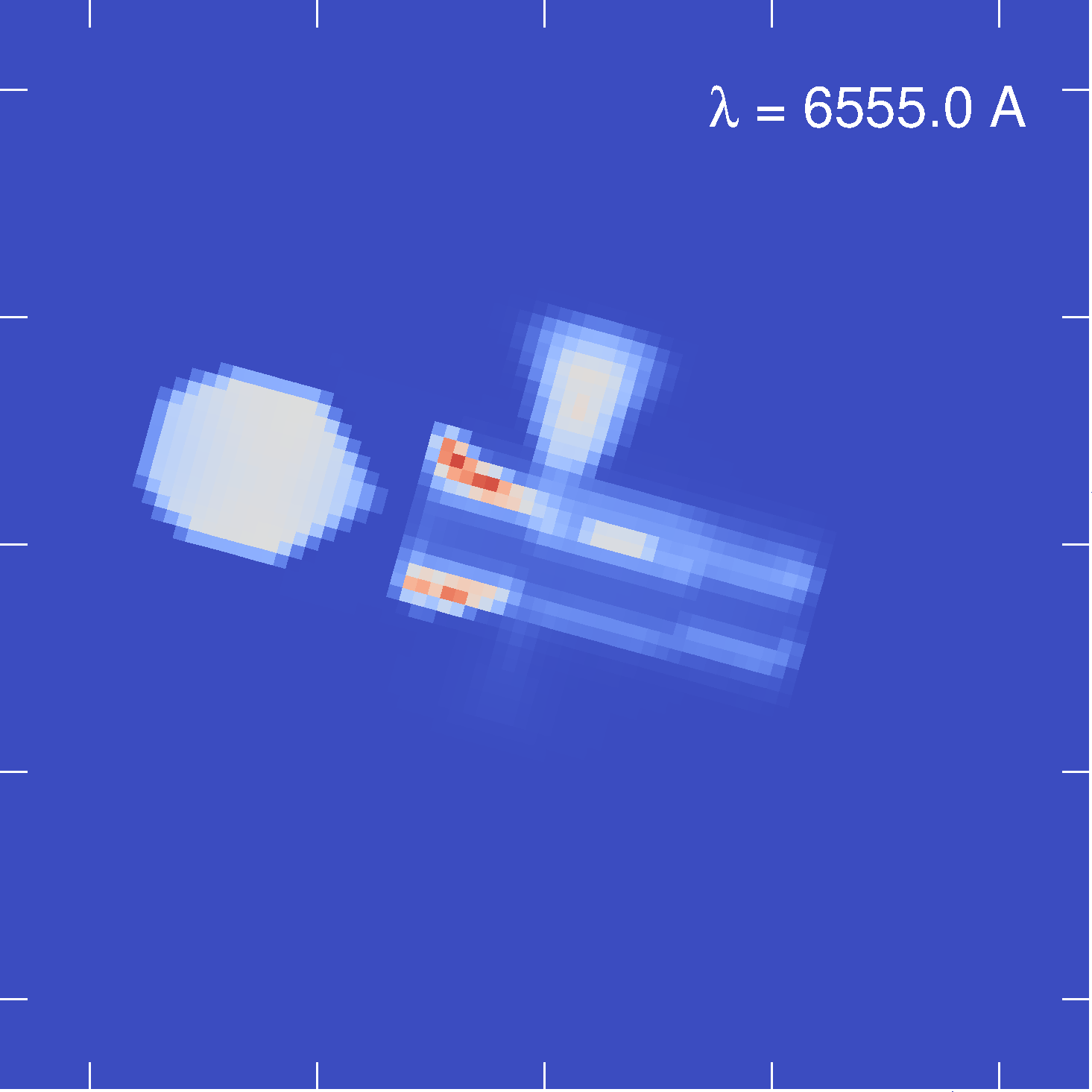} &
\includegraphics[width=\tmpdim]{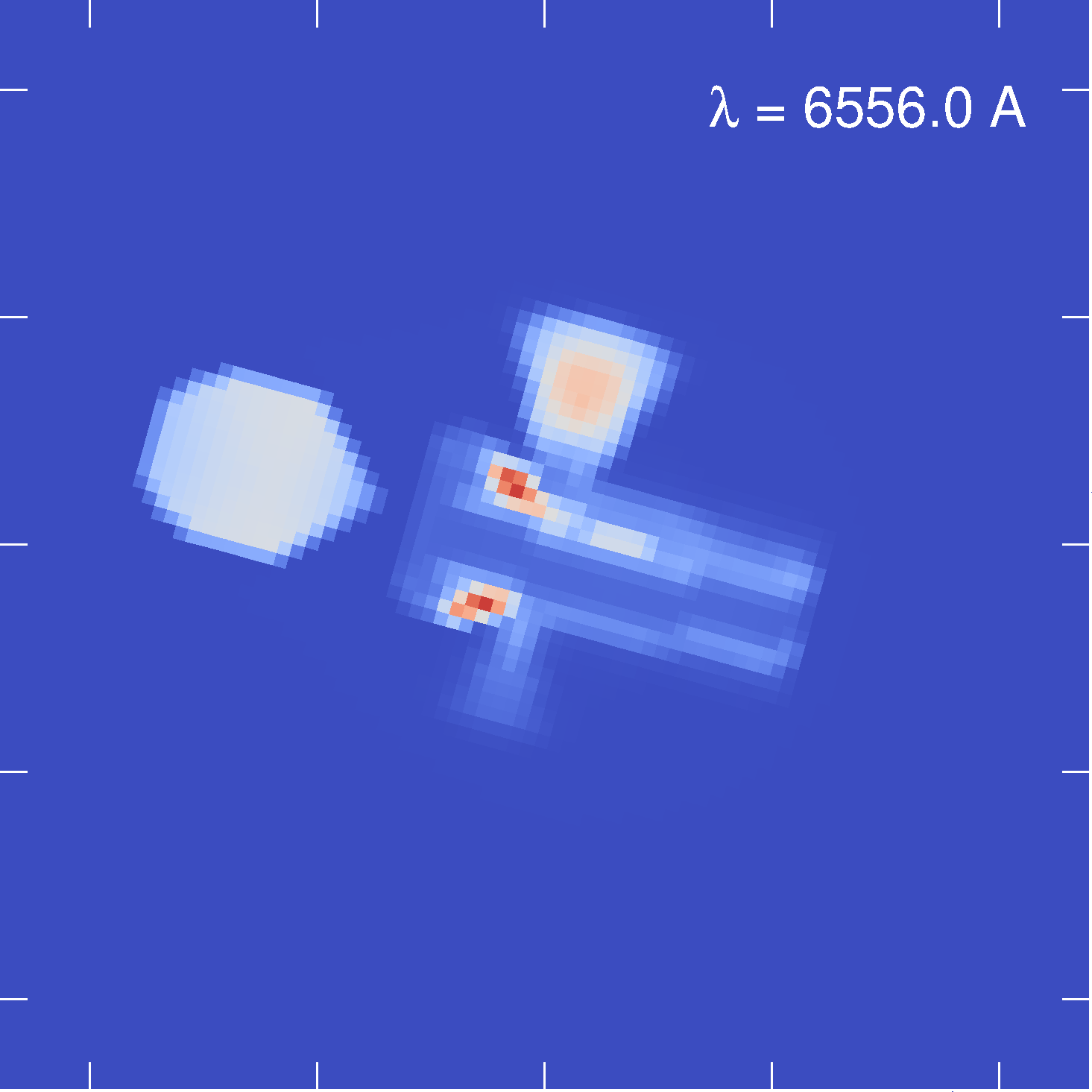} &
\includegraphics[width=\tmpdim]{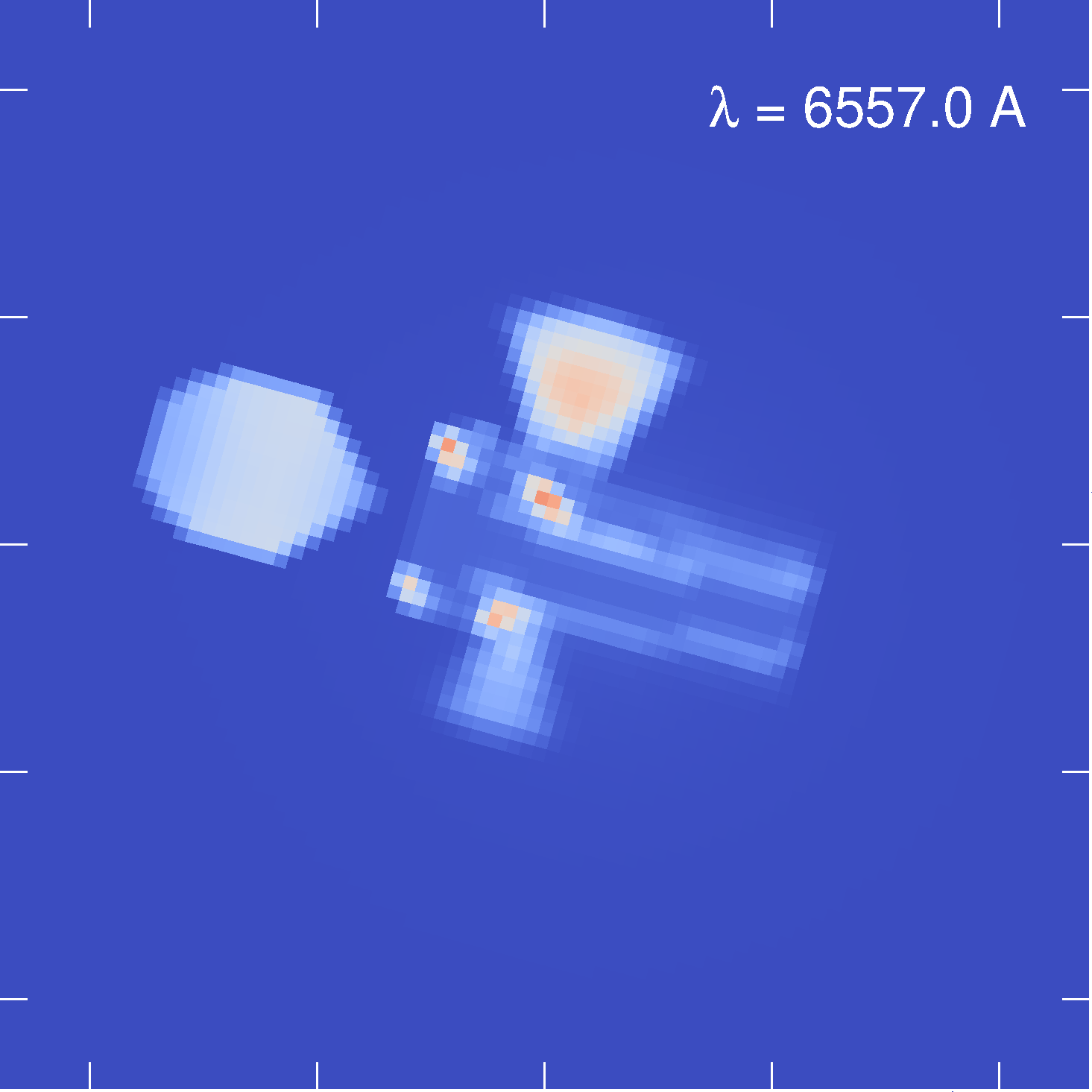} &
\includegraphics[width=\tmpdim]{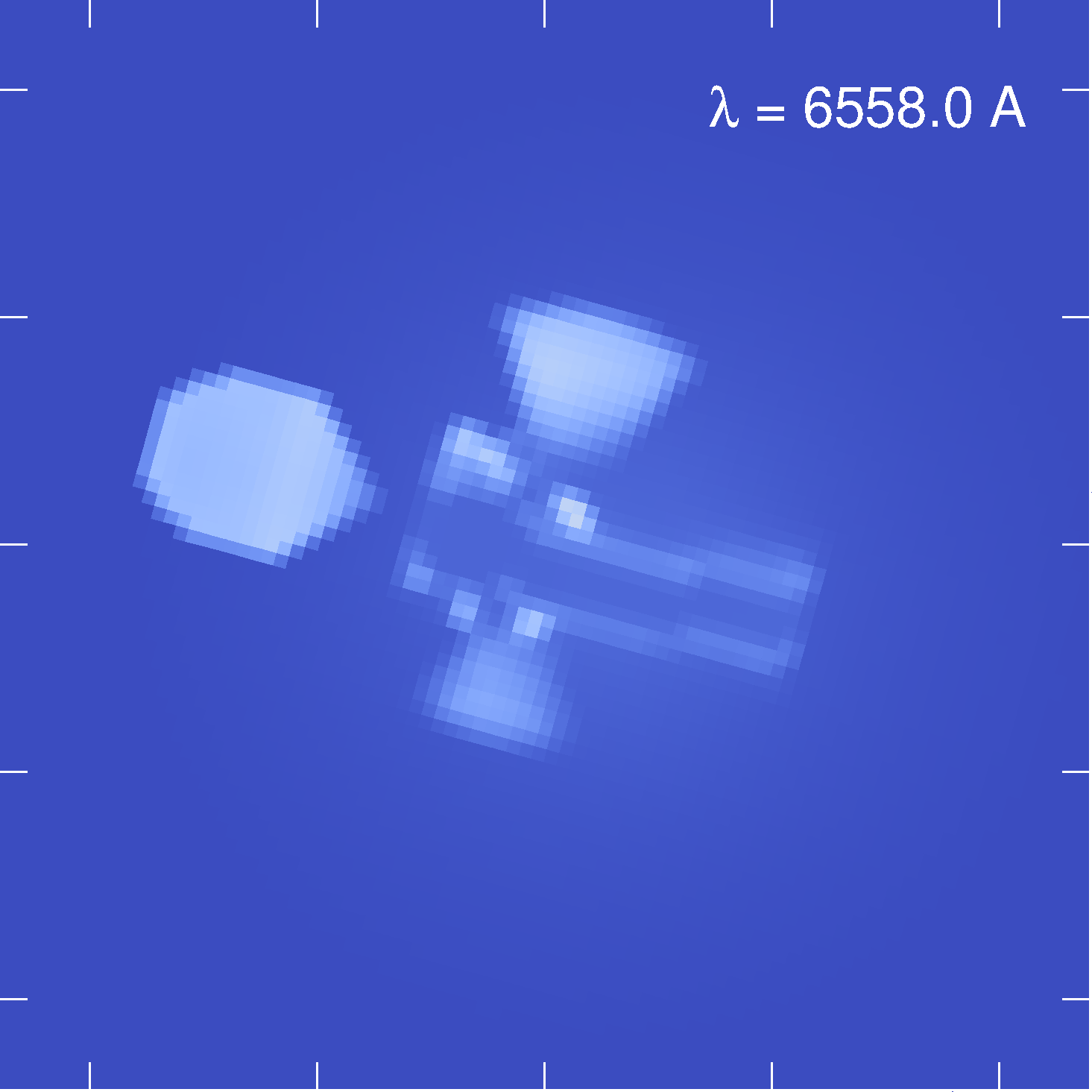} \\
\includegraphics[width=\tmpdim]{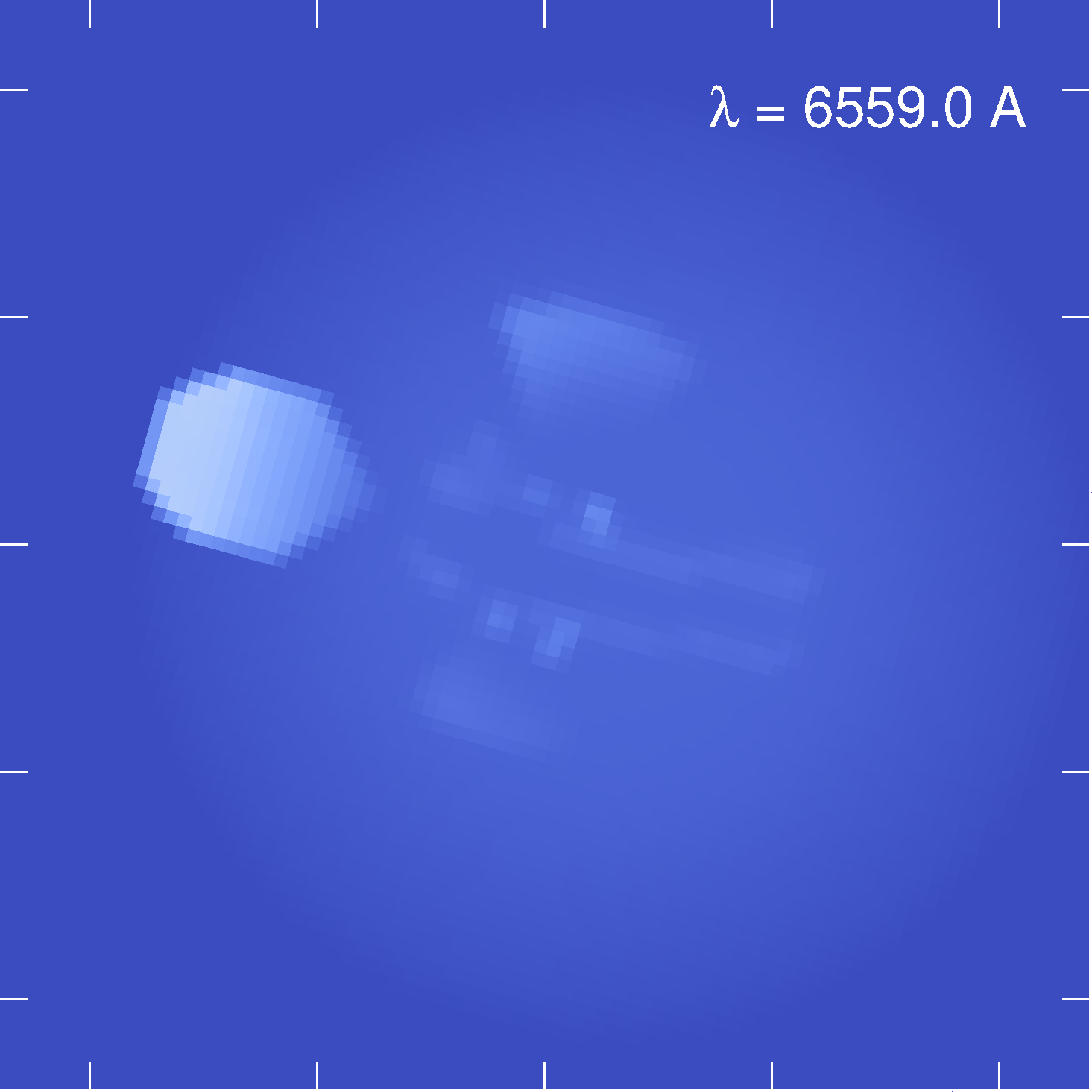} &
\includegraphics[width=\tmpdim]{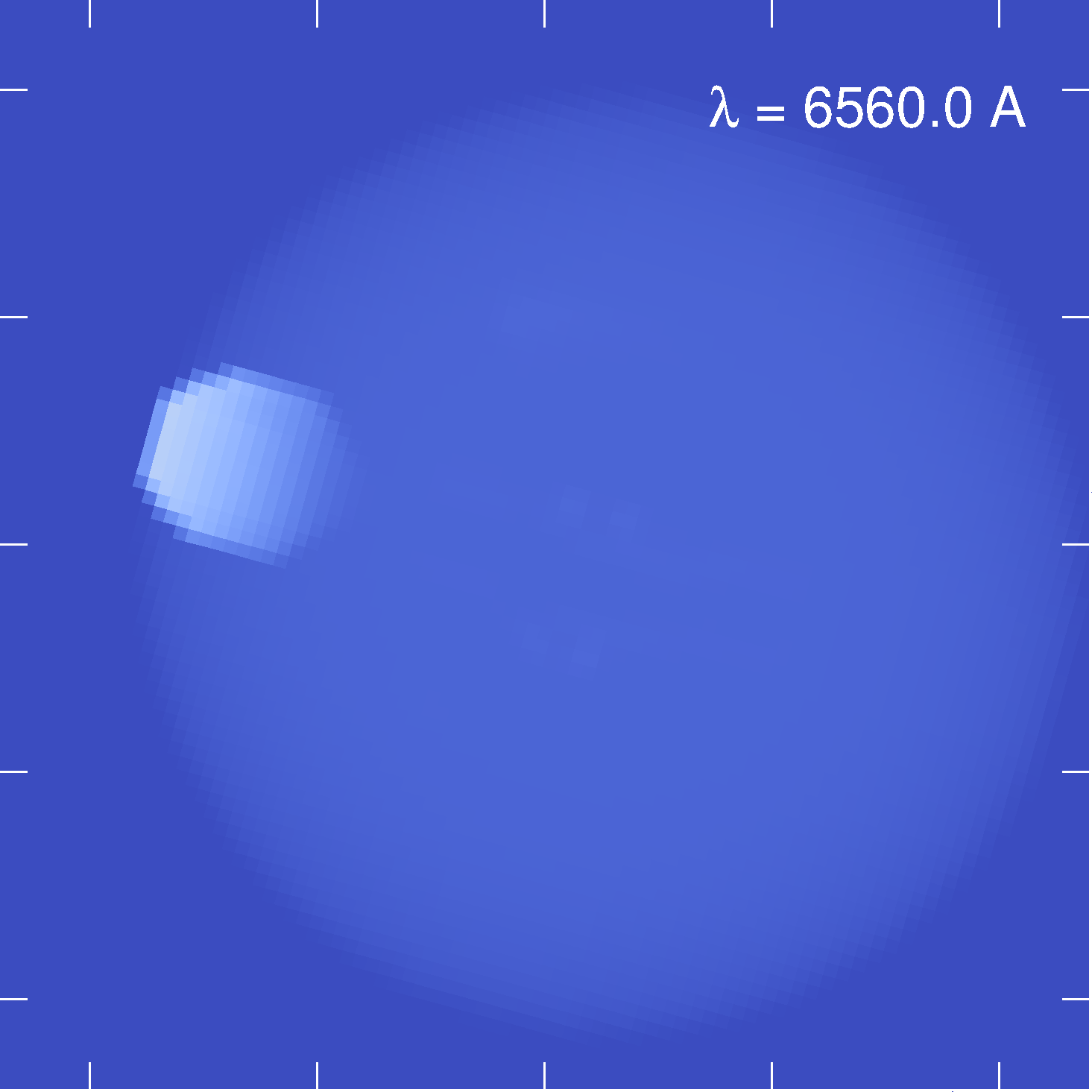} &
\includegraphics[width=\tmpdim]{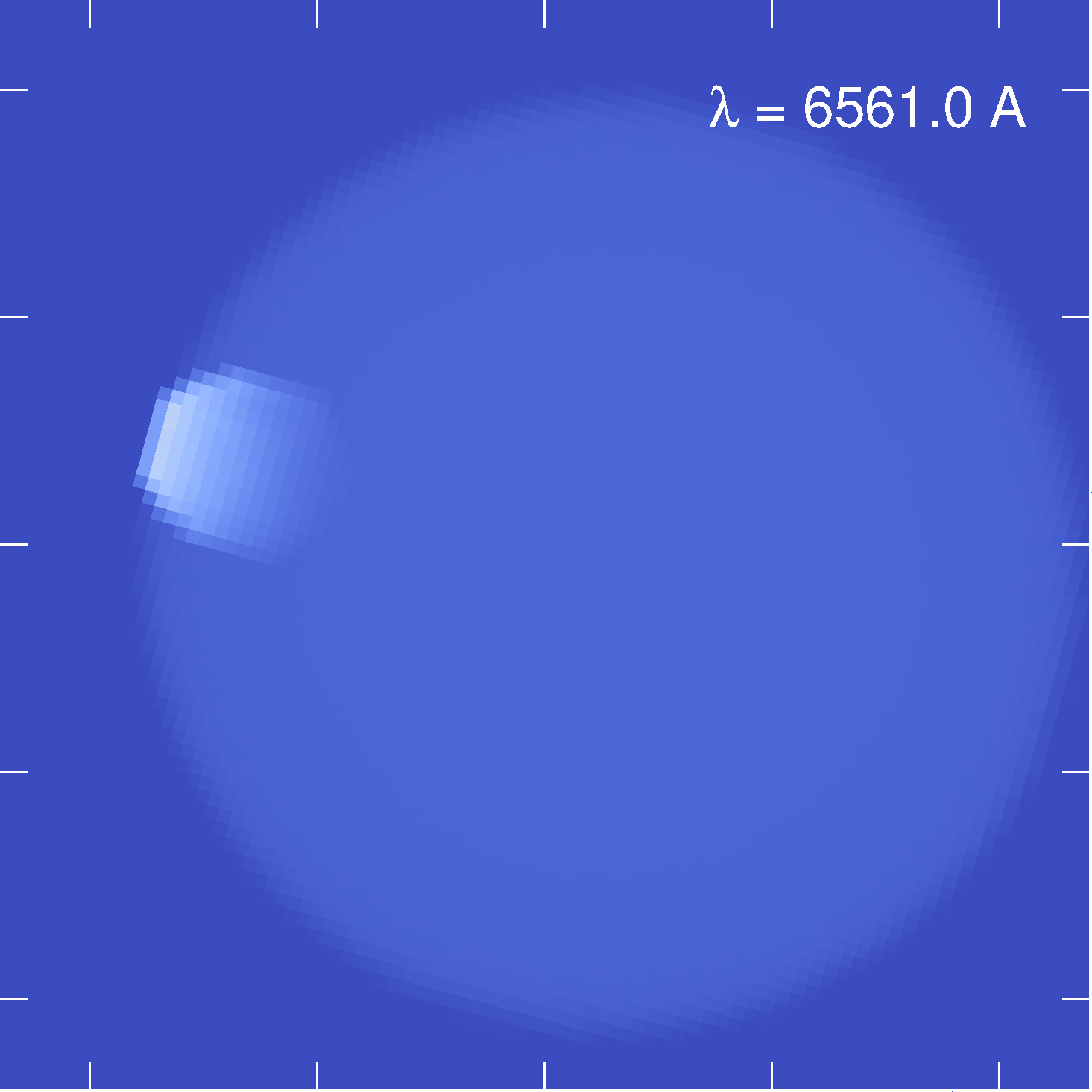} &
\includegraphics[width=\tmpdim]{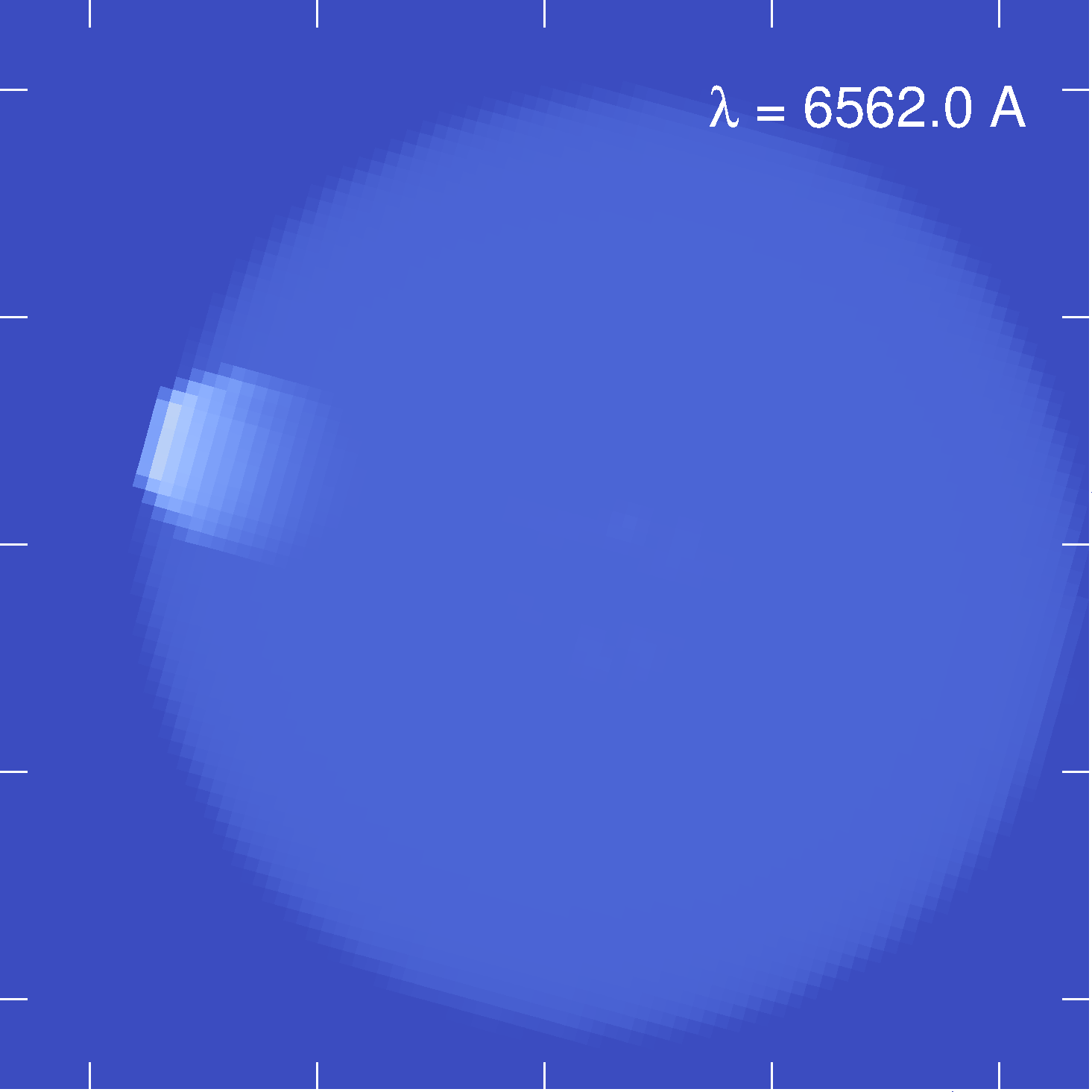} \\
\includegraphics[width=\tmpdim]{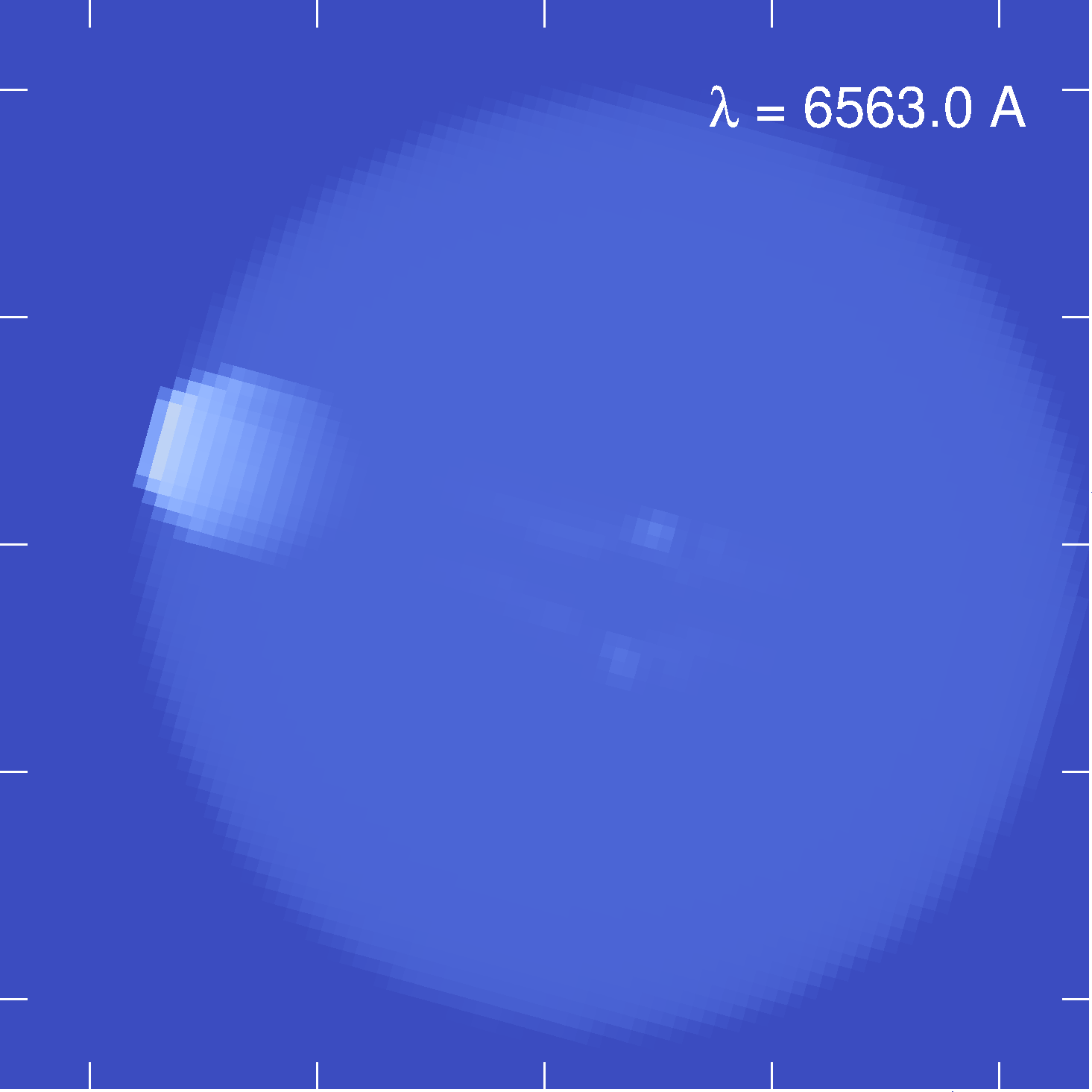} &
\includegraphics[width=\tmpdim]{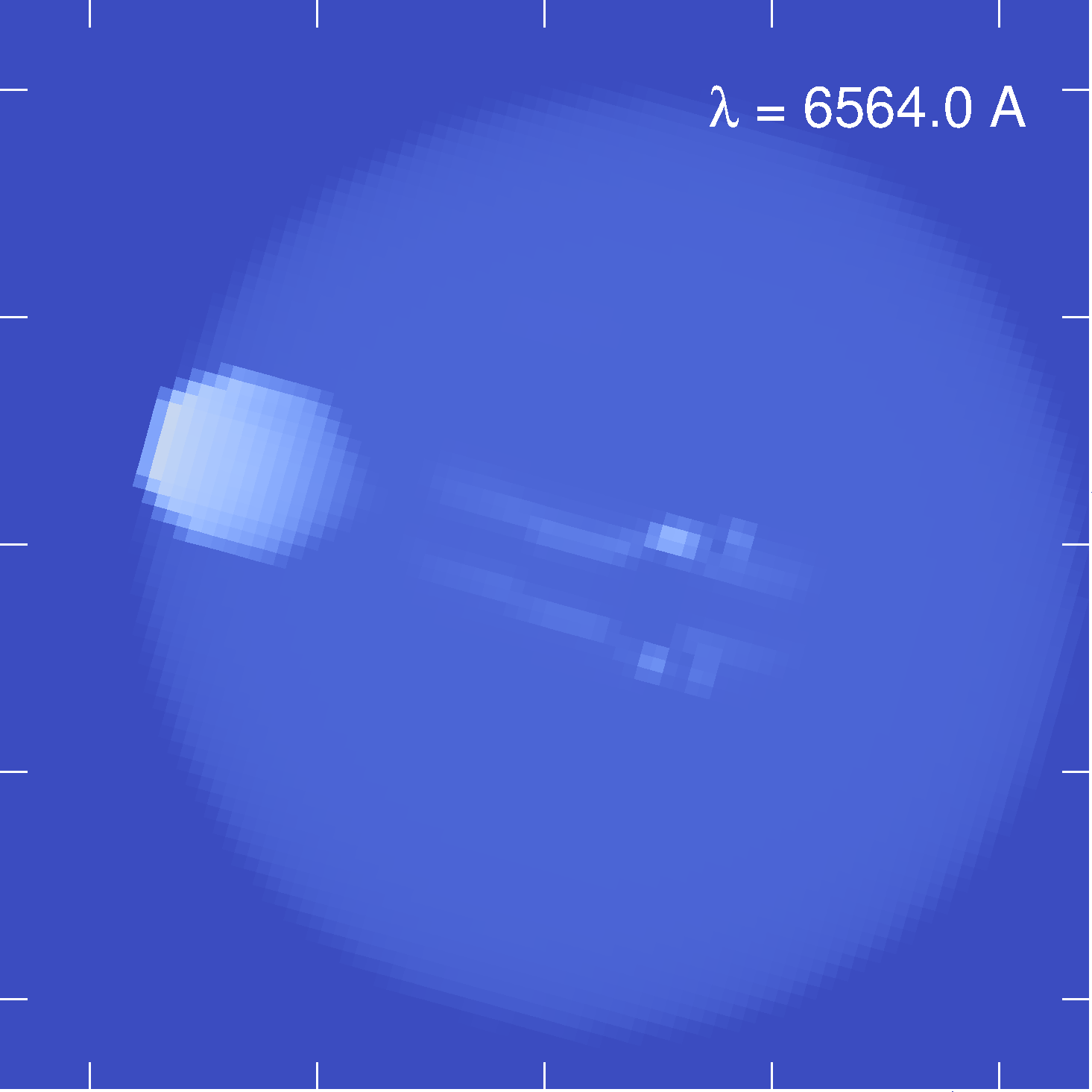} &
\includegraphics[width=\tmpdim]{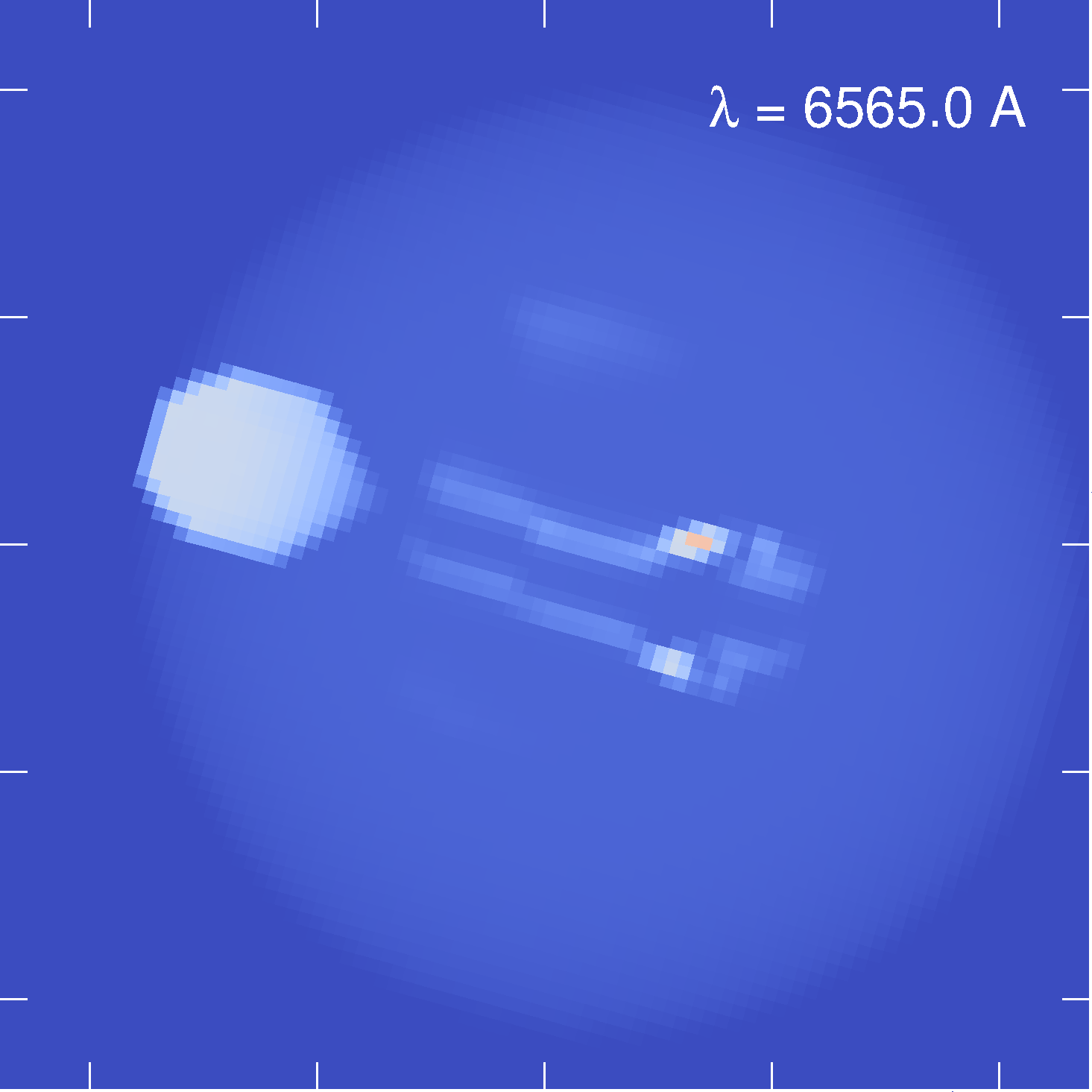} &
\includegraphics[width=\tmpdim]{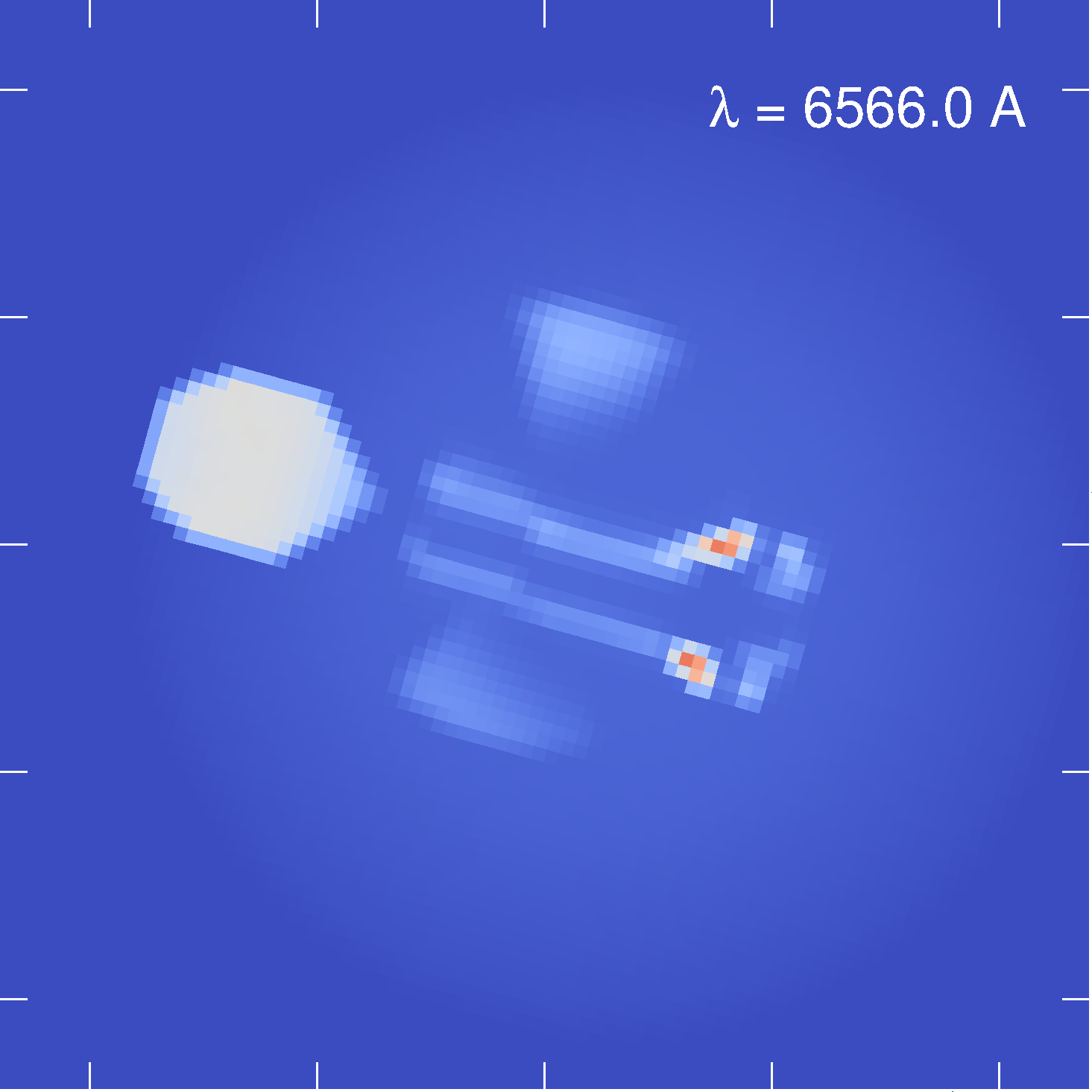} \\
\includegraphics[width=\tmpdim]{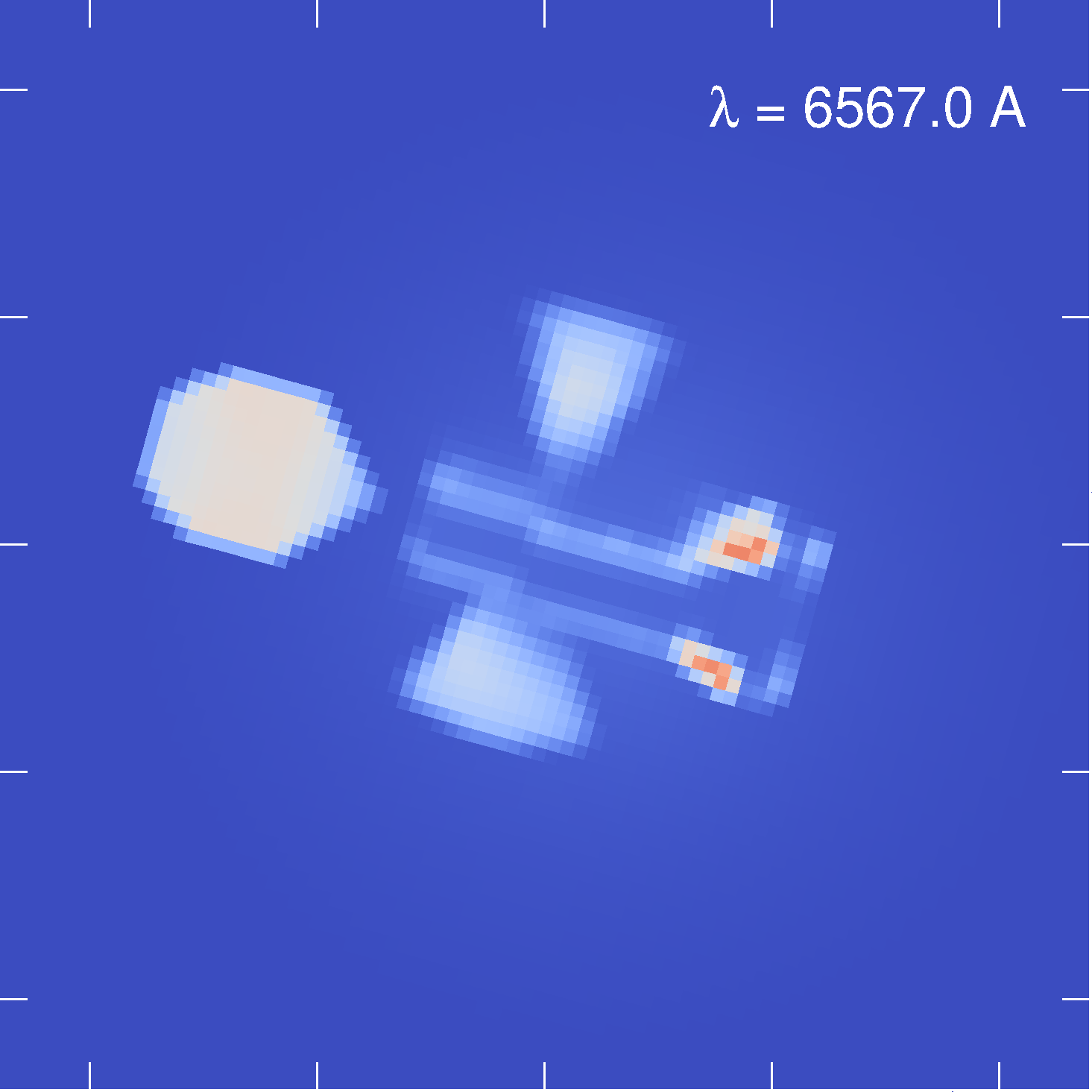} &
\includegraphics[width=\tmpdim]{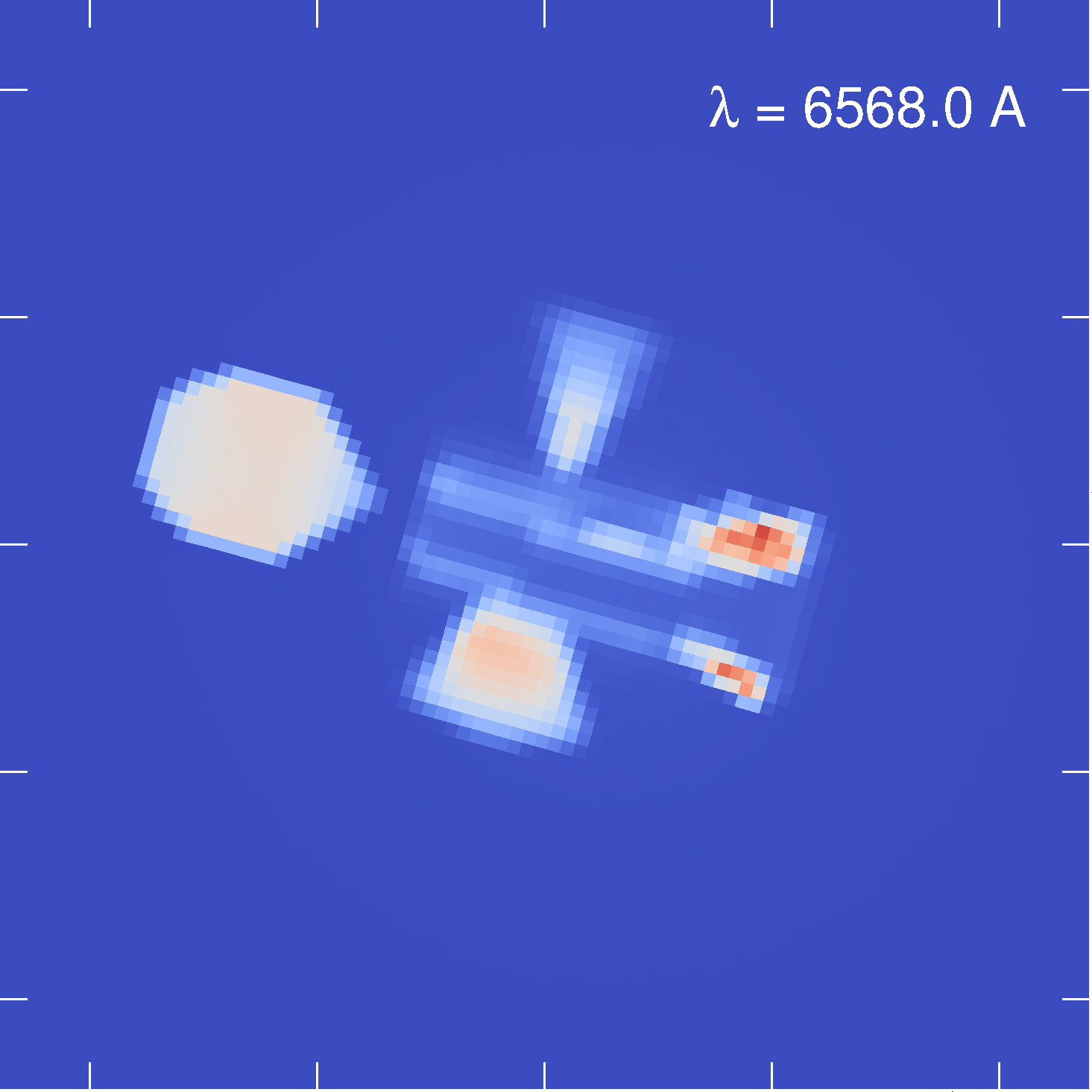} &
\includegraphics[width=\tmpdim]{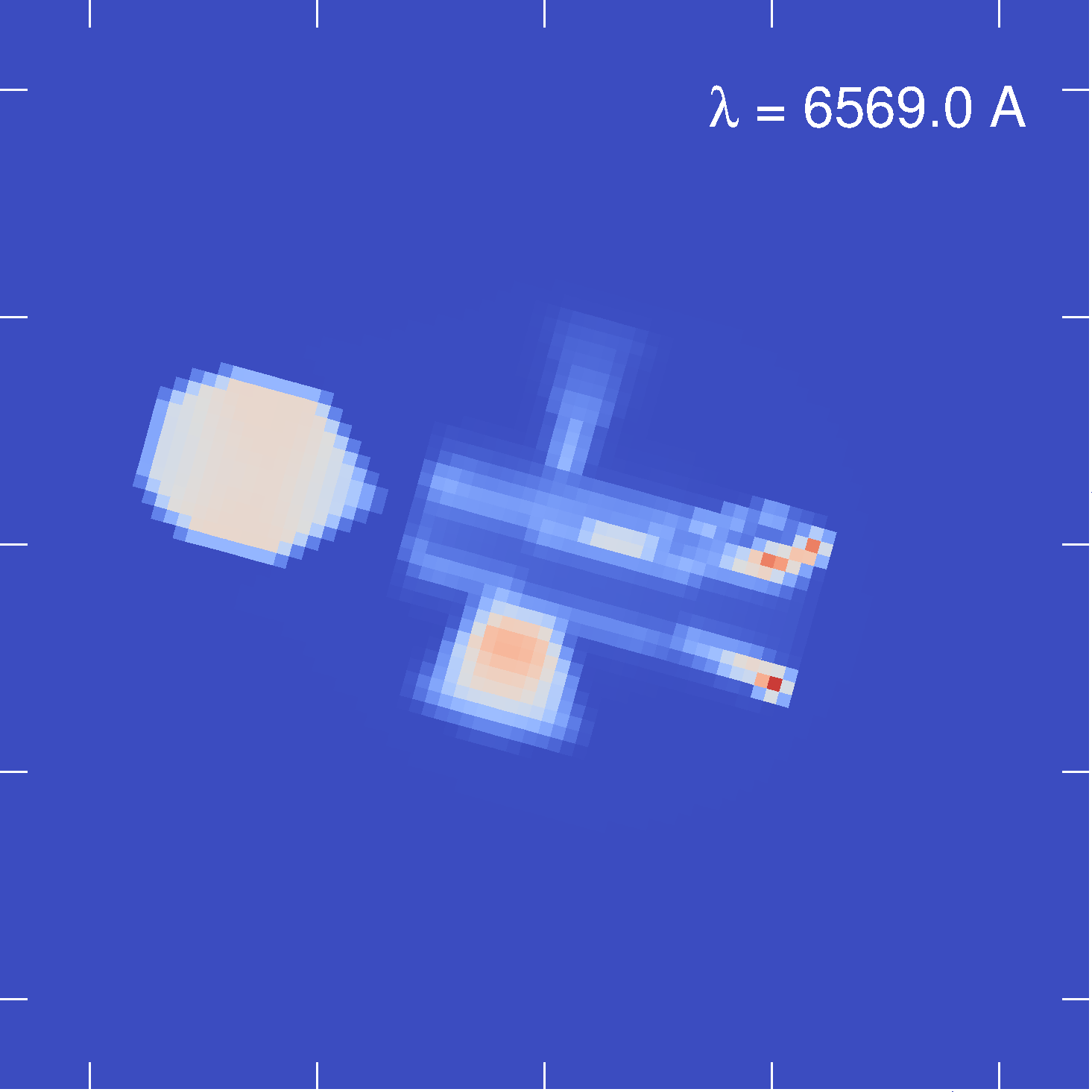} &
\includegraphics[width=\tmpdim]{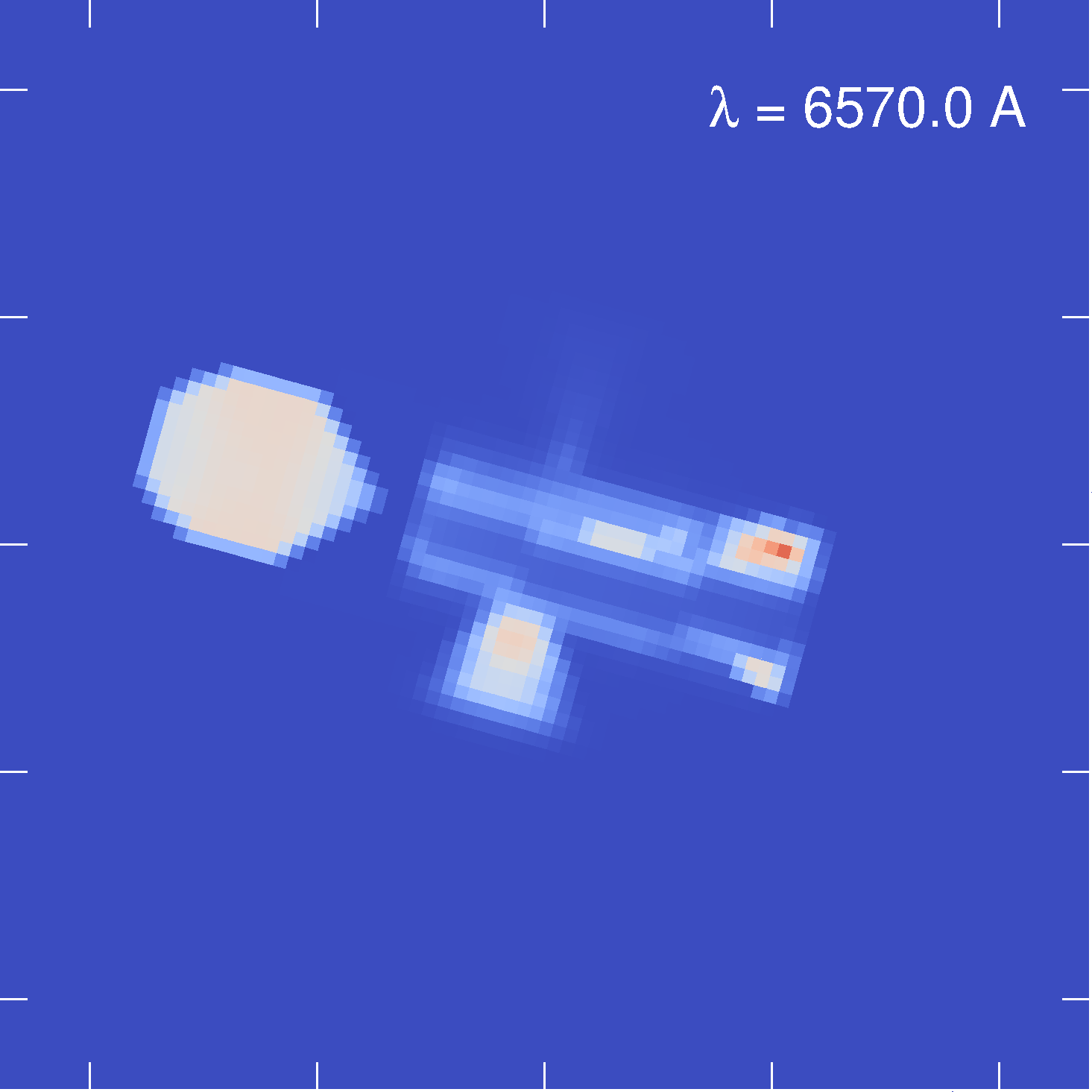} \\
\includegraphics[width=\tmpdim]{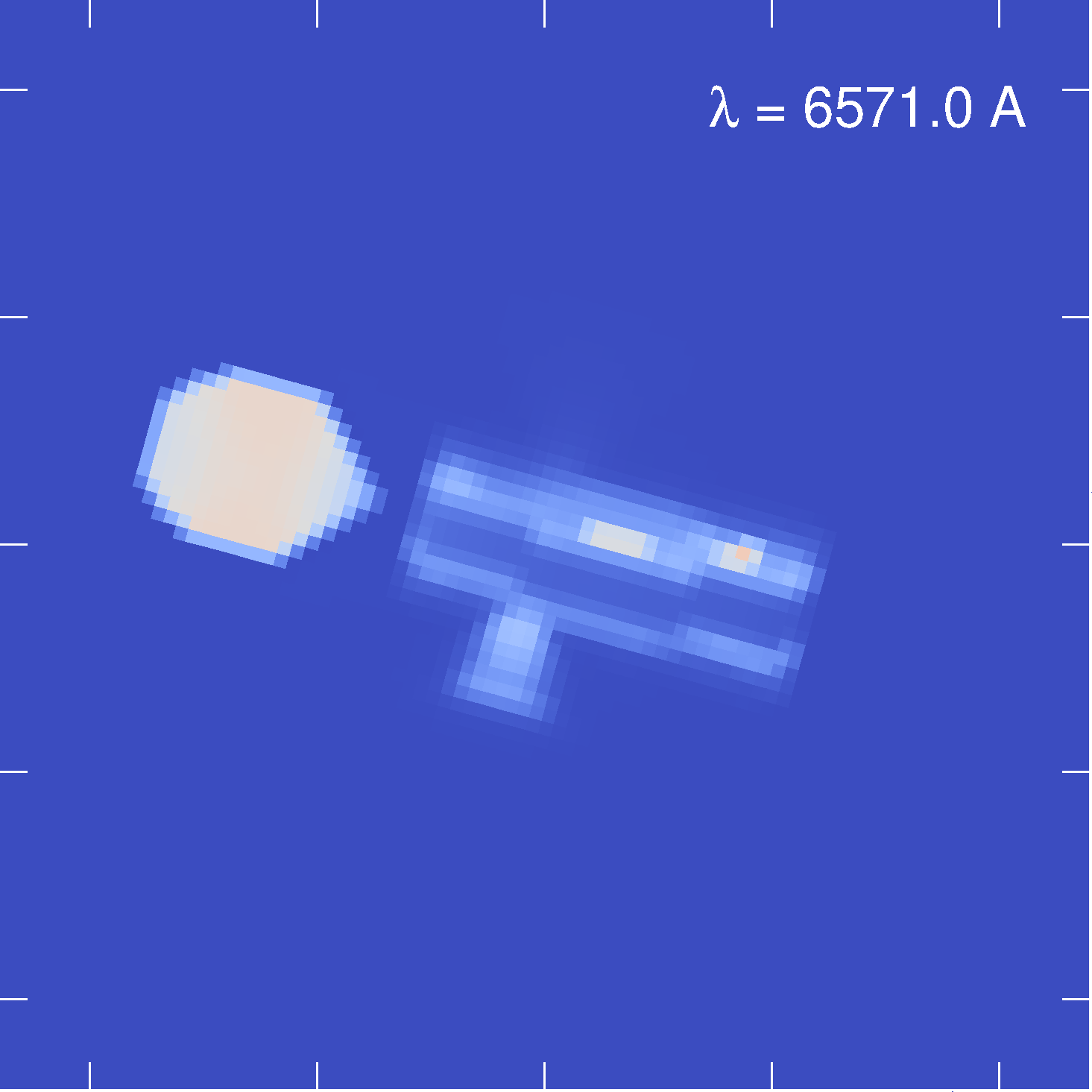} &
\includegraphics[width=\tmpdim]{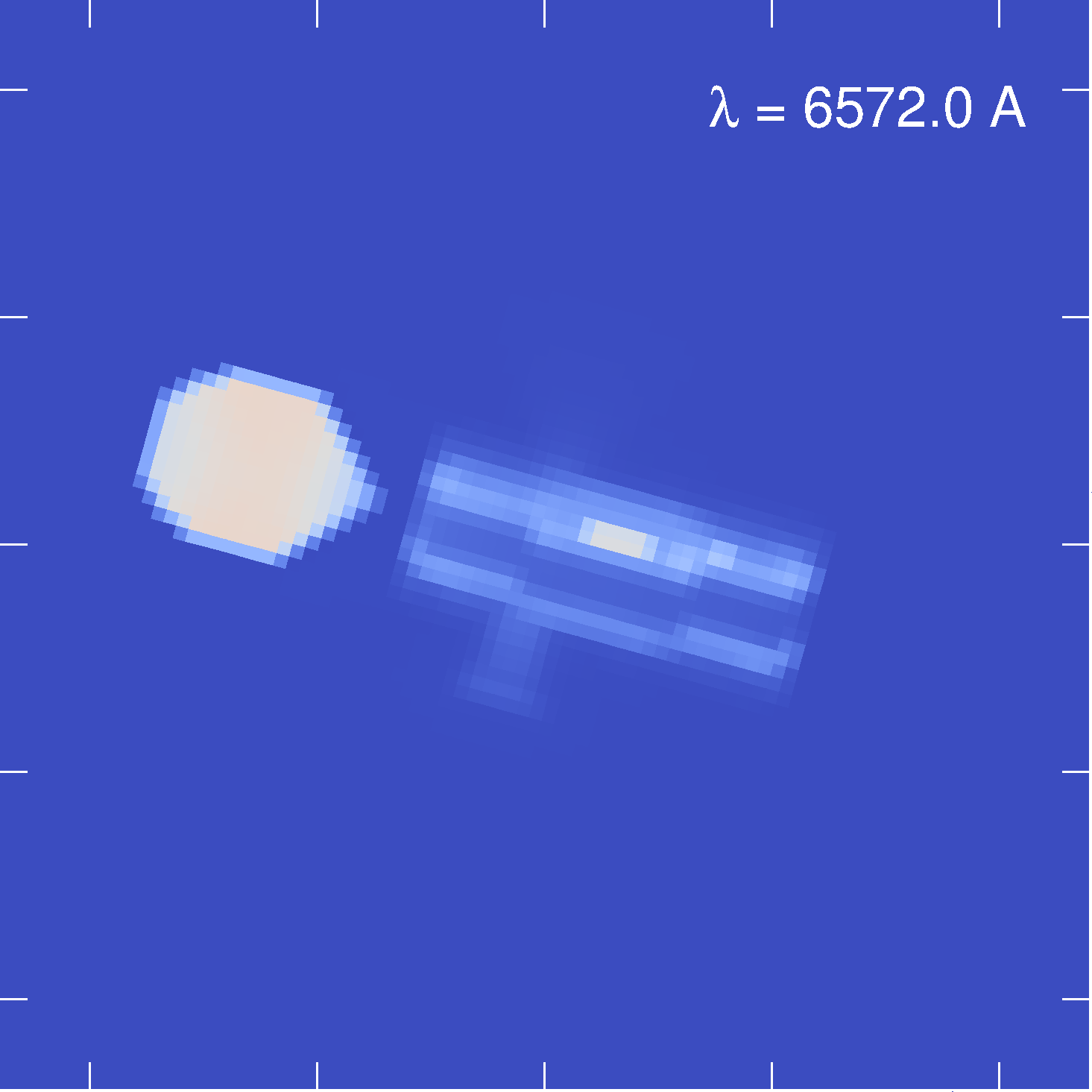} &
\includegraphics[width=\tmpdim]{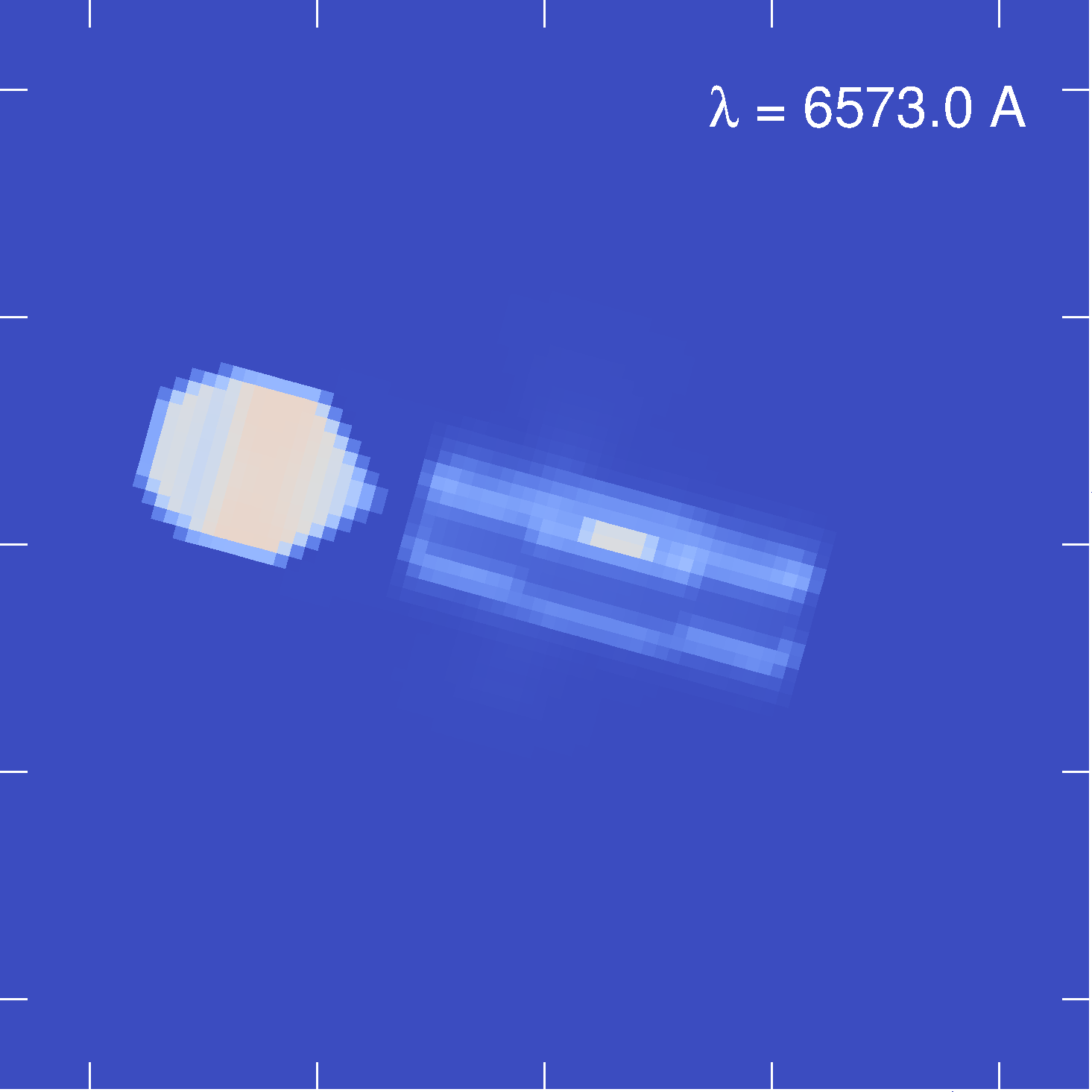} &
\includegraphics[width=\tmpdim]{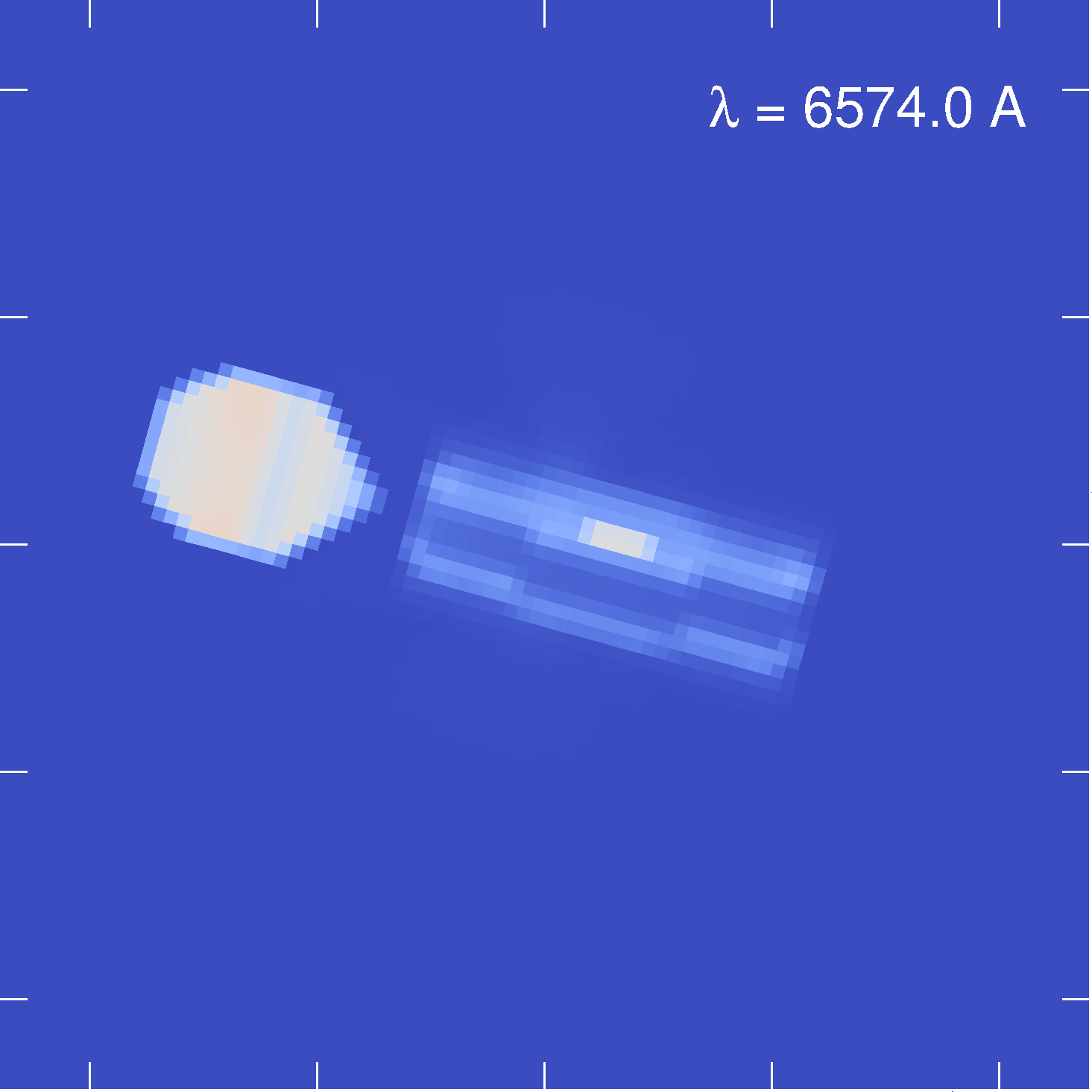} \\
\raise.5cm\hbox{\includegraphics[width=\tmpdim]{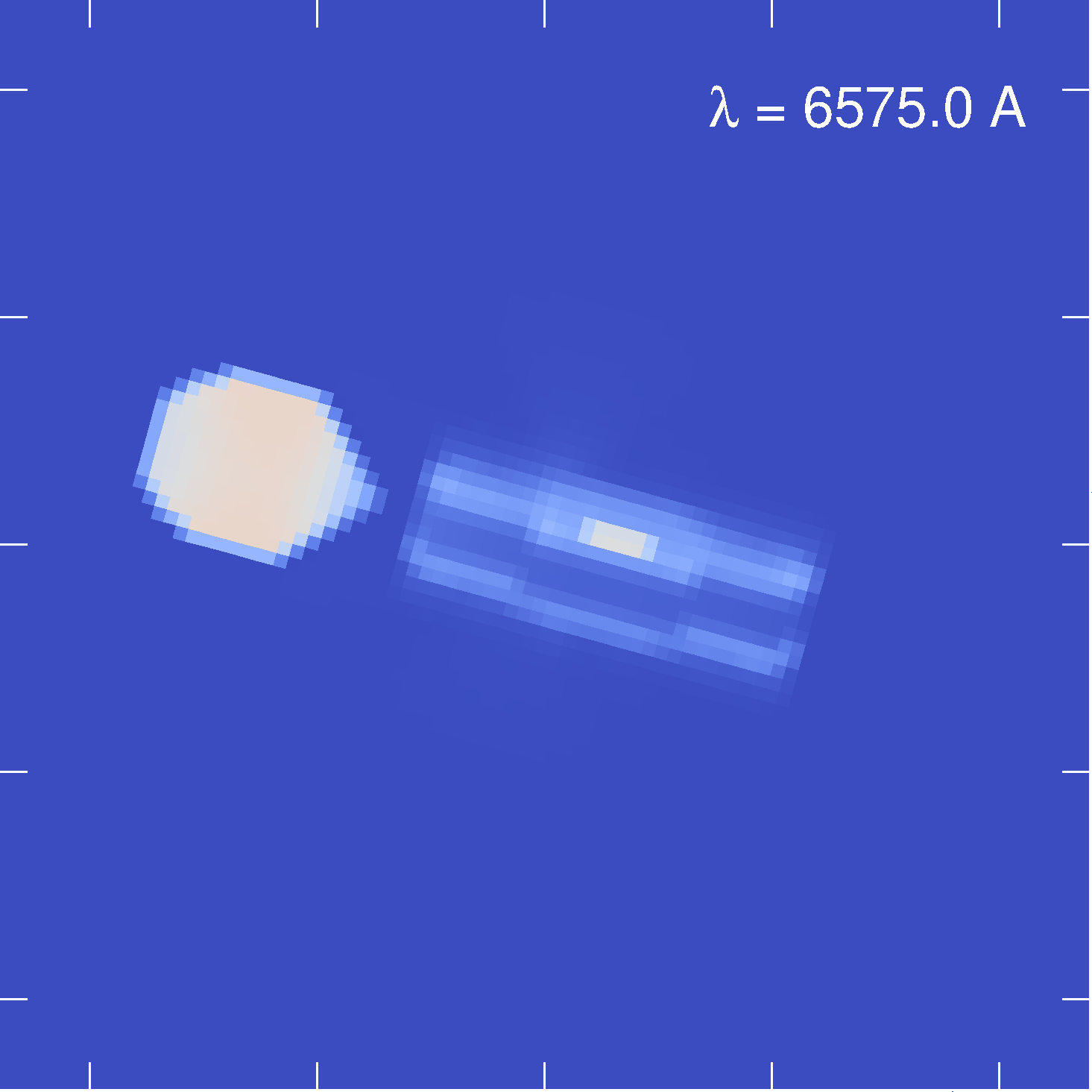}} &
\raise.5cm\hbox{\includegraphics[width=\tmpdim]{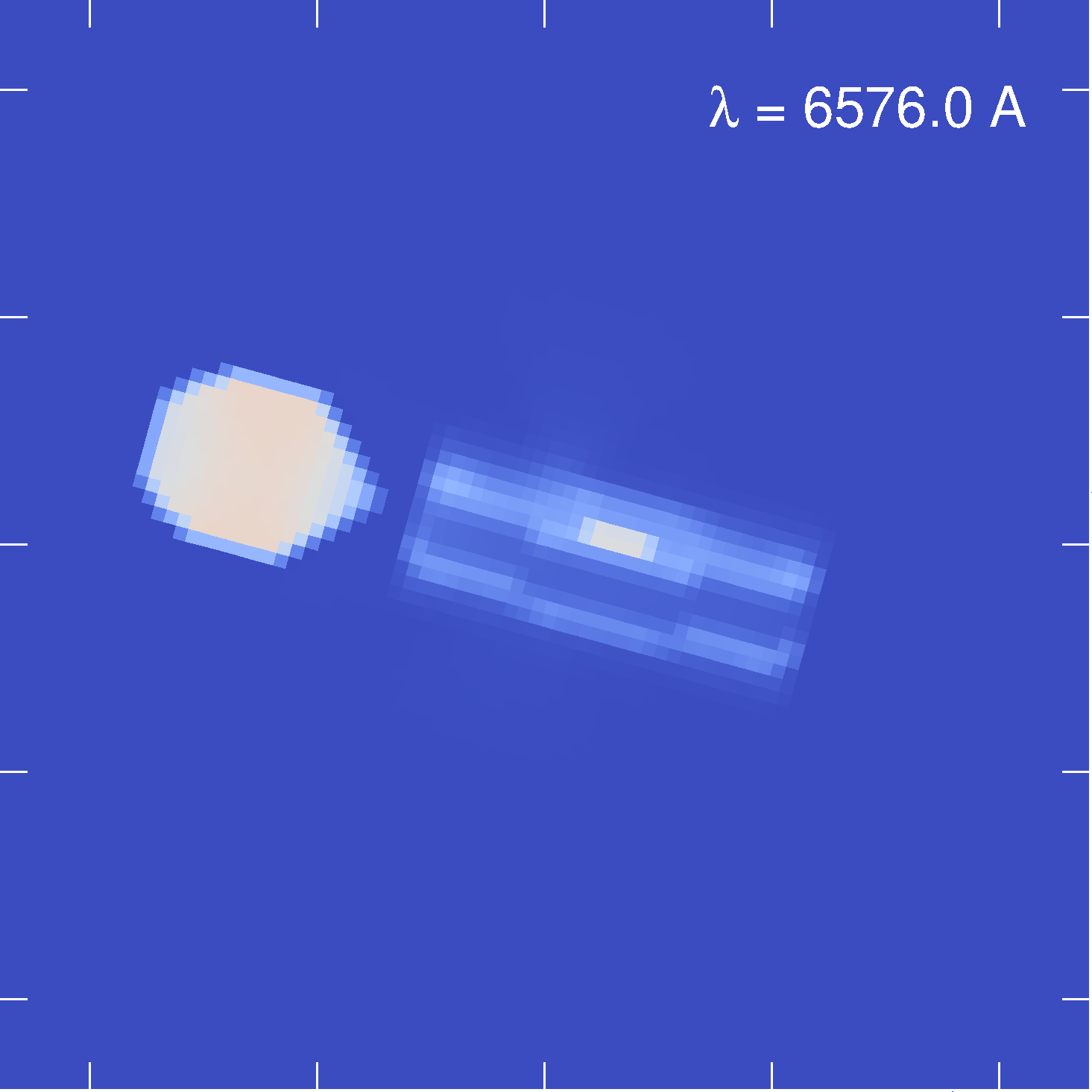}} &
\raise.5cm\hbox{\includegraphics[width=\tmpdim]{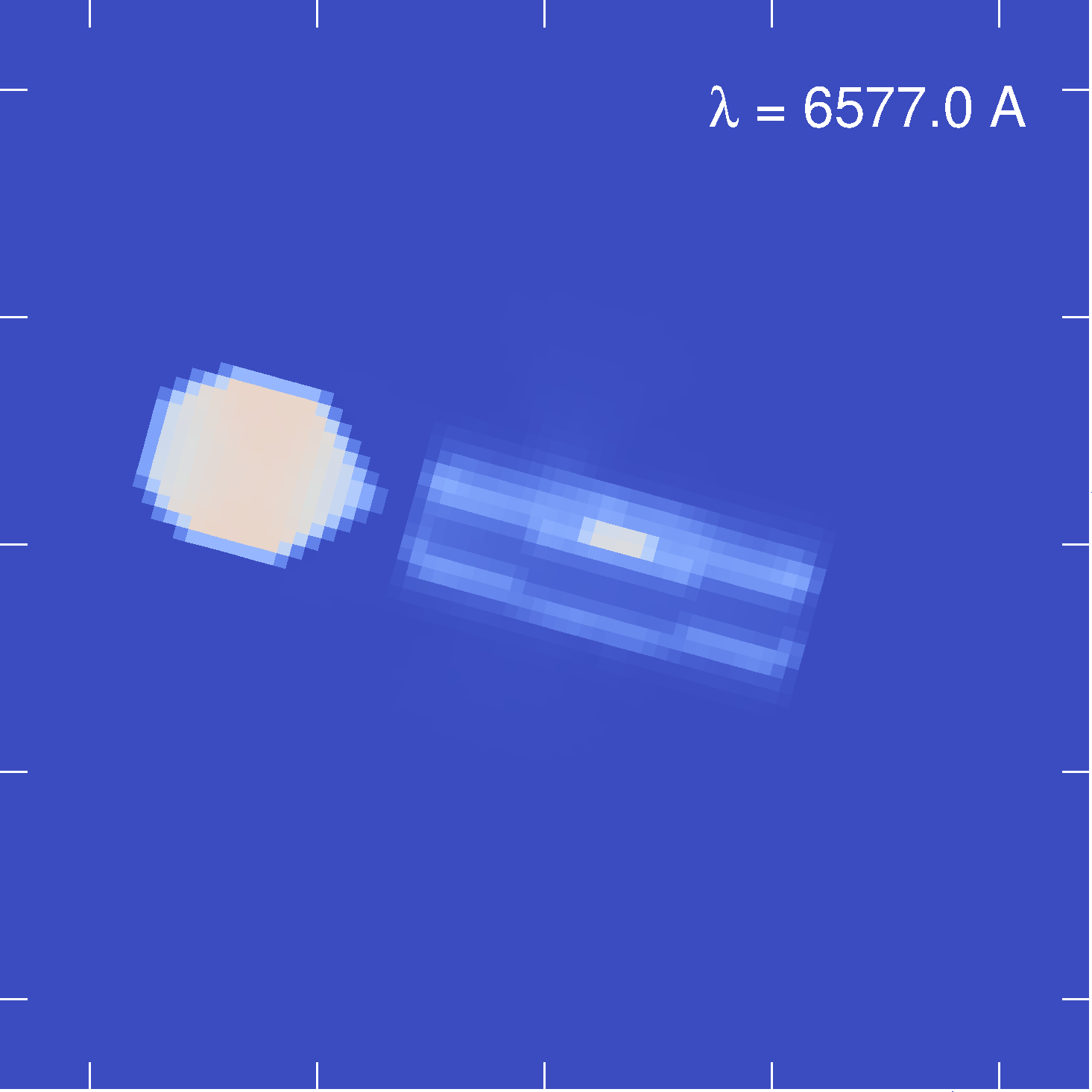}} &
\includegraphics[width=3.75cm]{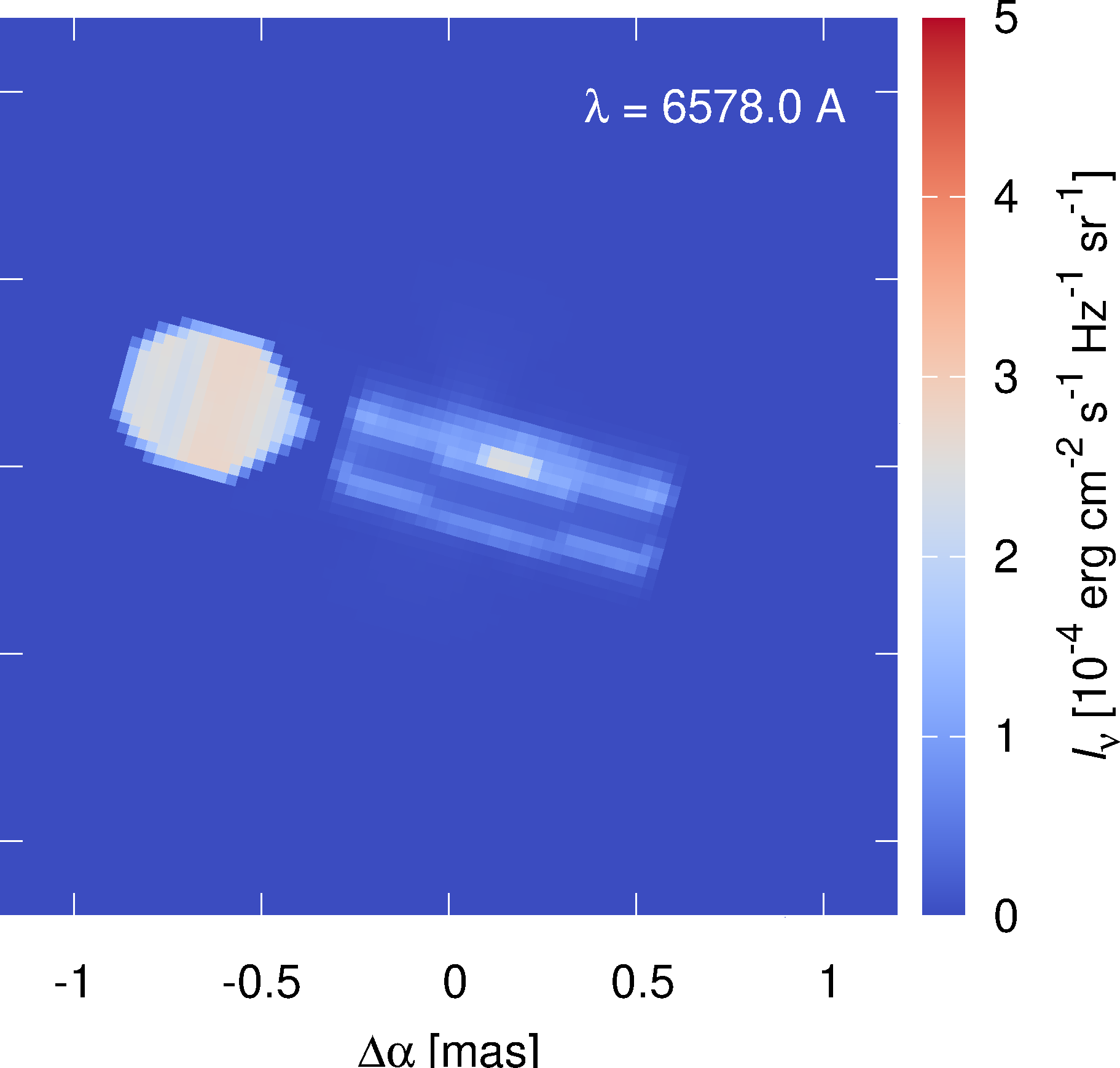}    \\
\end{tabular}
\caption{
Line-profile synthetic images of $\beta$~Lyr~A
computed for the wavelength range of H$\alpha$,
i.e., from 655.1 to 657.8\,nm,
at a {\em fixed\/} phase~0.25.
Optically-thin components are clearly visible.
The disk atmosphere appears first, because its
Keplerian velocities close to the inner rim produce the blue-shifted wing.
The jet inclined towards the observer appears second,
with high velocities being projected to the line of sight.
Finally, there is the spherical shell, with relatively
low velocities spanning the core of H$\alpha$, which obscures other
small-scale structures. For $\lambda > 6563\,\mbox{\AA}$, all objects
disappear in a reverse order.
An animated version is avaiable at
\protect\url{https://sirrah.troja.mff.cuni.cz/~mira/betalyr/}.
}
\label{img_6548.0}
\end{figure*}

A visual comparison of all observed and synthetic datasets is shown in 
Figures~\ref{fitting_shell8_20200415__20_chi2_LC_PHASE},
\ref{fitting_shell8_20200415__20_chi2_VIS},
\ref{fitting_shell8_20200415__20_chi2_CLO},
\ref{fitting_shell8_20200415__20_chi2_T3},
\ref{fitting_shell8_20200415__20_chi2_SED},
\ref{fitting_shell8_20200415__20_chi2_SPE_PHASE},
\ref{fitting_shell8_20200415__20_chi2_VAMP_null}, and
\ref{fitting_shell8_20200415__20_chi2_VPHI_null}.
The resulting geometrical model of $\beta$~Lyr~A in the continuum
is shown in Figure~\ref{img_1550.0}. In particular, we see
optically-thick objects --- the primary, the secondary and the disk ---
and partly also the jets, but not the tenuous shell.
The same model for the wavelength range of H$\alpha$ is shown in Figure~\ref{img_6548.0}.
We can clearly see optically-thin circumstellar matter emitting in H$\alpha$,
including the velocity field.
In the following we describe individual components
of our model, as inferred from the observations:

PRIMARY -- the gainer is an object, for which we fixed several
parameters (Table~\ref{tab1b}). It is mostly hidden in the disk,
but as one can see in the figures its polar region is visible.
It is the source of hot radiation which is also scattered by the
circumstellar medium (CSM) towards the observer.

SECONDARY -- the donor is filling the Roche lobe with
limb and gravity darkening. The limb darkening coefficient
is interpolated for given $\lambda$ from \cite{vanhamme1993} tables.
The gravity darkening parameter was set to a value
suitable for non-convective atmospheres of stars ($0.25$).
One can see that the regions near the L1 point are indeed
dimmer because of it. Its polar temperature is ${\sim}\,14000\,{\rm K}$
and inferred polar radius about $14.1\,R_\odot$.

DISK -- is an axially symmetric accretion disk centered on the gainer.
It has an~outer radius of $31.5\,R_\odot$ and almost fills the Roche lobe.
The density profile decreases with radius and its slope ($-0.57$)
is slightly less steep then in some Algols, where it
attains $-1.0$ \citep{budaj2005,stone2012}.
The temperature profile also decreases and its slope ($-0.73$)
is slightly steeper than a typical profile due to
irradiation ($-0.5$) but is in surprisingly good agreement
with the theoretical temperature profile of the steady viscous accretion disks 
($-0.75$; \citealt{pringle1981}).
The temperature at the inner rim of the disk reaches ${\sim}\,30000\,{\rm K}$, 
which is a very reasonable value, comparable with the temperature
of the gainer. The temperature inversion reaches $1.5$
which means that the temperature increases in the vertical direction
by this factor. This is most probably caused by the irradiation
of the disk atmosphere.
For this reason we see that the disk is brighter on the top and 
bottom and dimmer in the middle. Moreover, the atmosphere scatters
the radiation from the gainer.

The parameter $h_{\rm cnb} = 3.8$ means that the vertical scale height
is multiplied by this factor and is more extended than the expected
equilibrium value. This may be due to non-negligible hydrodynamic flows
within the disk.
There is a significant radial velocity component in the surface layers 
($v_{\rm nb} = 112\,{\rm km}/{\rm s}$) which might be due to stellar
wind or radiative acceleration. 
The turbulence is relatively low ($v_{\rm trbnb} = 11\,{\rm km}/{\rm s}$)
which means that Keplerian and radial components describe the velocity
field very well.

JET -- actually, two conical jets perpendicular to the orbital plane.
They do not seem to be associated with the polar regions of the gainer (cf.~$R_{\rm poljt}$).
The 'net' velocity $v_{\rm poljt}$ assigned to the jets was treated
as a free parameter and we thus have to discuss whether its value
$10\,{\rm km}\,{\rm s}^{-1}$ is reasonable or not.
Because the projected orbital velocities of the primary
and secondary are
$K_1 = 41\,{\rm km}\,{\rm s}^{-1}$ and
$K_2 = -186\,{\rm km}\,{\rm s}^{-1}$
(at the phase 0.25),
we consider it to be reasonable, although indicating that
the jets may not follow Keplerian velocities at the disk rim.

The terminal (expansion) velocity $v_{\rm jt}$ is almost $700\,{\rm km}/{\rm s}$
and the respective exponent ($e_{\rm eveljt} = 1.27$)
is slightly smaller than that of the shell or disk.
Turbulence is about 10~times smaller than the terminal velocity
which indicates that the velocity field is fitted reasonably well.
This object is optically thin in continuum so it is constrained
mainly by observations in the H$\alpha$ line.
Without this object the reduced $\chi^2_{\rm R}$ would increase
up to $59$ which justifies its role in our model
(although a re-convergence might decrease it again).

SHELL -- a spherical object which extends to more than $70\,R_\odot$.
The respective net velocity is low ($-5\,{\rm km}/{\rm s}$);
the shell may not be exactly centered and co-moving with the primary.
Its terminal velocity is very low, only $79\,{\rm km}/{\rm s}$,
but it is interesting that its velocity exponent is similar to that 
of the disk (${\sim}\,1.9$). On the other hand, turbulence is very high
($v_{\rm trbsh} = 102\,{\rm km}/{\rm s}$)
which indicates that the velocity field is not well described
by our formulation (Eq.~(\ref{vrsh})).
This is probably not surprising given that it fills
a broad spherical region in the vicinity of the orbiting stars
where gravitational potential is far from being isotropic and radial.
Surprising is that our data indicate only a small radial temperature gradient (cf.~$e_{\rm tmpsh}$).
It is also optically thin and constrained mainly by interferometry in H$\alpha$.
If this object were excluded from the model the reduced $\chi^2_{\rm R}$
would increase to $62$ which also justifies its role.

Generally, the new model seems to be compatible with our previous model \citep{Mourard_etal_2018A&A...618A.112M},
but we should emphasise that the major difference
is the firm detection of previously conjectured structures (the jets and the shell), which was possible thanks to spectro-interferometric and spectral observations in the \ha region.
There are minor differences, however:
the disk outer rim radius is larger ($R_{\rm outnb} = 31.5$ vs previous $30\,R_\odot$),
the disk thickness slightly smaller ($h_{\rm cnb} = 3.8$ vs $4.3$),
the orbital inclination also larger ($i = 96^\circ$ vs $93.5^\circ$).
All these differences may be enforced by the need of emission
in the H$\alpha$ line, which is enhanced if the disk is more extended
and more inclined. For jets perpendicular to the disk, larger~$i$
leads to larger line-of-sight velocities and also to larger asymmetry
due to the obscuration by the disk, as needed to explain the asymmetric
H$\alpha$~profile.
In this particular model, the asymmetries in the H$\alpha$ profile arise
mostly from overlapping velocity fields of objects with increasing priorities
(shell$\,\rightarrow\,$disk$\,\rightarrow\,$jet).

\begin{table*}
\centering
\caption{Free parameters, $\chi^2$ values for a joint model
and for observation-specific models.}
\label{tab1}

\begin{tabular}{llllllllllll}
\hline
\hline
\vrule width 0pt height 10pt depth 0pt
parameter &
unit &
joint &
LC &
VIS &
CLO &
T3 &
SED &
SPE &
VAMP &
VPHI &
$\sigma$
\\
\hline
\vrule width 0pt height 10pt depth 0pt
$T_{\rm cp}$        & K & 14334 & 14500 & 14353 & 14591 & 14378 & 14406 & 14375 & 14592 & 14580 & 1400 \\
$R_{\rm innb}$      & $R_\odot$ & 8.7 & 8.4 & 10.2 & 8.8 & 9.8 & 8.4 & 8.8 & 7.1 & 8.0 & 1.6  \\
$R_{\rm outnb}$     & $R_\odot$ & 31.5 & 29.2 & 31.2 & 32.8 & 32.2 & 32.0 & 31.0 & 29.7 & 30.6 & 0.3  \\
$h_{\rm invnb}$     & $H$ & 3.5 & 2.9 & 2.9 & 4.1 & 4.0 & 3.0 & 3.6 & 2.9 & 4.8 & 1.0 \\
$T_{\rm invnb}$     & 1 & 1.5 & 1.8 & 1.9 & 1.8 & 1.5 & 1.8 & 1.3 & 2.0 & 1.6 & 1.2 \\
$h_{\rm windnb}$    & $H$  & 3.0 & 3.1 & 3.0 & 3.1 & 3.0 & 3.0 & 3.0 & 3.1 & 4.2 & 3.4 \\
$h_{\rm cnb}$       & $H$  & 3.8 & 4.1 & 4.1 & 4.2 & 4.1 & 3.7 & 3.6 & 3.6 & 12.0 & 1.9 \\
$v_{\rm nb}$        & km/s & 112 & 112 & 197 & 114 & 176 & 96 & 115 & 199 & 117 & 40  \\
$e_{\rm velnb}$     & 1 & $1.91$ & $1.94$ & $1.95$ & $1.99$ & $1.98$ & $1.99$ & $1.99$ & $2.00$ & $1.95$ & 0.31 \\
$h_{\rm shdnb}$     & $H$ & 5.0 & 4.8 & 5.0 & 5.0 & 4.6 & 5.0 & 4.9 & 5.0 & 4.9 & 2.3 \\
$T_{\rm nb}$        & K  & 30345 & 32260 & 30068 & 30539 & 32968 & 29662 & 30716 & 30483 & 33449 & 3300 \\
$\varrho_{\rm nb}$ & $10^{-9}\,{\rm g}/{\rm cm}^3$ & 1.21 & 0.97 & 1.03 & 0.89 & 0.73 & 0.36 & 1.14 & 0.26 & 1.12 & 1.54 \\
$v_{\rm trbnb}$     & km/s  & 11 & 42 & 56 & 13 & 54 & 98 & 15 & 93 & 51 & 17 \\
$e_{\rm dennb}$     & 1 & $-0.57$ & $-0.77$ & $-0.53$ & $-0.69$ & $-0.66$ & $-0.52$ & $-0.60$ & $-0.50$ & $-0.65$ & 0.11 \\
$e_{\rm tmpnb}$     & 1 & $-0.73$ & $-0.71$ & $-0.86$ & $-0.71$ & $-0.71$ & $-1.08$ & $-0.72$ & $-1.04$ & $-0.72$ & 0.07 \\
$a_{\rm jet}$       & deg  & 28.8 & 23.5 & 47.4 & 29.2 & 32.0 & 30.0 & 28.9 & 33.4 & 26.9 & 7.6 \\
$R_{\rm injt}$      & $R_\odot$ & 5.6 & 5.0 & 7.1 & 5.3 & 8.7 & 5.8 & 5.6 & 5.0 & 4.8 & 0.9 \\
$R_{\rm outjt}$     & $R_\odot$ & 35.9 & 31.8 & 35.3 & 41.3 & 44.0 & 44.8 & 36.5 & 34.9 & 33.5 & 9.6 \\
$v_{\rm jt}$        & km/s  & 676 & 1193 & 491 & 1100 & 1386 & 1405 & 686 & 876 & 764 & $\sim$100 \\
$e_{\rm veljt}$     & km/s  & 1.27 & 1.81 & 1.71 & 1.31 & 1.33 & 1.38 & 1.28 & 2.00 & 1.90 & $\sim$0.1 \\
$T_{\rm jt}$        & K  & 15089 & 16989 & 20274 & 15200 & 15682 & 15714 & 14712 & 28182 & 23382 & 1600 \\
$\varrho_{\rm jt}$ & $10^{-12}\,{\rm g}/{\rm cm}^3$ & 5.52 & 12.54 & 4.24 & 4.83 & 4.61 & 4.14 & 5.41 & 2.52 & 6.39 & 4.21 \\
$v_{\rm trbjt}$     & km/s  & 66 & 239 & 101 & 144 & 67 & 278 & 63 & 92 & 166 & $\sim$10 \\
$R_{\rm poljt}$     & $R_\odot$ & 33.0 & 32.8 & 32.5 & 34.9 & 33.6 & 33.2 & 32.2 & 34.9 & 34.1 & 4.3 \\
$v_{\rm poljt}$     & km/s  & 10 & 9 & 18 & 56 & 71 & 27 & 12 & 16 & 60 & 5 \\
$\alpha_{\rm jt}$  & deg  & $-70$ & $149$ & $-34$ & $-43$ & $-26$ & $-6$ & $-71$ & $149$ & $-23$ & 26 \\
$R_{\rm insh}$      & $R_\odot$ & 7.4 & 7.4 & 8.1 & 9.3 & 7.5 & 17.3 & 7.4 & 7.0 & 7.2 & 1.8 \\
$R_{\rm outsh}$     & $R_\odot$ & 72.9 & 62.4 & 60.0 & 86.0 & 76.0 & 100.8 & 73.1 & 84.0 & 67.6 & 25.6 \\
$v_{\rm sh}$        & km/s  & 79 & 100 & 83 & 78 & 81 & 97 & 77 & 97 & 88 & $\sim$10 \\
$e_{\rm velsh}$     & 1 & $1.90$ & $1.38$ & $1.80$ & $1.77$ & $1.86$ & $1.06$ & $1.88$ & $1.96$ & $1.94$ & $\sim$0.1 \\
$v_{\rm ysh}$       & km/s  & $-5$ & $44$ & $-5$ & $-23$ & $18$ & $-22$ & $-9$ & $-12$ & $16$ & 19 \\
$T_{\rm sh}$        & K  & 5631 & 6638 & 5952 & 5705 & 6633 & 5852 & 5639 & 5562 & 6011 & 2300 \\
$\varrho_{\rm sh}$ & $10^{-11}\,{\rm g}/{\rm cm}^3$ & 2.86 & 3.01 & 3.30 & 3.08 & 2.85 & 4.34 & 2.99 & 1.94 & 3.03 & 1.72 \\
$v_{\rm trbsh}$     & km/s & 102 & 149 & 111 & 103 & 109 & 162 & 100 & 128 & 109 & $\sim$10 \\
$e_{\rm tmpsh}$     & 1 & $-0.01$ & $-0.17$ & $-0.14$ & $-0.02$ & $-0.05$ & $-0.07$ & $-0.01$ & $-0.01$ & $-0.06$ & $\sim$0.1 \\
$i$                  & deg & 96.3 & 96.0 & 95.8 & 96.2 & 96.7 & 96.6 & 96.3 & 96.4 & 96.4 & 0.8 \\
$\Omega$            & deg & 254.6 & 254.8 & 253.3 & 254.6 & 254.6 & 254.7 & 254.5 & 254.7 & 254.6 & 2.2 \\
$d$                  & pc & 328.4 & 327.1 & 322.7 & 329.9 & 327.8 & 328.0 & 328.6 & 329.5 & 329.5 & 7.0 \\
\hline\vrule width 0pt height 10pt depth 4pt$N_{\rm iter}$ & --& 2761& 1042& 2561& 1142& 2176& 1918& 1798& 1544& 1543\\
$N$ & -- & 45102 & 2305 & 14354 & 7717 & 2913 & 1815 & 13338 & 1330 & 1330\\
$\chi^2$ & -- & 767681 & 7083 & 56941 & 25910 & 16194 & 8578 & 588455 & 5959 & 58562\\
$\chi^2_{\rm R}$ & -- & 17.0 & 3.1 & 4.0 & 3.4 & 5.6 & 4.7 & 44.1 & 4.5 & 44.0\vrule width 0pt height 0pt depth 4pt\\
\hline\vrule width 0pt height 10pt depth 4pt$\chi^2$ (spec.) & -- & & 5176 & 63270 & 24604 & 21011 & 3866 & 557963 & 1977 & 31662\\
$\chi^2_{\rm R}$ (spec.) & -- & & 2.2 & 3.7 & 3.2 & 3.6 & 2.1 & 41.8 & 1.5 & 23.8\\

\hline
\end{tabular}

\tablefoot{
$T_{\rm cp}$ denotes the temperature at the pole of the secondary (donor).
{\bf Disk} (a.k.a. nebula):
$R_{\rm innb}$ inner radius,
$R_{\rm outnb}$ outer radius,
$h_{\rm invnb}$ inversion height,
$T_{\rm invnb}$ temperature inversion factor,
$h_{\rm windnb}$ wind region height,
$h_{\rm cnb}$ scale height factor,
$v_{\rm nb}$ terminal radial velocity,
$e_{\rm velnb}$ its slope,
$h_{\rm shdnb}$ shadowing height,
$T_{\rm nb}$ temperature at the inner radius,
$\varrho_{\rm nb}$ gas density (ditto),
$v_{\rm trbnb}$ turbulent velocity,
$e_{\rm dennb}$ density slope,
$e_{\rm tmpnb}$ temperature slope.
{\bf Jet:}
$a_{\rm jet}$ opening angle,
$R_{\rm injt}$ inner radius,
$R_{\rm outjt}$ outer radius,
$v_{\rm jt}$ terminal velocity,
$e_{\rm veljt}$ velocity slope,
$T_{\rm jt}$ temperature,
$\varrho_{\rm jt}$ density at the inner radius,  
$v_{\rm trbjt}$ turbulent velocity,
$R_{\rm poljt}$ radial offset,
$v_{\rm poljt}$ polar velocity,
$\alpha_{\rm jt}$ polar angle.
{\bf Shell:}
$R_{\rm insh}$ inner radius,
$R_{\rm outsh}$ outer radius,
$v_{\rm sh}$ terminal velocity,
$e_{\rm velsh}$ its slope,
$v_{\rm ysh}$ net velocity,
$T_{\rm sh}$ temperature,
$\varrho_{\rm sh}$ density,
$v_{\rm trbsh}$ turbulent velocity,
$e_{\rm tmpsh}$ temperature slope.
$i$ is orbital inclination,
$\Omega$ longitude of ascending node, and
$d$ distance.
}
\end{table*}

\begin{table}
\centering
\caption{Fixed parameters for the joint and observation-specific models.}
\label{tab1b}

\begin{tabular}{llllllllllll}
\hline
\hline
\vrule width 0pt height 10pt depth 4pt
parameter &
unit &
value
\\
\hline
\vrule width 0pt height 10pt depth 0pt
$R_\star$        & $R_\odot$ & 5.987   \\
$T_\star$        & K         & 30000   \\
$M_\star$        & $M_\odot$ & 13.048  \\
$q$              & 1         & 0.223   \\
$d_{\rm gcp}$    & 1         & 0.25    \\
$a_{\rm neb}$    & $H$       & 5.0     \\
$h_{\rm velnb}$  & $H$       & 3.0     \\
$a_{\rm symjt}$  & 1         & 0.0     \\
$a\sin I$        & $R_\odot$ & 58.19   \\
$\gamma$         & km/s      & $-18.0$ \\
\hline
\end{tabular}

\tablefoot{
$R_\star$ denotes the radius of the primary (gainer), 
$T_\star$ its effective temperature,
$M_\star$ mass,
$q$ mass ratio,
$d_{\rm gcp}$ gravity darkening coefficient,
$a_{\rm neb}$ extent of nebula,
$h_{\rm velnb}$ minimum height for radial velocity,
$a_{\rm asymjt}$ asymmetry of jet,
$a\sin i$ projected semimajor axis, and
$\gamma$ systemic velocity.
}
\end{table}

\subsection{Observation-specific models}\label{sec:obsspe}

\begin{figure}
\includegraphics[width=9cm]{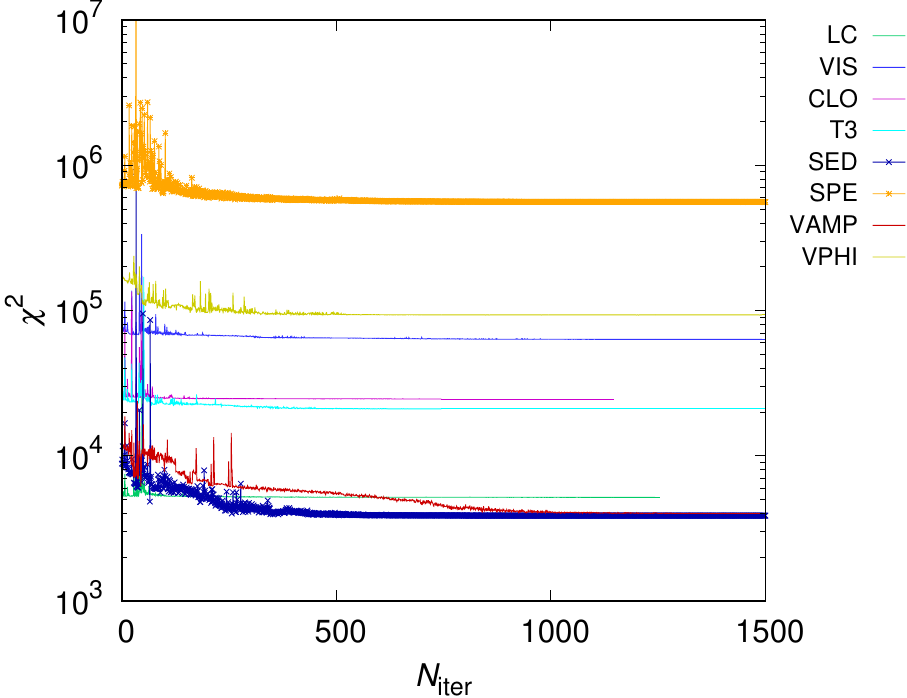}
\caption{The $\chi^2$ convergence for observation-specific models,
when starting from the best-fit joint model.
After performing up to $10^3$ iterations,
substantial improvements were achieved for some datasets
(LC, T3, SED, VAMP, \dots).
It confirms systematic differences between observational datasets.
}
\label{fitting_shell9_20200415_chi2_iter}
\end{figure}

Starting from the `compromise' model above,
we converged the model again
to fit individual datasets
to understand the trends
and potential disagreements.
From the $\chi^2$ convergence (Figure~\ref{fitting_shell9_20200415_chi2_iter}),
it is evident that our model is indeed capable to fit individual datasets better.
For example, the reduced $\chi^2_{\rm lc}$ can easily reach 2.2 (instead of 3.1).
For other $\chi^2$ values, see Table~\ref{tab1} (last row).
Consequently, we think there are either systematic differences between datasets,
or our (complex) model is still not complete.
We may miss additional objects,
some asymmetries,
or a temporal variability.
Alternatively, we may modify weights of individual datasets (and use, e.g., $w_{\rm lc} = 10$),
but this does not `solve' the problem, of course.

\begin{figure*}
\centering
\tmpdim=3.0cm
\begin{tabular}{l@{\kern1mm}l@{\kern1mm}l@{\kern1mm}l}
LC & VIS & CLO & T3 \\
\includegraphics[width=\tmpdim]{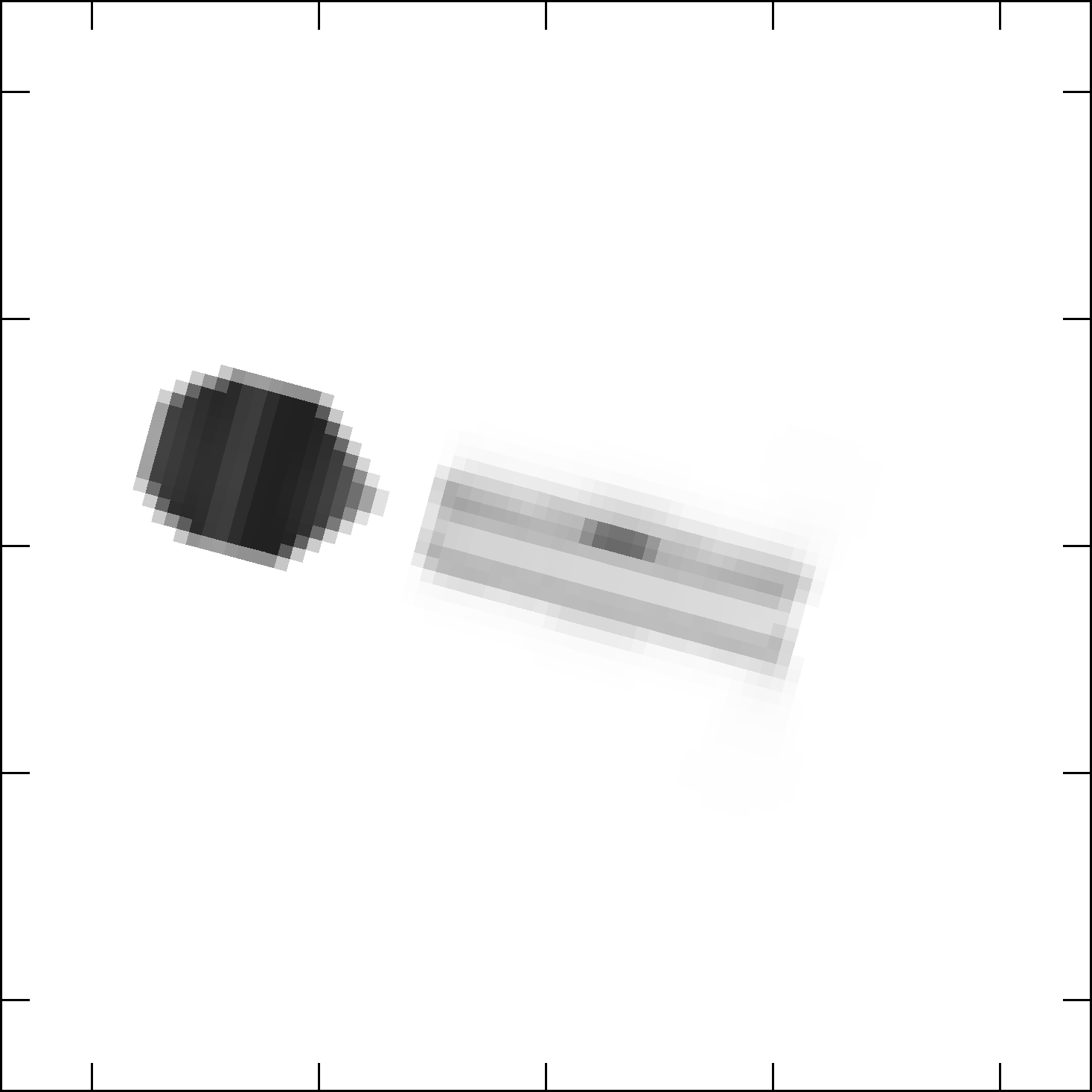} &
\includegraphics[width=\tmpdim]{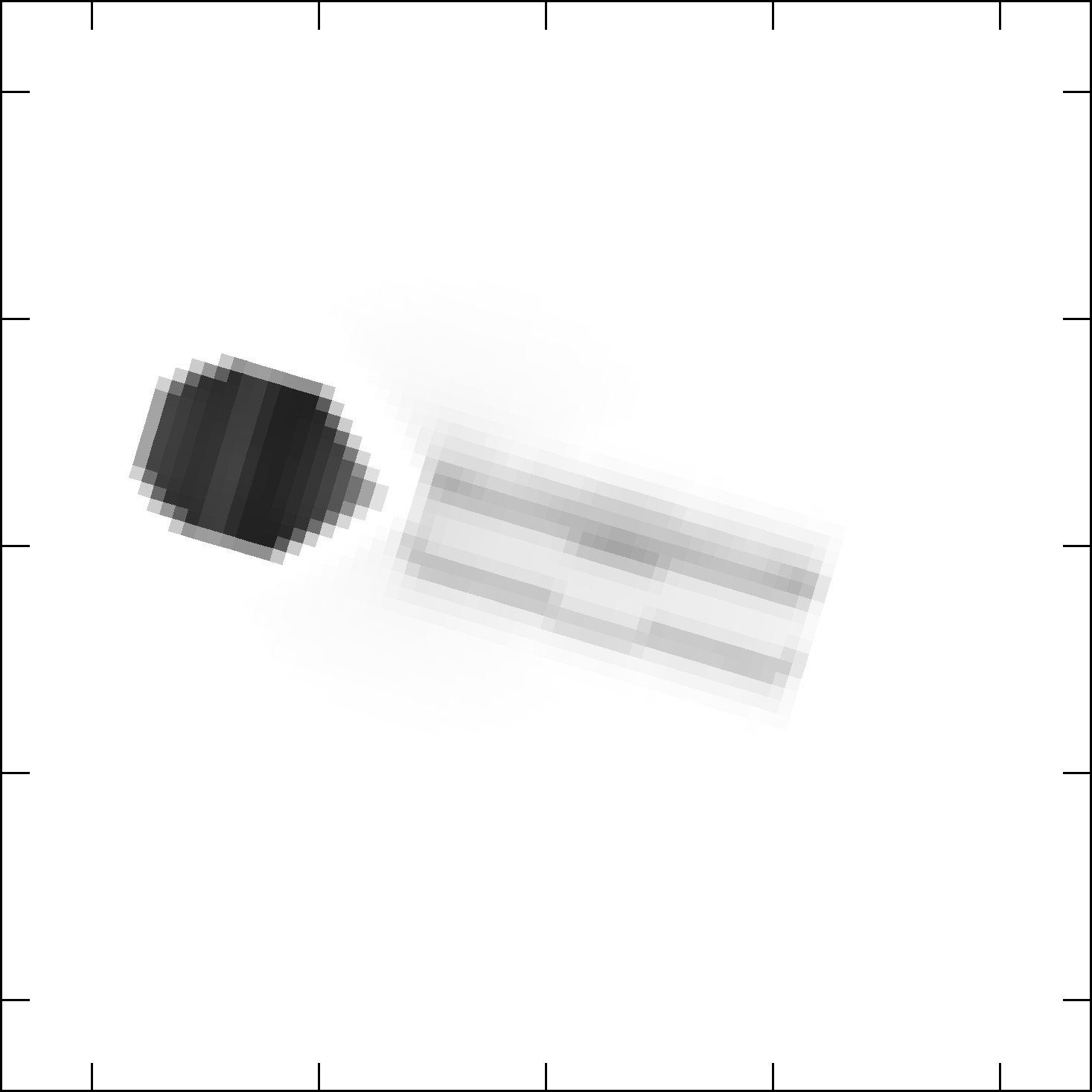} &
\includegraphics[width=\tmpdim]{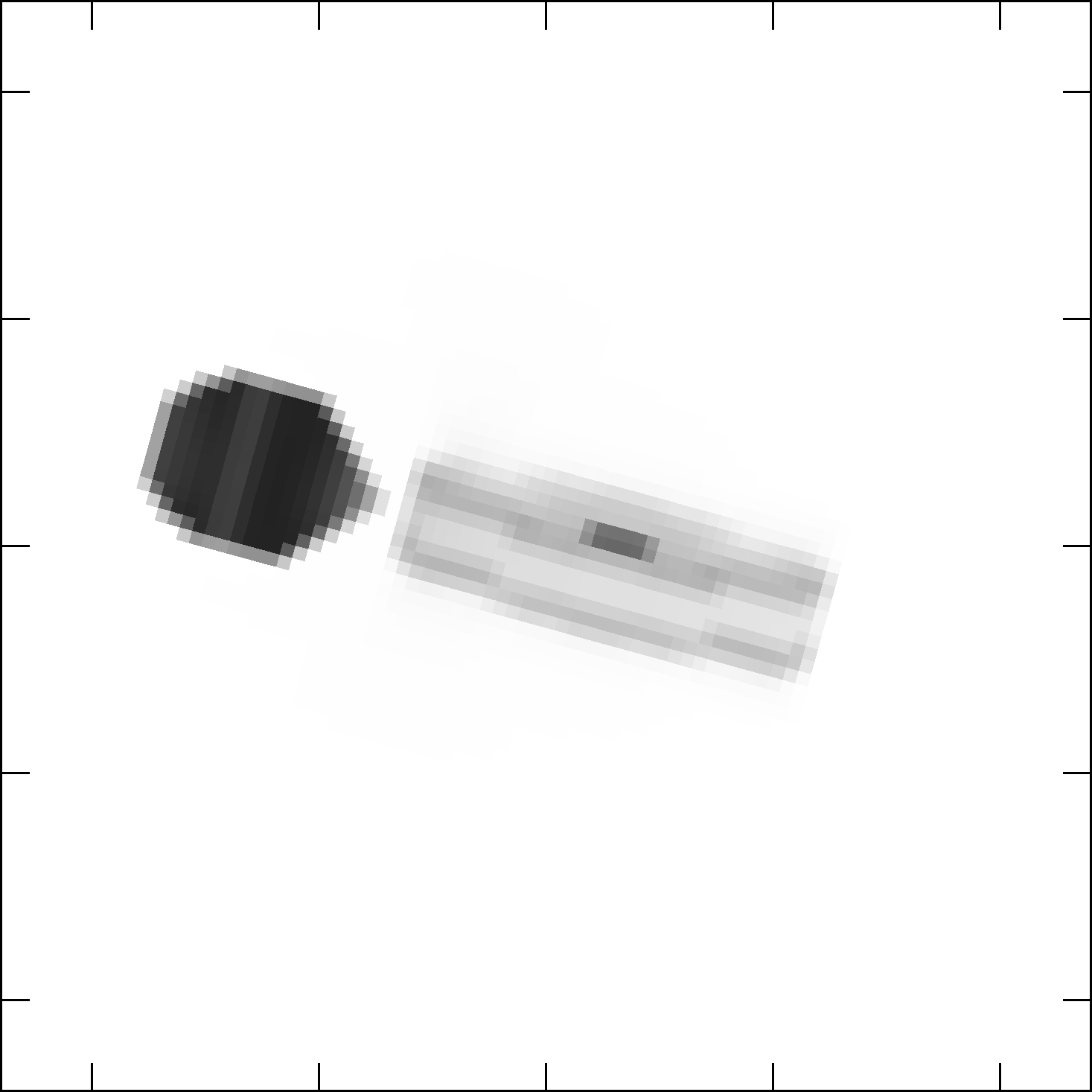} &
\includegraphics[width=\tmpdim]{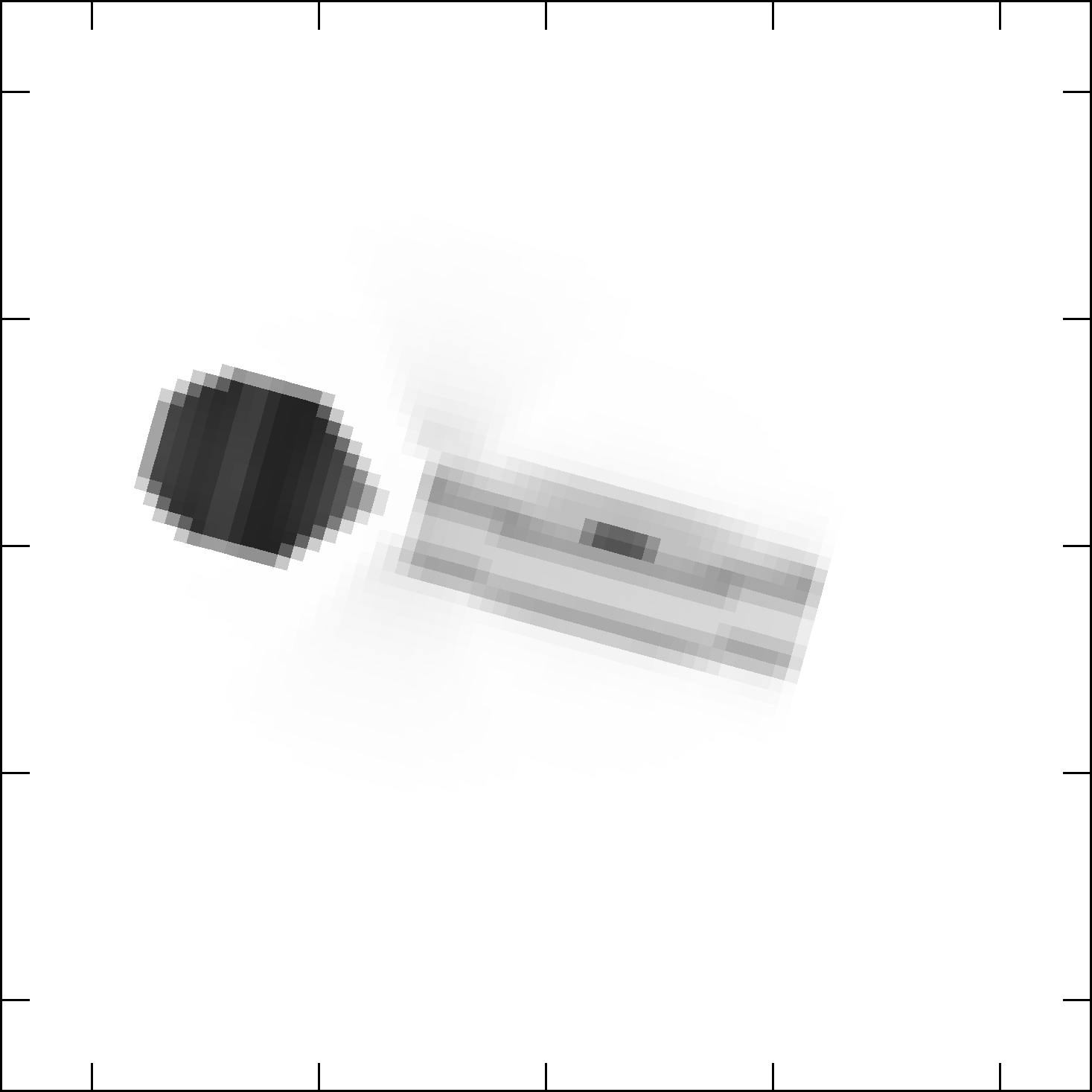} \\
SED & SPE & VAMP & VPHI \\
\raise0.5cm\hbox{\includegraphics[width=\tmpdim]{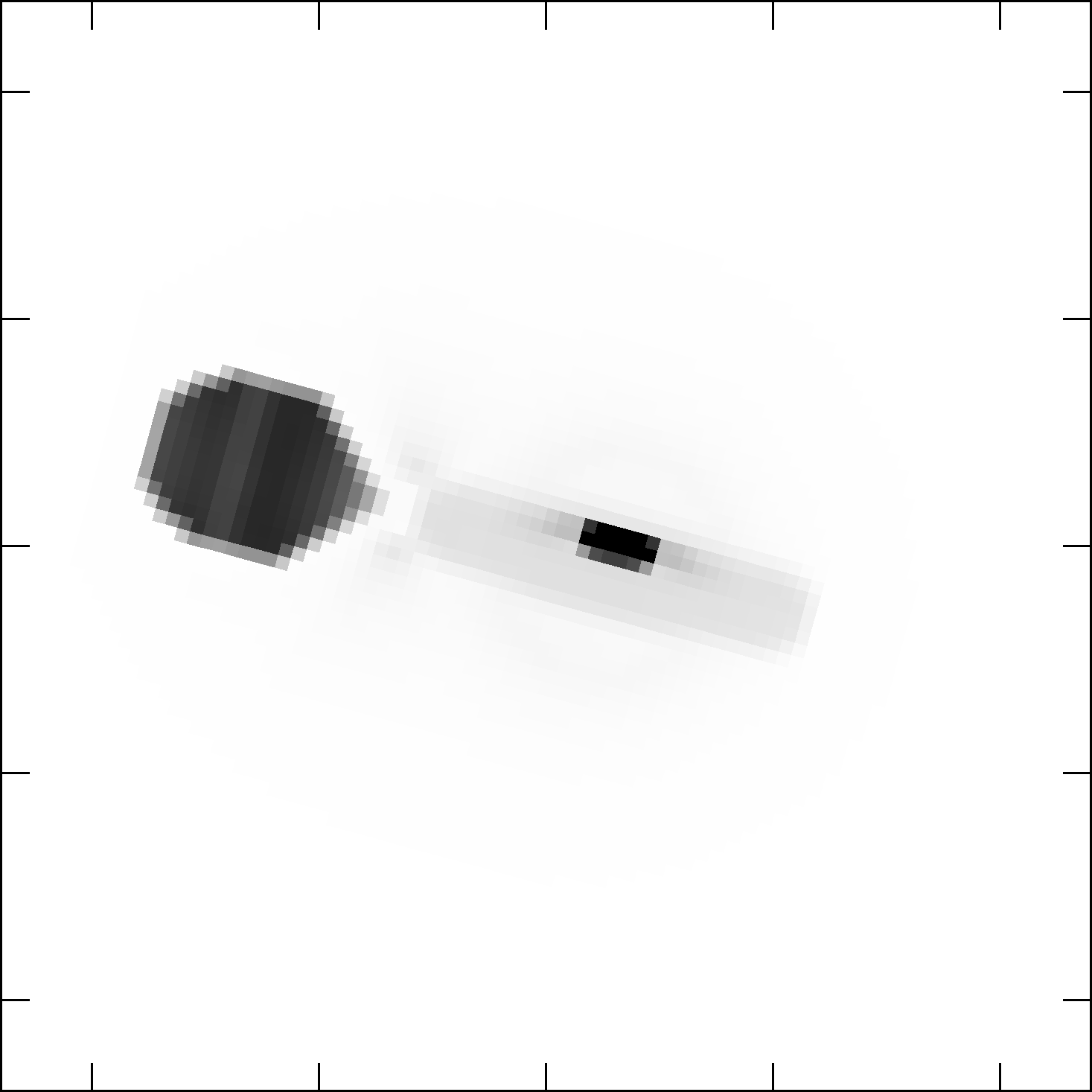}} &
\raise0.5cm\hbox{\includegraphics[width=\tmpdim]{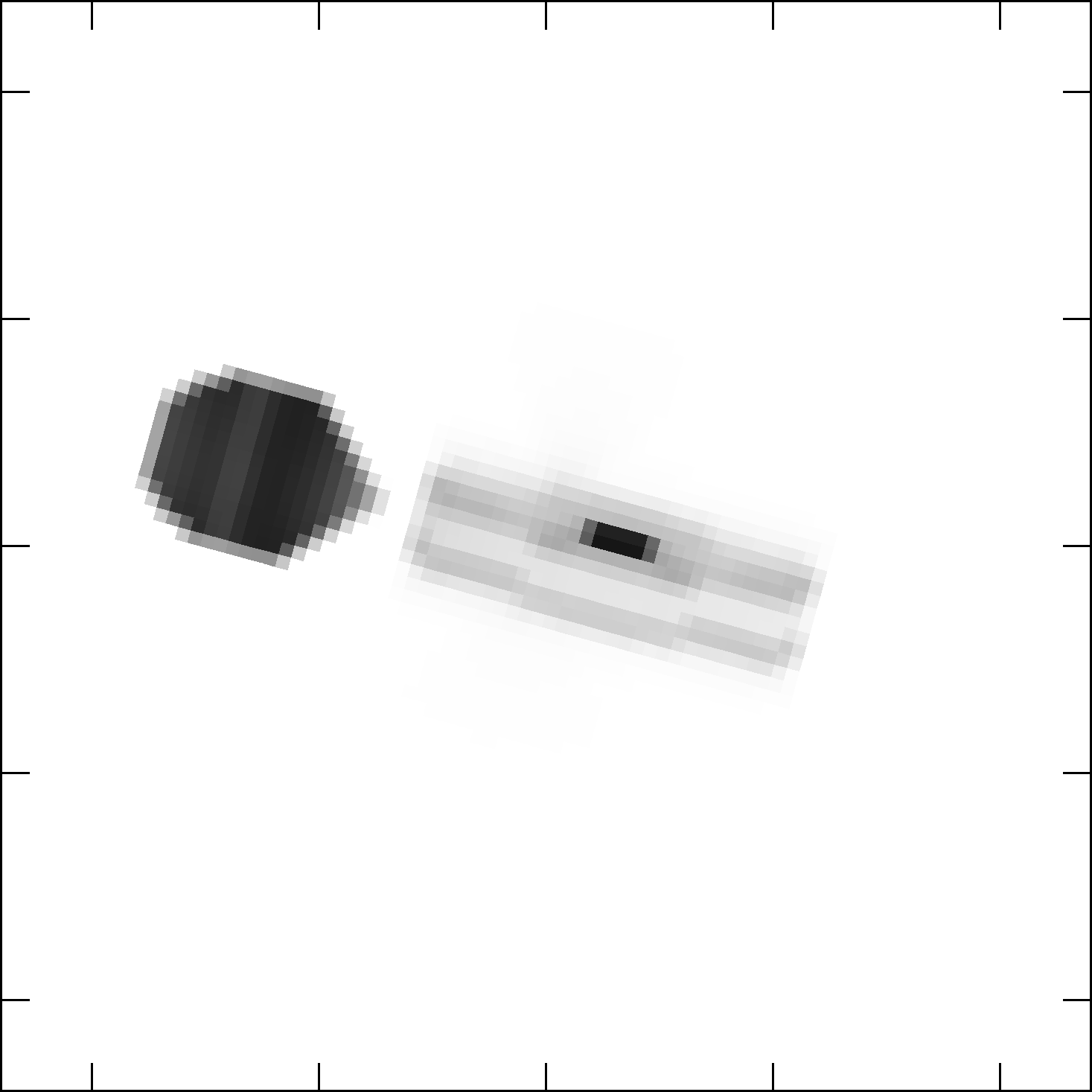}} &
\raise0.5cm\hbox{\includegraphics[width=\tmpdim]{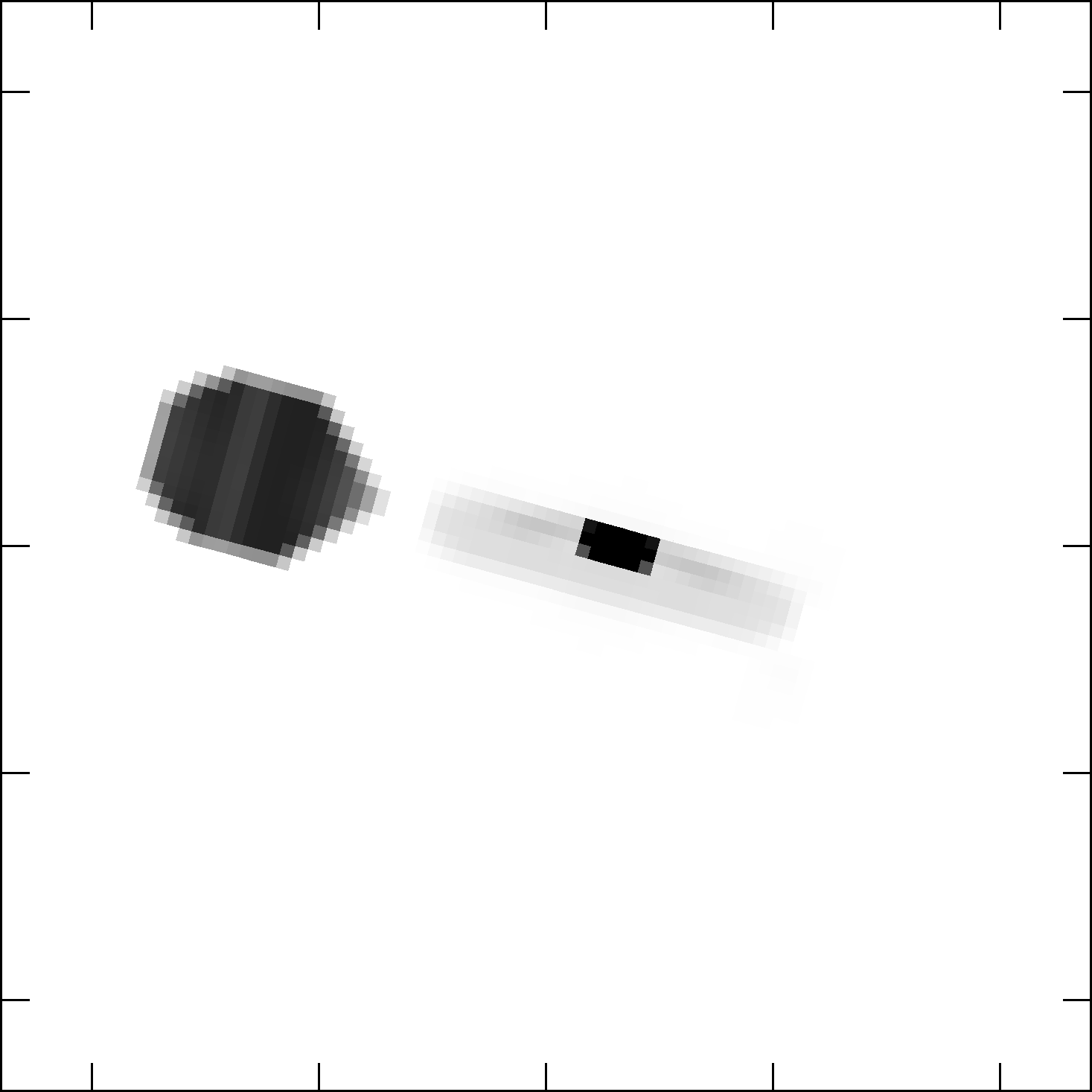}} &
\includegraphics[width=4.275cm]{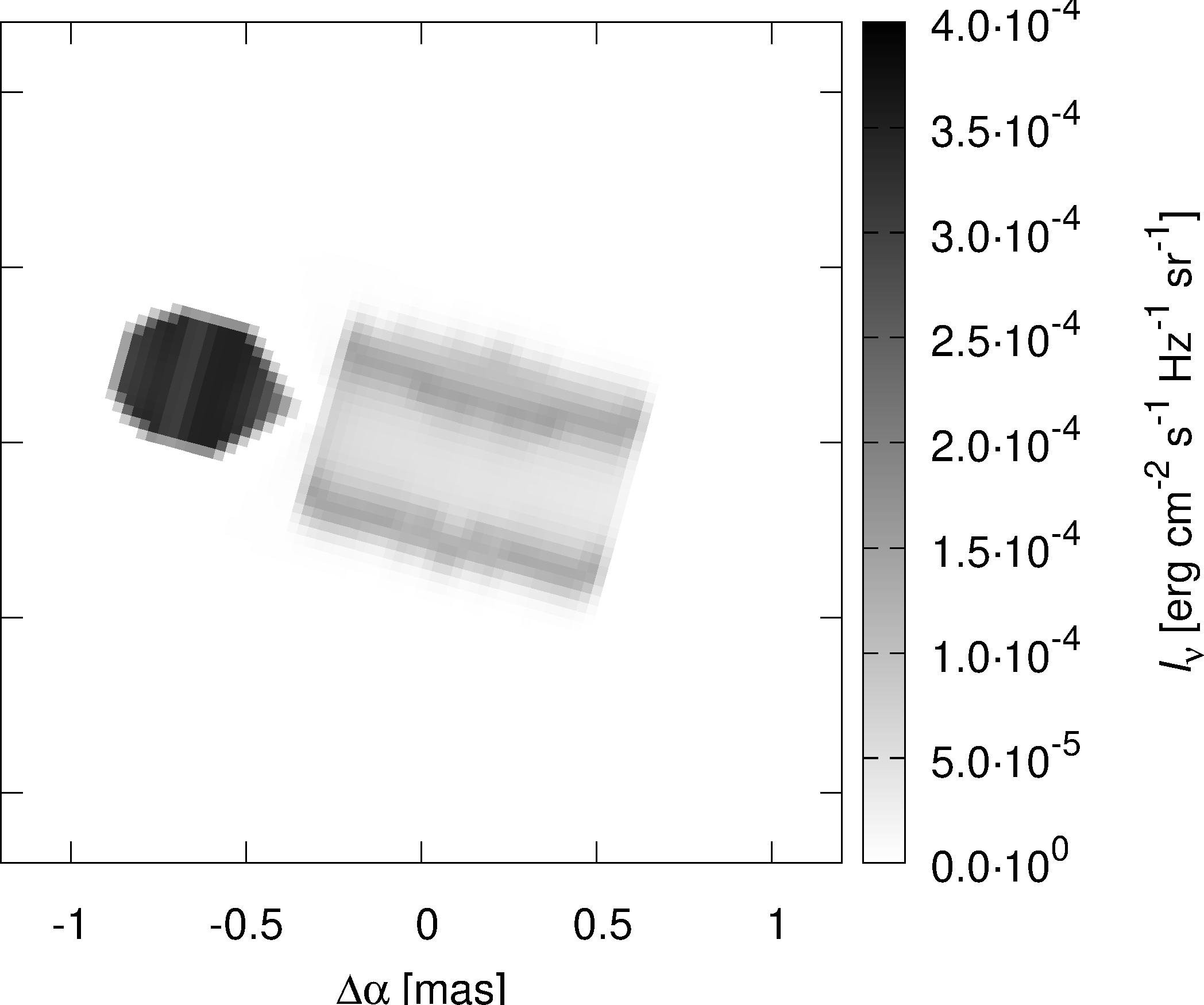} \\
\end{tabular}
\caption{
Continuum synthetic images for observation-specific models
(datasets LC, VIS, CLO, T3, SED, SPE, VAMP, VPHI)
for the wavelength 545\,nm (V).
The apparent differences
(e.g., the thickness of the disk,
the appearance of the primary)
demonstrate systematics between datasets.
Alternately, some datasets (e.g., VPHI) do not constrain certain parameters.
}
\label{itting_shell4_SYNSLIPS_LC_img_5450.0}
\end{figure*}

The results of all observation-specific models are summarized in Table~\ref{tab1}
(columns 'LC' to 'VPHI'). For the V band, it is also possible to compare the models
visually, in Figure~\ref{itting_shell4_SYNSLIPS_LC_img_5450.0}.
The differences are demonstrated as the disk thickness and the intensity
of its outer edge, which is proportional to the temperature profile~$T(r)$.
In most models, the primary is directly visible, but the LC dataset
tends to produce a continuum emission from a more extended hot area.

Given these results, uncertainties of model parameters (cf. Tab.~\ref{tab1};
last column) were determined as the maximum differences
between the joint model and {\em relevant\/} observation-specific models,
because they are almost certainly dominated by systematics,
not by the extent of local $\chi^2$ minima,
not even by the global one.
To be more specific, $\sigma$'s for all velocities
can be only constrained by the joint model and the SPE, VAMP and VPHI datasets;
similarly, $\sigma$'s for flux-related quantities ($\varrho$, $T$, $R$)
can be hardly constrained by relative measurements.

\section{Stellar evolution modelling}\label{evolution}

\begin{figure}
\centering
\includegraphics[width=8.5cm]{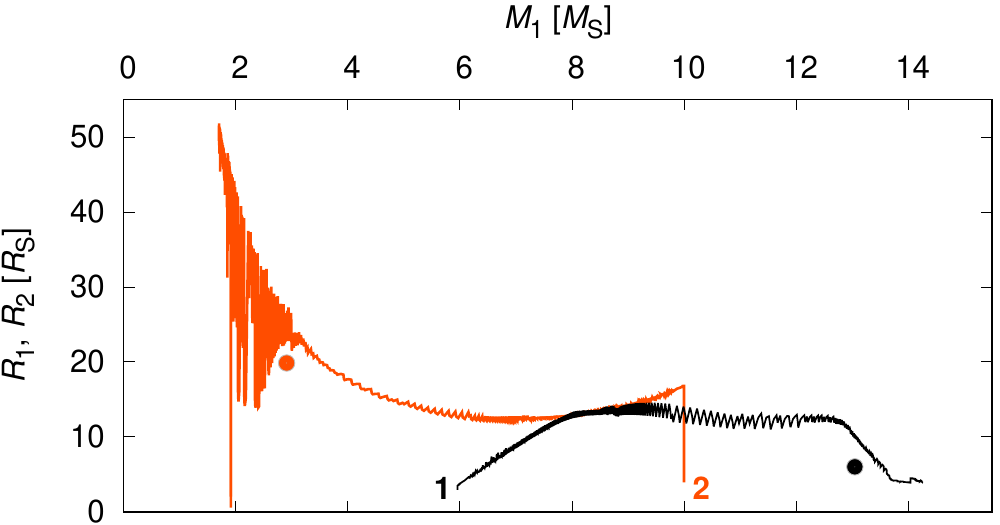}
\includegraphics[width=8.5cm]{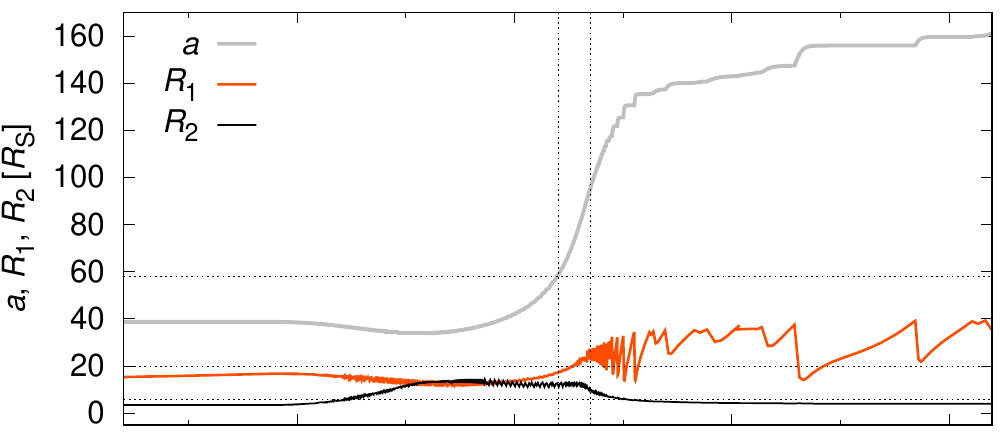}
\includegraphics[width=8.5cm]{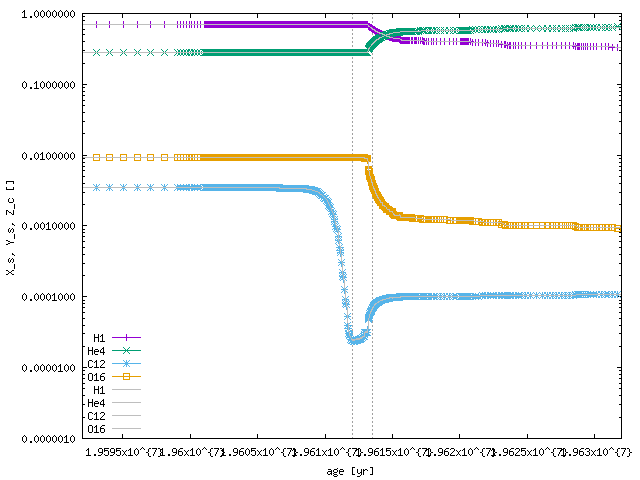}
\caption{
Stellar evolution of a binary corresponding to $\beta$~Lyr~A.
The component radii $R_1$, $R_2$ vs the primary mass $M_1$ are plotted (top),
with observed values indicated by full circles,
together with the distance $a$ and $R_1$, $R_2$ vs time~$t$ (middle),
cf. horizontal dashed lines,
and surface abundances $X$, $Y$, $A_{\rm C12}$, $A_{\rm O16}$ vs time~$t$ (bottom).
At $t = 19.613\cdot 10^6\,{\rm yr}$, the mass ratio~$q$ corresponds
to the observed one.
}
\label{M1_10.0_M2_6.0_P_07.0_at}
\end{figure}

To understand the evolutionary stage of $\beta$~Lyr~A binary,
and its relation to the observations,
we performed a simplified 1D modelling with the MESA stellar evolution program
\citep{Paxton_etal_2011ApJS..192....3P,Paxton_etal_2015ApJS..220...15P}.
In Figure~\ref{M1_10.0_M2_6.0_P_07.0_at}, we present a nominal evolution
of a binary with the initial masses
$M_1 = 10\,M_\odot$,
$M_2 = 6\,M_\odot$,
and the orbital period
$P = 7\,{\rm d}$.
We assumed the solar composition.
We used an explicit Ritter scheme
and we restricted the mass accretion rate~$\dot M$ up to
$10^{-3}\,M_\odot\,{\rm yr}^{-1}$.
Although we performed a small survey of parameters
($P = 5$ to $50\,{\rm d}$,
$M_1 = 8$ to $10\,M_\odot$,
while keeping $M_1 + M_2 = {\rm const.}$),
we restrict our discussion to the nominal case,
because other values lead to binaries incompatible with $\beta$~Lyr~A,
or to a common-envelope phase which is difficult to describe in 1D.
The mass ratio~$q$ is already reversed,
so that index 1 corresponds to the observed secondary (donor),
and 2 to the primary (gainer).

A comparison to the observed values of $a$, $R_1$, $R_2$ shows
some differences: an offset $10^3\,{\rm yr}$ between the time of
best-fit for~$a$ and the best-fit for $M_1$, $M_2$ (see dotted vertical lines).
The synthetic radii are a factor of 1.2 and 2 larger at this moment.
However, let us recall it is only a 1D model, without an accretion disk.
Consequently, we consider these differences to be acceptable.
The model of \cite{vanrensbergen2016} produced a qualitatively
similar HRD for the gainer, but their initial period was shorter,
$P = 2.36\,{\rm d}$.

At this stage, the surface chemical composition is already modified.
At $t = 19.612\cdot 10^{6}\,{\rm yr}$, there is a low C abundance
(by a factor of $10^2$) and a high N abundance (by a factor of~5).
At $t = 19.613\cdot 10^{6}\,{\rm yr}$, even He abundance is
increased up to 0.36 (and H correspondingly decreased).
For our modelling, it means we should also test models
with a substantially modified chemical composition.
This is in accord with \cite{balach86} who suggested
He enrichment
$N({\rm H}) = 0.4$,
$N({\rm He}) = 0.6$ (by number),
and also
N to be overabundant, as well as
C, O underabundant, namely
${\rm C}/{\rm N} \le 0.11$,
${\rm O}/{\rm N} \le 0.25$.


Interestingly, further evolution would lead to a detached system
with a stripped He dwarf (secondary).
A hot subdwarf of the sdB or sdO type is expected  
\citep{Ulrich_2009ARA&A..47..211H,Lei_etal_2018ApJ...868...70L}.
$\phi$~Per binary might be
just in this (late) evolutionary stage \citep{Mourard_etal_2015A&A...577A..51M}
and a dedicated comparative study might be very useful.

\section{Alternative models}

\subsection{High-resolution model}

For a spatial resolution increased twice to $1\,R_\odot$,
some datasets are fitted even better,
e.g., $\chi^2_{\rm lc}$ (not reduced) decreased from 7083 to 6720,
because the primary and the disk rim are better resolved
and thus contribute more to FUV, NUV fluxes.
On contrary, $\chi^2_{\rm sed}$ increased from 8578 to 9832,
due to the same reasons.
The most sensitive term seems to be $\chi^2_{\rm spe}$
which increased substantially from 588454 to 682016,
because H$\alpha$ line profiles are slightly 'sharper'
and the \ion{He}{i} 6678 emission is enhanced;
although the profiles remain qualitatively very similar.
Other contributions are slightly decreased.
This conclusion is preliminary, though, without a repeated convergence.
In principle, even the high-resolution model could be converged again
which would decrease the increased~$\chi^2$.
We conclude the model is resolution dependent.
However, this is not necessarily a bad thing;
a low-resolution model may simply represent shallower gradients,
or not so sharp transitions between objects.



\subsection{Distance fixed to $294\,{\rm pc}$}\label{294pc}

First, let us compare our distance with \cite{Bastian_2019A&A...630L...8B}, who discovered the Gaia~8 cluster,
with parallaxes around $3.4\,{\rm mas}$ or distance $294\,{\rm pc}$,
and the intrinsic dispersion of only $0.06\,{\rm mas}$
(i.e., $5\,{\rm pc}$ radial, $1^\circ$ spatial).
$\beta$~Lyr~A is located in the middle (spatially);
if it is also in the centre of mass,
then our photo-spectro-interferometric value
$d = (328\pm 7)\,{\rm pc}$
is substantially larger.

If we fix the distance in our model to $d = 294\,{\rm pc}$ instead,
and converge the model, we obtain different parameters, of course.
In particular, in Tab.~\ref{tab2} (column '294pc') we can see that 
secondary temperature $T_{\rm cp} = 13512\,{\rm K}$ is lower, and
disk outer radius $R_{\rm outnb} = 35.2\,R_\odot$ is larger.
While the overall fit seems better,
$\chi^2_{\rm R} = 16.9$ (as compared to 17.0),
mostly because $\chi^2_{\rm spe}$, $\chi^2_{\rm vamp}$
contributions were improved,
it is at the expense of other terms being much worse!
Especially $\chi^2_{\rm sed}$ was increased three times
which is unacceptable for us,
and $\chi^2_{\rm vis}$, $\chi^2_{\rm clo}$ were increased too.
Moreover, disk outer radius is too large and overshoots
not only the tidal cutoff radius $26.3\,R_\odot$
(possible in principle as the disk is not an isolated system during ongoing mass transfer), but also its Roche lobe.
This is the reason why we still prefer the original model.



\subsection{The mass ratio}

We cannot make the primary~$M_\star$ and secondary mass $M_{\rm cp}$ free
when we keep $a\sin i$ and the period~$P$ 'fixed'.
Nevertheless, if we free the mass ratio $q$ (and also all other parameters),
it affects mainly the size of the donor, which is the major source of light.
We obtained parameters shown in Tab.~\ref{tab2} (column 'QRATIO').
While a majority of them remained close to the previous (local) minimum, the value of $q = 0.2177$ is lower.
Because we introduced one more parameter,
it is logical that the overall fit is better,
with $\chi^2_{\rm R} = 16.5$.
However, we do not consider the respective changes of parameters
to be substantial.



\subsection{Spot-like asymmetry}\label{SPOT}

Similarly as in \cite{Mourard_etal_2018A&A...618A.112M}, we introduced
a spot in our model, which represents an additional spherical object.
We converged not only spot parameters, but also all other parameters,
to be sure that all objects can adapt to new geometrical constraints.
The results are listed in Tab.~\ref{tab3} (for the spot itself)
as well as in Tab.~\ref{tab2} (column 'SPOT').
We see a minor improvement of the light curve, spectra and
differential interferometry, at the expense of other datasets though
(the reduced $\chi^2_{\rm R}$ decreased to $16.8$).
The position of the spot converged close to the donor--gainer line
and the distance corresponds to the outer radius of the disk.
Consequently, such a spot may represent either
a part of the flow from the donor,
the base of the jets, or
an asymmetry of the disk rim.
We consider the existence of this spot likely,
although not so prominent as before (cf.~\citealt{Mourard_etal_2018A&A...618A.112M}),
because there are additional objects in our model.

\begin{table}
\centering
\caption{Free parameters related to the spot.}
\label{tab3}

\begin{tabular}{llll}
\hline
\hline
\vrule width 0pt height 10pt depth 0pt
parameter &
unit &
SPOT &
\\
\hline
\vrule width 0pt height 10pt depth 0pt
$R_{\rm sp}$        & $R_\odot$ & $5.28$ &  \\
$T_{\rm sp}$        & K  & 7146 &  \\
$\varrho_{\rm sp}$ & $10^{-9}\,{\rm g}/{\rm cm}^3$ & 9.97 &  \\
$R_{\rm polsp}$     & $R_\odot$ & 29.9 &  \\
$v_{\rm polsp}$     & km/s  & 2 &  \\
$\alpha_{\rm sp}$  & deg  & $9$ &  \\
\hline
\end{tabular}

\tablefoot{
$R_{\rm sp}$~denotes the spot radius,
$T_{\rm sp}$~temperature,
$\rho_{\rm sp}$~density,
$R_{\rm polsp}$~radial offset,
$v_{\rm polsp}$~polar velocity, and
$\alpha_{\rm sp}$~polar angle.
The remaining parameters were included in Tab.~\ref{tab2}, column 'SPOT'.
}
\end{table}



\subsection{Asymmetric jets}

Our stringent geometrical constraints may be partly relaxed
by using the parameter $a_{\rm symjt}$ which allows for an
asymmetry of the jets (Eq.~(\ref{rhojt})). The resulting model
is shown in Tab.~\ref{tab2}, column 'ASYMJT'.
As before, we introduced one more parameter
and it is not surprising that the fit is better,
with $\chi^2_{\rm R} = 16.5$.
However, the resulting value $a_{\rm symjt} = 0.02$ is not
far from zero. It seems that this additional parameter actually
allowed for tiny adjustments of other parameters
and we cannot conclude that jets are asymmetric.



\subsection{Jet temperature gradients}\label{ETMPJT}

The fit of the \ion{He}{i} 6678 line is far from being perfect.
To improve it, we used a model with a substantial temperature
gradient $T(r)$ in the jets. We obtained a significantly better fit
($\chi^2_{\rm R} = 16.4$; see Tab.~\ref{tab2}, column 'ETMPJT')
with the slope $e_{\rm tmpjt} = -0.55$
and the temperature at the base of the jets up to
$T_{\rm jt} \doteq 30000\,{\rm K}$. 
It seems to improve both
H$\alpha$ and \ion{He}{i} line profiles. Although the systematics
in the core of \ion{He}{i} line remained qualitatively the same,
there are improvements in the wings of both lines.
The parameters of these jets are reasonable because their temperature
corresponds very well to the temperature of the gainer
and the exponent to the heating by irradiation from the gainer.
Consequently, it may be considered as our preferred model.



\subsection{Shell temperature gradients}\label{ETMPSH}

We also tried to enforce the temperature gradient in the shell
by initially decreasing the slope~$e_{\rm tmpsh} = -0.5$
and adjusting the temperature $T_{\rm sh} = 20000\,{\rm K}$
accordingly (to end up with 6000\,K at $R_{\rm outsh}$).
All parameters were converged again and the result is shown
in Tab.~\ref{tab2} (column 'ETMPSH'). We can see the gradient
is preserved, but the fit is worse with $\chi^2_{\rm R} = 18.1$,
especially the $\chi^2_{\rm spe}$ term.
The problem is that the gradient affects also all other lines
(H$\alpha$, \ion{Si}{ii}, \ion{Ne}{i}),
creates an excess emission which prevents further convergence.



\subsection{Low C abundance}\label{LOWC}

\begin{figure}
\includegraphics[width=8.5cm]{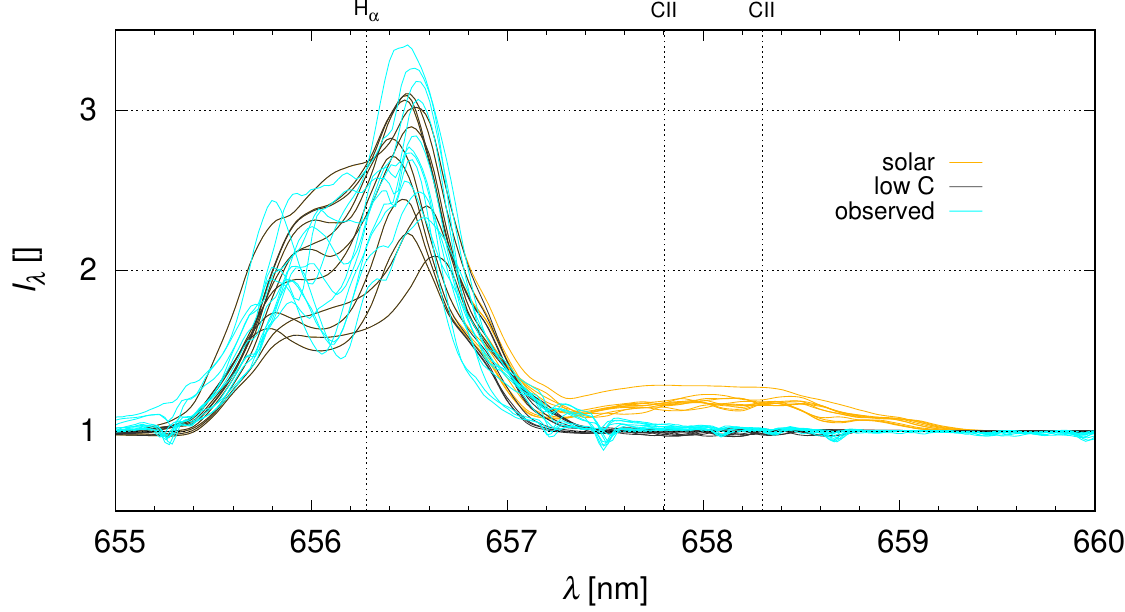}
\caption{Synthetic spectra for the solar abundances (yellow)
and $10^{-2}$ lower abundance of~C (gray); observed spectra (blue)
are plotted for comparison. For the solar composition,
\ion{C}{ii} 6578 and 6583 emission is too strong.}
\label{fitting_shell7_LOWC__51_chi2_SPE}
\end{figure}

We tried to use higher abundances for Si and Ne 
in order to explain the depth of the respective lines.
On the other hand, according to Section~\ref{evolution},
there might be 100 times lower abundance of~C
in surface layers and consequently in the CSM.
Our spectra do contain the region of \ion{C}{ii} 6578 and 6583 lines,
but they are too weak.
Nevertheless, we checked synthetic spectra with these transitions
and solar abundance ($3.31\cdot10^{-4}$ by the number of atoms;
\citealt{Grevesse_Sauval_1998SSRv...85..161G}).
It turned out it would create so strong \ion{C}{ii} emission,
similarly wide as H$\alpha$, that the abundance must be low
(see Figure~\ref{fitting_shell7_LOWC__51_chi2_SPE}).
The abundance $10^{-2}$ of the solar value is fully compatible
with observations; the upper limit is about~$10^{-1}$.
Subsequently, these low abundances were also applied
in the 'joint' model.


\subsection{He-rich abundance}\label{H0.4_He0.6}

If we would like to use the non-solar abundances of
\cite{balach86}, the situation is much more complicated.
We must not use standard atmospheric models,
because the abundances $N({\rm H})=0.4$, $N({\rm He})=0.6$
do change their hydrostatic profiles and the emerging
spectra must be different.

\begin{figure}
\centering
\includegraphics[width=6cm]{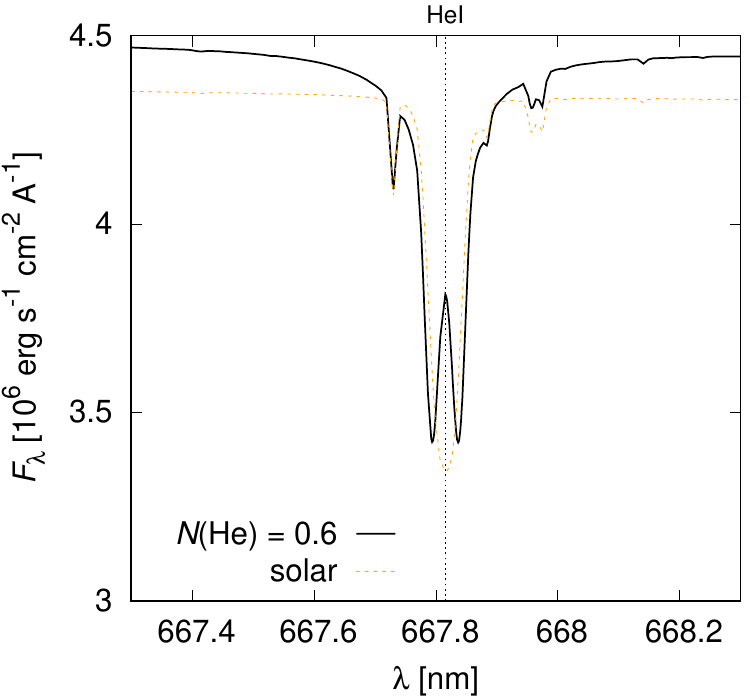}
\caption{Non-rotated NLTE synthetic spectra computed for stellar
atmospheres with He-rich ($N({\rm H} = 0.4$), $N({\rm He}) = 0.6$; black) and solar (orange) compositions.
The effective temperature was 14000\,K in both cases.
For the He-rich case, the profile exhibits a central absorption.}
\label{n14000g23henl.7_HeI_}
\end{figure}

As a preliminary check, we computed several NLTE models
with the program Tlusty 
\citep{Hubeny_Lanz_1995ApJ...439..875H,Hubeny_Lanz_2017arXiv170601859H}.
For the modified chemical composition, it was necessary
to improve the convergence by adjusting several parameters.%
\footnote{namely ${\rm ND} = 70$,
${\rm NITER} = 50$,
${\rm ITEK} = 20$,
${\rm TAUDIV} = 1.0$}
The output of Tlusty was then used as an input for
the program Synspec \citep{lanz2007,Hubeny_Lanz_2017arXiv170601859H},
to obtain a detailed synthetic spectrum;
with the line list of Kurucz.
We computed spectra only for the effective temperature
$14000\,{\rm K}$ and a limited wavelength range 6330 to 6700\,\AA\
(see Figure~\ref{n14000g23henl.7_HeI_}).
The level of continuum is by 3\,\% higher for He-rich abundances.
The (non-rotated) H$\alpha$ line profile shows Lorentzian wings
deeper by 6\,\% and \ion{He}{i} 6678 line exhibits a significant central absorption. 
Luckily, after rotational broadening, the profiles will not be so
different from standard ones and we proceed without an extensive
computation of a new grid of synthetic spectra.
Moreover, one of our stars (primary, gainer)
is partly hidden within the disk and its radiation
is reprocessed by circumstellar matter.

For simplicity, we thus use He-rich abundances only for the CSM.
The result is shown in Tab.~\ref{tab2} (column 'H0.4\_He0.6').
The model does converge, but the overall
$\chi^2_{\rm R} = 19.5$ remained high,
with $\chi^2_{\rm spe}$ being the dominant term.
On the basis of our modelling we thus cannot confirm
that abundances are He-rich.

\subsection{Differential visibility in \ion{He}{i} 6678}

\begin{figure}
\centering
\includegraphics[width=9cm]{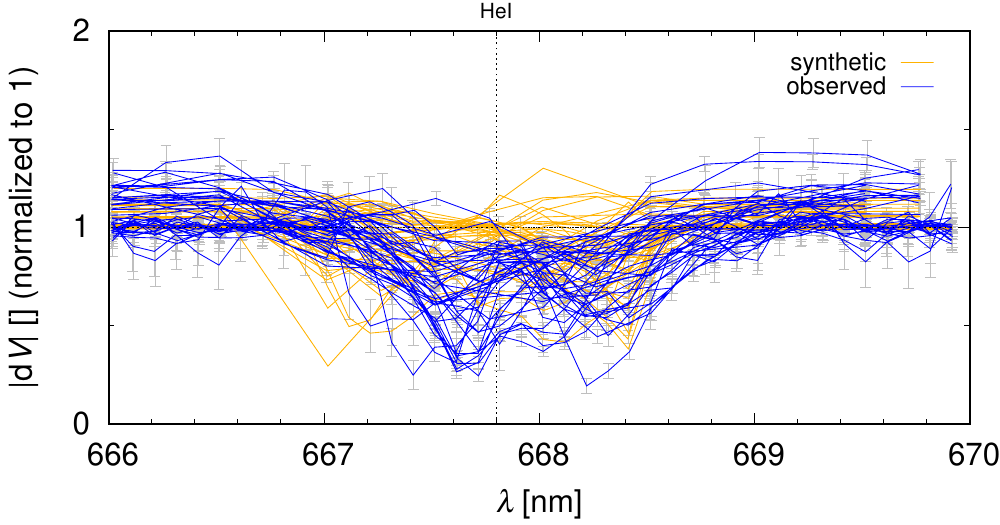}
\caption{
Differential visibility amplitude $|{\rm d}V|$.
for the \ion{He}{i} 6678 line.
The joint model was used, but no convergence of the respective \ion{He}{i} VAMP dataset.
}
\label{fitting_shell9_20200415__test_HE1_chi2_VAMP_null}
\end{figure}

\begin{figure}
\centering
\includegraphics[width=9cm]{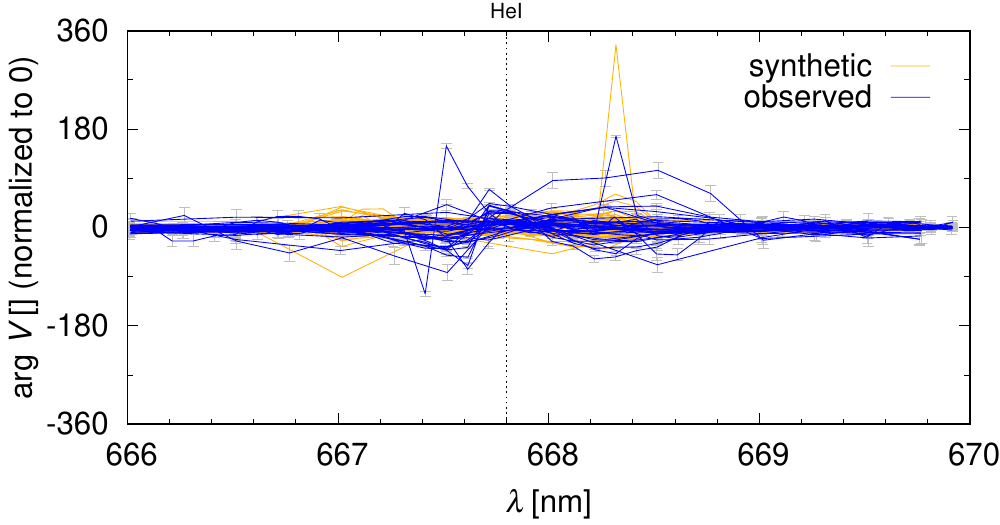}
\caption{
Differential visibility phase $\arg{\rm d}V$
for the \ion{He}{i} line.
Again, the joint model, but no convergence.
}
\label{fitting_shell9_20200415__test_HE1_chi2_VPHI_null}
\end{figure}

There are additional interferometric datasets,
namely differential visibilities in \ion{He}{i} 6678 line.
We performed a comparison only, not convergence.
According to Figure~\ref{fitting_shell9_20200415__test_HE1_chi2_VAMP_null} and Figure~\ref{fitting_shell9_20200415__test_HE1_chi2_VPHI_null}, there are systematic differences, with synthetic $|{\rm d}V|$'s being often flatter
than observed $|{\rm d}V|$'s.
The problem is likely the same as in Section~\ref{ETMPSH},
i.e., unidentified temperature gradients which would create
extended hot emission regions of \ion{He}{i}
(but not of H$\alpha$).
The comparison of these two figures with the similar ones for H$\alpha$ (Figure~\ref{fitting_shell8_20200415__20_chi2_VAMP_null} and Figure~\ref{fitting_shell8_20200415__20_chi2_VPHI_null}) is however very instructive. It is easily seen that the \ion{He}{i} data are firstly less resolved in the core of the line than in H$\alpha$ and secondly that almost no phase signal is detected. This result is in favor of a behavior dominated by the shell structure without any role for the jets here.

%

\subsection{Additional high-resolution spectroscopy}

There are also additional spectroscopic datasets,
particularly the one obtained at the Ritter Observatory and used by
\citet{Ignace_etal_2018AJ....156...97I}.
It has a~higher resolution than our spectra 
($\lambda/\Delta\lambda = 26000$) and a very good phase coverage.
The spectra were acquired in different seasons (1996 to 2000).
When we performed a comparison, not convergence, of 11 representative spectra there were systematic differences,
mainly in the observed emission peak around the primary eclipse
(i.e., the donor eclipsed, not the gainer)
which is substantially higher than in our model.
The observed H$\alpha$ profiles also do contain smaller features
which are not reproduced by our model. 
Nevertheless, the model can easily be adapted to the overall emission
(e.g., by adjusting densities $\rho_{\rm nb}$, $\rho_{\rm jt}$, $\rho_{\rm sh}$,
or by moving the jet along with $\alpha_{\rm jt} \simeq -140^\circ$
where a 2nd local minimum of $\chi^2_{\rm spe}$ is located).
It may be an indication that the distribution of optically-thin CSM
had been evolving on the time scale of ${>}10$ years (or ${>}300$ orbits).
Several investigators, most recently \citet{rucin2019}, noted
also cycle-to-cycle changes in the shape of the binary light curve.
If true, our model should be treated as a 'time-averaged' snapshot
of the system.



\begin{table*}
\centering
\caption{Free parameters, fixed parameters, $\chi^2$ values for a joint model
and for several alternative models.}
\label{tab2}

\begin{tabular}{lllllllllll}
\hline
\hline
\vrule width 0pt height 10pt depth 0pt
parameter &
unit &
joint &
294pc &
QRATIO &
SPOT &
ASYMJT &
{\bf ETMPJT} &
ETMPSH &
H0.4\_He0.6 &
\\
\hline
\vrule width 0pt height 10pt depth 0pt
$T_{\rm cp}$        & K & 14334 & 13512 & 14525 & 14580 & 14566 & 14085 & 14580 & 14293 &  \\
$R_{\rm innb}$      & $R_\odot$ & 8.7 & 8.2 & 8.8 & 8.7 & 8.7 & 10.2 & 9.5 & 10.4 &  \\
$R_{\rm outnb}$     & $R_\odot$ & 31.5 & 35.2 & 31.0 & 30.3 & 31.2 & 30.3 & 32.8 & 31.0 &  \\
$h_{\rm invnb}$     & $H$ & 3.5 & 4.4 & 3.6 & 3.5 & 3.8 & 4.4 & 3.9 & 3.7 &  \\
$T_{\rm invnb}$     & 1 & 1.5 & 1.8 & 1.5 & 1.5 & 1.5 & 1.5 & 1.5 & 1.8 &  \\
$h_{\rm windnb}$    & $H$  & 3.0 & 4.0 & 3.0 & 3.0 & 3.0 & 3.0 & 3.1 & 3.1 &  \\
$h_{\rm cnb}$       & $H$  & 3.8 & 3.1 & 3.6 & 3.7 & 3.6 & 3.8 & 2.9 & 2.9 &  \\
$v_{\rm nb}$        & km/s & 112 & 106 & 111 & 114 & 107 & 123 & 102 & 120 &  \\
$e_{\rm velnb}$     & 1 & $1.91$ & $1.57$ & $1.95$ & $1.91$ & $1.96$ & $1.69$ & $1.94$ & $1.99$ &  \\
$h_{\rm shdnb}$     & $H$ & 5.0 & 5.0 & 4.8 & 3.9 & 4.8 & 5.0 & 4.3 & 4.9 &  \\
$T_{\rm nb}$        & K  & 30345 & 30558 & 30443 & 30435 & 30836 & 29398 & 31233 & 32619 &  \\
$\varrho_{\rm nb}$ & $10^{-9}\,{\rm g}/{\rm cm}^3$ & 1.21 & 4.86 & 1.21 & 1.25 & 1.20 & 0.91 & 1.62 & 1.15 &  \\
$v_{\rm trbnb}$     & km/s  & 11 & 99 & 13 & 12 & 12 & 25 & 12 & 16 &  \\
$e_{\rm dennb}$     & 1 & $-0.57$ & $-0.55$ & $-0.56$ & $-0.57$ & $-0.54$ & $-0.57$ & $-0.59$ & $-0.59$ &  \\
$e_{\rm tmpnb}$     & 1 & $-0.73$ & $-0.70$ & $-0.73$ & $-0.73$ & $-0.73$ & $-0.73$ & $-0.73$ & $-0.71$ &  \\
$a_{\rm jet}$       & deg  & 28.8 & 32.7 & 28.8 & 28.7 & 28.9 & 28.8 & 21.4 & 32.0 &  \\
$R_{\rm injt}$      & $R_\odot$ & 5.6 & 6.6 & 5.6 & 5.6 & 5.6 & 5.4 & 5.8 & 5.5 &  \\
$R_{\rm outjt}$     & $R_\odot$ & 35.9 & 34.3 & 35.9 & 36.0 & 36.0 & 37.3 & 35.0 & 40.3 &  \\
$v_{\rm jt}$        & km/s  & 676 & 535 & 679 & 671 & 674 & 661 & 880 & 627 &  \\
$T_{\rm jt}$        & K  & 15089 & 22702 & 15150 & 15440 & 14668 & 30014 & 15822 & 13902 &  \\
$\varrho_{\rm jt}$ & $10^{-12}\,{\rm g}/{\rm cm}^3$ & 5.52 & 6.74 & 5.48 & 5.53 & 5.47 & 5.11 & 7.97 & 5.55 &  \\
$v_{\rm trbjt}$     & km/s  & 66 & 31 & 60 & 65 & 61 & 61 & 97 & 69 &  \\
$R_{\rm poljt}$     & $R_\odot$ & 33.0 & 3.6 & 32.4 & 33.1 & 32.5 & 33.0 & 33.3 & 34.0 &  \\
$v_{\rm poljt}$     & km/s  & 10 & 3 & 11 & 10 & 14 & 14 & 28 & 17 &  \\
$\alpha_{\rm jt}$        & deg  & $-70$ & $-28$ & $-70$ & $-70$ & $-70$ & $-70$ & $-55$ & $-104$ &  \\
$R_{\rm insh}$      & $R_\odot$ & 7.4 & 7.3 & 7.4 & 7.4 & 7.4 & 7.4 & 7.1 & 11.1 &  \\
$R_{\rm outsh}$     & $R_\odot$ & 72.9 & 77.0 & 72.7 & 73.1 & 72.8 & 71.1 & 69.4 & 75.7 &  \\
$v_{\rm sh}$        & km/s  & 79 & 98 & 79 & 80 & 78 & 70 & 90 & 87 &  \\
$e_{\rm velsh}$     & 1 & $1.90$ & $1.97$ & $1.95$ & $1.93$ & $1.96$ & $1.99$ & $1.89$ & $1.93$ &  \\
$v_{\rm ysh}$       & km/s  & $-5$ & $-37$ & $-3$ & $-3$ & $-4$ & $-5$ & $10$ & $-23$ &  \\
$T_{\rm sh}$        & K  & 5631 & 5549 & 5628 & 5620 & 5637 & 5631 & 18888 & 5678 &  \\
$\varrho_{\rm sh}$ & $10^{-11}\,{\rm g}/{\rm cm}^3$ & 2.86 & 1.42 & 2.92 & 2.94 & 2.91 & 2.86 & 0.90 & 3.05 &  \\
$v_{\rm trbsh}$     & km/s & 102 & 134 & 101 & 102 & 99 & 95 & 96 & 96 &  \\
$I$                  & deg & 96.3 & 96.2 & 96.3 & 96.3 & 96.3 & 96.4 & 96.9 & 96.4 &  \\
$\Omega$            & deg & 254.6 & 254.9 & 254.6 & 254.6 & 254.6 & 254.0 & 255.0 & 254.6 &  \\
$d$                  & pc & 328.4 & 294.0 & 328.6 & 328.5 & 328.6 & 325.7 & 329.5 & 328.6 &  \\
$M_\star$            & $M_\odot$ & 13.048 & 13.048 & 13.260 & 13.048 & 13.048 & 13.048 & 13.048 & 13.048 &  \\
$q$                  & 1 & 0.2230 & 0.2230 & 0.2177 & 0.2230 & 0.2230 & 0.2230 & 0.2230 & 0.2230 &  \\
$a_{\rm symjt}$     & 1 & $0.00$ & $0.00$ & $0.00$ & $0.00$ & $0.02$ & $0.00$ & $0.00$ & $0.00$ &  \\
$e_{\rm veljt}$     & 1 & $1.27$ & $1.29$ & $1.24$ & $1.29$ & $1.21$ & $1.39$ & $1.67$ & $1.25$ &  \\
$e_{\rm tmpjt}$     & 1 & $0.00$ & $0.00$ & $0.00$ & $0.00$ & $0.00$ & $-0.55$ & $0.00$ & $0.00$ &  \\
$e_{\rm tmpsh}$     & 1 & $-0.01$ & $-0.01$ & $-0.01$ & $-0.01$ & $-0.01$ & $-0.00$ & $-0.64$ & $-0.00$ &  \\
\hline\vrule width 0pt height 10pt depth 4pt$N_{\rm iter}$      & -- & 2761 & 965 & 946 & 228 & 1300 & 1721 & 1615 & 879 &  \\
$N$                  & -- & 45102 & 45102 & 45102 & 45102 & 45102 & 45102 & 45102 & 45102 &  \\
$\chi^2$             & -- & 767681 & 763215 & 743809 & 755684 & 743503 & 739977 & 818502 & 878773 &  \\
\vrule width 0pt height 0pt depth 4pt$\chi^2_{\rm R}$    & -- & 17.0 & 16.9 & 16.5 & 16.8 & 16.5 & 16.4 & 18.1 & 19.5 &  \\
\hline\vrule width 0pt height 10pt depth 4pt$\chi^2_{\rm lc}$   & -- & 7083 & 11545 & 6937 & 6707 & 6964 & 6954 & 8489 & 8323 &  \\
$\chi^2_{\rm vis}$  & -- & 56941 & 77843 & 57526 & 57771 & 57837 & 56017 & 56773 & 69041 &  \\
$\chi^2_{\rm clo}$  & -- & 25910 & 41057 & 25632 & 25792 & 25598 & 25786 & 27332 & 27355 &  \\
$\chi^2_{\rm t3}$   & -- & 16194 & 11845 & 17183 & 18141 & 17455 & 14690 & 16291 & 24981 &  \\
$\chi^2_{\rm sed}$  & -- & 8578 & 24490 & 8989 & 10018 & 9289 & 10035 & 14463 & 13188 &  \\
$\chi^2_{\rm spe}$  & -- & 588455 & 537458 & 565051 & 573972 & 564484 & 565680 & 640013 & 671153 &  \\
$\chi^2_{\rm vamp}$ & -- & 5959 & 4478 & 5951 & 5884 & 6017 & 5833 & 6958 & 5016 &  \\
\vrule width 0pt height 0pt depth 4pt$\chi^2_{\rm vphi}$ & -- & 58562 & 54498 & 56541 & 57398 & 55859 & 54982 & 48183 & 59716 &  \\
\hline
\end{tabular}

\tablefoot{The quantities are the same as in Tab.~\ref{tab1}.
The model with jet temperature gradients (denoted 'ETMPJT')
is our preferred model, as explained in Section~\ref{ETMPJT}.
}
\end{table*}






\section{Conclusions (and problems)}

We presented a geometrically-constrained model of $\beta$~Lyr~A
which takes into account all types of available observational data.
It contains the primary,
the Roche-filling secondary,
the optically thick disk,
its optically thin atmosphere,
the jets,
and the shell.
We determined absolute sizes of all the components,
physical properties ($\rho$, $T$, $v$ profiles),
and the distance to the system.
They are summarized in Table~\ref{tab1}.

Some of the parameters of the joint model are close to their
maximum/minimum values, in particular
$T_{\rm cp}$,
$R_{\rm outnb}$,
$e_{\rm velsh}$.
It is an independent indication that our model is not yet complete
and we may miss some features.
For example, the outer radius $R_{\rm outnb}$ of the disk (a.k.a. nebula)
almost touches the Roche lobe.
This may induce perturbations and a precession of the disk.
These instabilities are not accounted for in our model.
Additionally, the outer rim may not be in an exact equilibrium,
because of the ongoing mass transfer,
and the secondary may induce spiral arms,
i.e., azimuthal variations in the disk
\citep{Panoglou_etal_2019MNRAS.486.5139P}.

The mass loss rate from jets is substantial.
Given the surface area at the beginning of the cone,
$S = 2\pi R_{\rm injt}^2(1-\cos a_{\rm jet})$,
and the respective expansion velocity, we get
$\dot M_{\rm jt} = 2 S v_{\rm r} \rho_{\rm jt} 
\simeq 8.3\cdot 10^{-7}\,M_\odot\,{\rm yr}^{-1}$ 
which is about 4\,\% of the mass transfer rate
$\dot M \simeq 2\cdot10^{-5}\,M_\odot\,{\rm yr}^{-1}$.
Consequently, the mass transfer is not conservative,
but it is not far from being conservative.
The time scale related to the jets is
$\tau \simeq R_{\rm outjt}/v_{\rm jt} \simeq 0.5\,{\rm d}$
which is shorter than the orbital period.
The jets are continuously replenished as they follow the orbital motion.
It is interesting to integrate the mass loss over long time scales
and check observationally where the (expanded and cooled-down) CSM is located.
According to \cite{umana2000} measurements of the radio emission,
the CSM is very extended ($145\,{\rm mas}\,\hat=\, 10^4\,R_\odot$ at our $d$)
and the integrated mass is $M \simeq 0.015\,M_\odot$.
Consequently, the time scale would be of the order of
$\tau \simeq M/\dot M_{\rm jt} \simeq 10^5\,{\rm yr}$
which agrees with the binary evolution time scale
\citep{vanrensbergen2016}.

We kept masses of both components more-or-less fixed, in accord with
previous spectroscopic analyses of individual absorption lines (\ion{Si}{ii}, \ion{Ne}{i}).
The distance is then determined mainly from interferometry and the SED,
but spectroscopy (SPE) is also affected,
because Keplerian velocity fields are determined by central masses.
Let us recall that our preferred distance $d = (328\pm 7)\,{\rm pc}$
is larger than 294\,pc inferred by \cite{Bastian_2019A&A...630L...8B}
(see Section~\ref{294pc}).

Looking at synthetic profiles of the \ion{He}{i} 6678 line
in detail, it is in a broad emission
(due to the inner hot edge of the disk and Keplerian broadening),
but the observed profile is steeper,
double-peaked,
with a red peak being stronger
and a blue-shifted central absorption
(i.e., very similar to H$\alpha$).
Consequently, the inner edge should be even more visible
and should exhibit some absorption due to winds.
Our model cannot easily adapt to this,
because the emission in H$\alpha$ would be increased immediately
and we already match its EW.
Moreover, the differences in the H$\alpha$ and \ion{He}{i} are
of the same order, which is an indication of a compromise.

On contrary, the synthetic \ion{Si}{ii} 6347 and 6371 lines tend to produce
a broad emission and a weak absorption, even at an increased metallicity,
but the observed \ion{Si}{ii}'s exhibit deeper absorptions.
Similarly, the synthetic \ion{Ne}{i} 6402 line is in a broad (disk) emission,
but observed profile is rather flat.

According to a standard stellar evolution, the metallicity of the
stellar surfaces -- as well as of the CSM --
can be substantially different from normal (solar),
if the mass transfer have reached chemically modified layers.
For the parameters corresponding to our model,
we expect a low abundance of C (by a factor of $10^2$).
A likely final outcome would be a detached system with a He-rich dwarf
(similar to $\phi$~Per; \citealt{Mourard_etal_2015A&A...577A..51M}).

From the technical point of view, our model is somewhat resolution-dependent.
The peak densities or temperatures are not necessarily well resolved;
in other words, the profiles are effectively shallower/smoother.
A higher resolution may be also needed to obtain smooth P~Cygni profiles
in thin layers with velocity gradients, such as in expanding atmospheres.

In order to improve the convergence of our model, it may be useful
to use the least correlated parameters. For example,
the total mass of the shell
(instead of $R_{\rm insh}$, $R_{\rm routsh}$, and $\rho_{\rm sh}$),
or a suitable reference radius
(between $R_{\rm innb}$, $R_{\rm outnb}$; inclusive),
in accord with observational datasets, which are sensitive
either to outer, or inner radii; in our case, outer seems better.

\vskip\baselineskip

Nevertheless, in spite of the remaining limitations discussed above,
it is encouraging that our modelling of an~extended set of different types of observational data led to a generally consistent {\em quantitative\/} picture
of the system. We note that the principal physical properties obtained
from several specific considered models are numerically quite stable,
not too different from one model to another.
Our models nicely confirm the conjecture that the so called B spectrum (introduced and discussed in the classical early studies of \ble) originates mainly in the jets, the disk atmosphere or other circumstellar matter above/below the orbital plane,
including their blueshift of about 
$50$ to $100\,{\rm km}\,{\rm s}^{-1}$ \citep{hec92,hec96,bonneau2011}.
Also the presence of an extended shell, observationally
detected by \citet{ak2007}, is required by our model and data. 
Another important result of this study is a~convincing confirmation of 
the strong carbon underabundance, in accord with the models of the
large-scale mass exchange in binaries. We thus believe that the present study constitutes a good starting point to even more sophisticated
modelling of the system.


\begin{acknowledgements}
We thank J.A.~Nemravov\'a for the initial development of the Pyshellspec.
We also thank P.~Koubsk\'y, R.~K\v{r}\'\i\v{c}ek, D.~Kor\v{c}\'akov\'a as well as J.A.N., who obtained some of the spectra used in this study.
We thank R. Ignace for providing us with his reduced spectra
from the Ritter Observatory in phase space and for illuminating comments on his study. We also acknowledge the Ritter Observatory for making the archive of spectra public.
M.B. and P.H. were supported by the Czech Science Foundation grant~19-01995S.
J.B. was supported by the VEGA 2/0031/18 grant and by the Slovak Research and Development Agency under the contract No. APVV-15-0458.
H.B. acknowledges financial support from the Croatian Science Foundation
under the project 6212 ``Solar and Stellar Variability''.
The CHARA Array is supported by the National Science Foundation under Grant No. AST-1636624 and
AST-1715788.
\end{acknowledgements}


\bibliographystyle{aa}
\bibliography{references}

\begin{thebibliography}{45}
\expandafter\ifx\csname natexlab\endcsname\relax\def\natexlab#1{#1}\fi

\bibitem[{{Ak} {et~al.}(2007){Ak}, {Chadima}, {Harmanec}, {Demircan}, {Yang},
  {Koubsk{\'y}}, {{\v S}koda}, {{\v S}lechta}, {Wolf}, {Bo{\v z}i{\'c}}, {Ru{\v
  z}djak}, \& {Sudar}}]{ak2007}
{Ak}, H., {Chadima}, P., {Harmanec}, P., {et~al.} 2007, \aap, 463, 233

\bibitem[{{Armstrong} {et~al.}(1998){Armstrong}, {Mozurkewich}, {Rickard},
  {Hutter}, {Benson}, {Bowers}, {Elias}, {Hummel}, {Johnston}, {Buscher},
  {Clark}, {Ha}, {Ling}, {White}, \& {Simon}}]{npoi}
{Armstrong}, J.~T., {Mozurkewich}, D., {Rickard}, L.~J., {et~al.} 1998, ApJ,
  496, 550

\bibitem[{{Atwood-Stone} {et~al.}(2012){Atwood-Stone}, {Miller}, {Richards},
  {Budaj}, \& {Peters}}]{stone2012}
{Atwood-Stone}, C., {Miller}, B.~P., {Richards}, M.~T., {Budaj}, J., \&
  {Peters}, G.~J. 2012, \apj, 760, 134

\bibitem[{{Balachandran} {et~al.}(1986){Balachandran}, {Lambert}, {Tomkin}, \&
  {Parthasarathy}}]{balach86}
{Balachandran}, S., {Lambert}, D.~L., {Tomkin}, J., \& {Parthasarathy}, M.
  1986, \mnras, 219, 479

\bibitem[{{Bastian}(2019)}]{Bastian_2019A&A...630L...8B}
{Bastian}, U. 2019, \aap, 630, L8

\bibitem[{{Bonneau} {et~al.}(2011){Bonneau}, {Chesneau}, {Mourard},
  {B{\'e}rio}, {Clausse}, {Delaa}, {Marcotto}, {Perraut}, {Roussel}, {Spang},
  {Stee}, {Tallon-Bosc}, {McAlister}, {ten Brummelaar}, {Sturmann}, {Sturmann},
  {Turner}, {Farrington}, \& {Goldfinger}}]{bonneau2011}
{Bonneau}, D., {Chesneau}, O., {Mourard}, D., {et~al.} 2011, \aap, 532, A148

\bibitem[{{Budaj}(2011)}]{budaj2011b}
{Budaj}, J. 2011, \aj, 141, 59

\bibitem[{{Budaj} \& {Richards}(2004)}]{budaj2004}
{Budaj}, J. \& {Richards}, M.~T. 2004, Contributions of the Astronomical
  Observatory Skalnate Pleso, 34, 167

\bibitem[{{Budaj} {et~al.}(2005){Budaj}, {Richards}, \& {Miller}}]{budaj2005}
{Budaj}, J., {Richards}, M.~T., \& {Miller}, B. 2005, \apj, 623, 411

\bibitem[{{Burnashev} \& {Skulskii}(1978)}]{burnashev78}
{Burnashev}, V.~I. \& {Skulskii}, M.~Y. 1978, Bull. Crimean Astrophys. Obs.,
  58, 53

\bibitem[{{Green} {et~al.}(2015){Green}, {Schlafly}, {Finkbeiner}, {Rix},
  {Martin}, {Burgett}, {Draper}, {Flewelling}, {Hodapp}, {Kaiser}, {Kudritzki},
  {Magnier}, {Metcalfe}, {Price}, {Tonry}, \&
  {Wainscoat}}]{Green_etal_2015ApJ...810...25G}
{Green}, G.~M., {Schlafly}, E.~F., {Finkbeiner}, D.~P., {et~al.} 2015, \apj,
  810, 25

\bibitem[{{Grevesse} \& {Sauval}(1998)}]{Grevesse_Sauval_1998SSRv...85..161G}
{Grevesse}, N. \& {Sauval}, A.~J. 1998, \ssr, 85, 161

\bibitem[{{Harmanec}(1992)}]{hec92}
{Harmanec}, P. 1992, \aap, 266, 307

\bibitem[{{Harmanec}(2002)}]{hec2002}
{Harmanec}, P. 2002, AN, 323, 87

\bibitem[{{Harmanec} {et~al.}(1996){Harmanec}, {Morand}, {Bonneau}, {Jiang},
  {Yang}, {Guinan}, {Hall}, {Mourard}, {Hadrava}, {Bozic}, {Sterken},
  {Tallon-Bosc}, {Walker}, {McCook}, {Vakili}, {Stee}, \& {Le Contel}}]{hec96}
{Harmanec}, P., {Morand}, F., {Bonneau}, D., {et~al.} 1996, \aap, 312, 879

\bibitem[{{Heber}(2009)}]{Ulrich_2009ARA&A..47..211H}
{Heber}, U. 2009, \araa, 47, 211

\bibitem[{{Horn} {et~al.}(1996){Horn}, {Kub\'at}, {Harmanec}, {Koubsk\'y},
  {Hadrava}, {\v{S}imon}, {\v{S}tefl}, \& {\v{S}koda}}]{sef0}
{Horn}, J., {Kub\'at}, J., {Harmanec}, P., {et~al.} 1996, \aap, 309, 521

\bibitem[{{Huben\'y} \& {Lanz}(1995)}]{Hubeny_Lanz_1995ApJ...439..875H}
{Huben\'y}, I. \& {Lanz}, T. 1995, \apj, 439, 875

\bibitem[{{Huben\'y} \& {Lanz}(2017)}]{Hubeny_Lanz_2017arXiv170601859H}
{Huben\'y}, I. \& {Lanz}, T. 2017, arXiv e-prints, arXiv:1706.01859

\bibitem[{{Husser} {et~al.}(2013){Husser}, {Wende-von Berg}, {Dreizler},
  {Homeier}, {Reiners}, {Barman}, \& {Hauschildt}}]{husser2013}
{Husser}, T.-O., {Wende-von Berg}, S., {Dreizler}, S., {et~al.} 2013, \aap,
  553, A6

\bibitem[{{Ignace} {et~al.}(2018){Ignace}, {Gray}, {Magno}, {Henson}, \&
  {Massa}}]{Ignace_etal_2018AJ....156...97I}
{Ignace}, R., {Gray}, S.~K., {Magno}, M.~A., {Henson}, G.~D., \& {Massa}, D.
  2018, \aj, 156, 97

\bibitem[{{Krpata}(2008)}]{spefo3}
{Krpata}, J. 2008, http://astro.troja.mff.cuni.cz/ftp/hec/SPEFO/

\bibitem[{{Lanz} \& {Huben\'y}(2003)}]{lanz2003}
{Lanz}, T. \& {Huben\'y}, I. 2003, \apjs, 146, 417

\bibitem[{{Lanz} \& {Huben\'y}(2007)}]{lanz2007}
{Lanz}, T. \& {Huben\'y}, I. 2007, \apjs, 169, 83

\bibitem[{{Lei} {et~al.}(2018){Lei}, {Zhao}, {N{\'e}meth}, \&
  {Zhao}}]{Lei_etal_2018ApJ...868...70L}
{Lei}, Z., {Zhao}, J., {N{\'e}meth}, P., \& {Zhao}, G. 2018, \apj, 868, 70

\bibitem[{{Monnier} {et~al.}(2004){Monnier}, {Berger}, {Millan-Gabet}, \& {ten
  Brummelaar}}]{mirc}
{Monnier}, J.~D., {Berger}, J.-P., {Millan-Gabet}, R., \& {ten Brummelaar},
  T.~A. 2004, in New Frontiers in Stellar Interferometry, ed. W.~A. {Traub},
  Vol. 5491, 1370

\bibitem[{{Mourard} {et~al.}(2011){Mourard}, {B{\'e}rio}, {Perraut}, {Ligi},
  {Blazit}, {Clausse}, {Nardetto}, {Spang}, {Tallon-Bosc}, {Bonneau},
  {Chesneau}, {Delaa}, {Millour}, {Stee}, {Le Bouquin}, {ten Brummelaar},
  {Farrington}, {Goldfinger}, \& {Monnier}}]{vega2}
{Mourard}, D., {B{\'e}rio}, P., {Perraut}, K., {et~al.} 2011, \aap, 531, A110

\bibitem[{{Mourard} {et~al.}(2018){Mourard}, {Bro{\v{z}}}, {Nemravov{\'a}},
  {Harmanec}, {Budaj}, {Baron}, {Monnier}, {Schaefer}, {Schmitt},
  {Tallon-Bosc}, {Armstrong}, {Baines}, {Bonneau}, {Bo{\v{z}}i{\'c}},
  {Clausse}, {Farrington}, {Gies}, {Jury{\v{s}}ek}, {Kor{\v{c}}{\'a}kov{\'a}},
  {McAlister}, {Meilland}, {Nardetto}, {Svoboda}, {{\v{S}}lechta}, {Wolf}, \&
  {Zasche}}]{Mourard_etal_2018A&A...618A.112M}
{Mourard}, D., {Bro{\v{z}}}, M., {Nemravov{\'a}}, J.~A., {et~al.} 2018, \aap,
  618, A112

\bibitem[{{Mourard} {et~al.}(2009){Mourard}, {Clausse}, {Marcotto}, {Perraut},
  {Tallon-Bosc}, {B{\'e}rio}, {Blazit}, {Bonneau}, {Bosio}, {Bresson},
  {Chesneau}, {Delaa}, {H{\'e}nault}, {Hughes}, {Lagarde}, {Merlin}, {Roussel},
  {Spang}, {Stee}, {Tallon}, {Antonelli}, {Foy}, {Kervella}, {Petrov},
  {Thiebaut}, {Vakili}, {McAlister}, {Ten Brummelaar}, {Sturmann}, {Sturmann},
  {Turner}, {Farrington}, \& {Goldfinger}}]{vega}
{Mourard}, D., {Clausse}, J.~M., {Marcotto}, A., {et~al.} 2009, \aap, 508, 1073

\bibitem[{{Mourard} {et~al.}(2015){Mourard}, {Monnier}, {Meilland}, {Gies},
  {Millour}, {Benisty}, {Che}, {Grundstrom}, {Ligi}, {Schaefer}, {Baron},
  {Kraus}, {Zhao}, {Pedretti}, {Berio}, {Clausse}, {Nardetto}, {Perraut},
  {Spang}, {Stee}, {Tallon-Bosc}, {McAlister}, {ten Brummelaar}, {Ridgway},
  {Sturmann}, {Sturmann}, {Turner}, \&
  {Farrington}}]{Mourard_etal_2015A&A...577A..51M}
{Mourard}, D., {Monnier}, J.~D., {Meilland}, A., {et~al.} 2015, \aap, 577, A51

\bibitem[{{Nemravov{\'a}} {et~al.}(2016){Nemravov{\'a}}, {Harmanec}, {Bro{\v
  z}}, {Vokrouhlick{\'y}}, {Mourard}, {Hummel}, {Cameron}, {Matthews},
  {Bolton}, {Bo{\v z}i{\'c}}, {Chini}, {Dembsky}, {Engle}, {Farrington},
  {Grunhut}, {Guenther}, {Guinan}, {Kor{\v c}{\'a}kov{\'a}}, {Koubsk{\'y}},
  {K{\v r}{\'{\i}}{\v c}ek}, {Kuschnig}, {Mayer}, {McCook}, {Moffat},
  {Nardetto}, {Pr{\v s}a}, {Ribeiro}, {Rowe}, {Rucinski}, {{\v S}koda}, {{\v
  S}lechta}, {Tallon-Bosc}, {Votruba}, {Weiss}, {Wolf}, {Zasche}, \&
  {Zavala}}]{jn2016}
{Nemravov{\'a}}, J.~A., {Harmanec}, P., {Bro{\v z}}, M., {et~al.} 2016, \aap,
  594, A55

\bibitem[{{Panoglou} {et~al.}(2019){Panoglou}, {Borges Fernandes}, {Baade},
  {Faes}, {Rivinius}, {Carciofi}, \&
  {Okazaki}}]{Panoglou_etal_2019MNRAS.486.5139P}
{Panoglou}, D., {Borges Fernandes}, M., {Baade}, D., {et~al.} 2019, \mnras,
  486, 5139

\bibitem[{{Paxton} {et~al.}(2011){Paxton}, {Bildsten}, {Dotter}, {Herwig},
  {Lesaffre}, \& {Timmes}}]{Paxton_etal_2011ApJS..192....3P}
{Paxton}, B., {Bildsten}, L., {Dotter}, A., {et~al.} 2011, \apjs, 192, 3

\bibitem[{{Paxton} {et~al.}(2015){Paxton}, {Marchant}, {Schwab}, {Bauer},
  {Bildsten}, {Cantiello}, {Dessart}, {Farmer}, {Hu}, {Langer}, {Townsend},
  {Townsley}, \& {Timmes}}]{Paxton_etal_2015ApJS..220...15P}
{Paxton}, B., {Marchant}, P., {Schwab}, J., {et~al.} 2015, \apjs, 220, 15

\bibitem[{{Pringle}(1981)}]{pringle1981}
{Pringle}, J.~E. 1981, \araa, 19, 137

\bibitem[{{Rowan}(1990)}]{Rowan_1990}
{Rowan}, N. 1990, Ph.D. thesis, Univ. Texas Austin

\bibitem[{{Rucinski} {et~al.}(2019){Rucinski}, {Pigulski}, {Kuschnig},
  {Moffat}, {Popowicz}, {Pablo}, {Wade}, {Weiss}, \& {Zwintz}}]{rucin2019}
{Rucinski}, S.~M., {Pigulski}, A., {Kuschnig}, R., {et~al.} 2019, \aj, 158, 148

\bibitem[{{Sahade}(1966)}]{sahade66}
{Sahade}, J. 1966, Transactions of the International Astronomical Union, Series
  B, 12, 494

\bibitem[{{Schlafly} \&
  {Finkbeiner}(2011)}]{Schlafly_Finkbeiner_2011ApJ...737..103S}
{Schlafly}, E.~F. \& {Finkbeiner}, D.~P. 2011, \apj, 737, 103

\bibitem[{{Skulskii}(2020)}]{skul2020}
{Skulskii}, M.~Y. 2020, arXiv e-prints, arXiv:2005.10802

\bibitem[{{ten Brummelaar} {et~al.}(2005){ten Brummelaar}, {McAlister},
  {Ridgway}, {Bagnuolo}, {Turner}, {Sturmann}, {Sturmann}, {Berger}, {Ogden},
  {Cadman}, {Hartkopf}, {Hopper}, \& {Shure}}]{chara}
{ten Brummelaar}, T.~A., {McAlister}, H.~A., {Ridgway}, S.~T., {et~al.} 2005,
  \apj, 628, 453

\bibitem[{{Tereshchenko} \& {Kharitonov}(1972)}]{teres72}
{Tereshchenko}, V.~M. \& {Kharitonov}, A.~V. 1972, Trudy Astrofizicheskogo
  Instituta Alma-Ata, 21

\bibitem[{{Umana} {et~al.}(2000){Umana}, {Maxted}, {Trigilio}, {Fender},
  {Leone}, \& {Yerli}}]{umana2000}
{Umana}, G., {Maxted}, P.~F.~L., {Trigilio}, C., {et~al.} 2000, \aap, 358, 229

\bibitem[{{van Hamme}(1993)}]{vanhamme1993}
{van Hamme}, W. 1993, \aj, 106, 2096

\bibitem[{{van Rensbergen} \& {De Greve}(2016)}]{vanrensbergen2016}
{van Rensbergen}, W. \& {De Greve}, J.~P. 2016, \aap, 592, A151

\end{thebibliography}


\appendix

\section{Figures for observation-specific models}

\begin{figure}
\includegraphics[width=9cm]{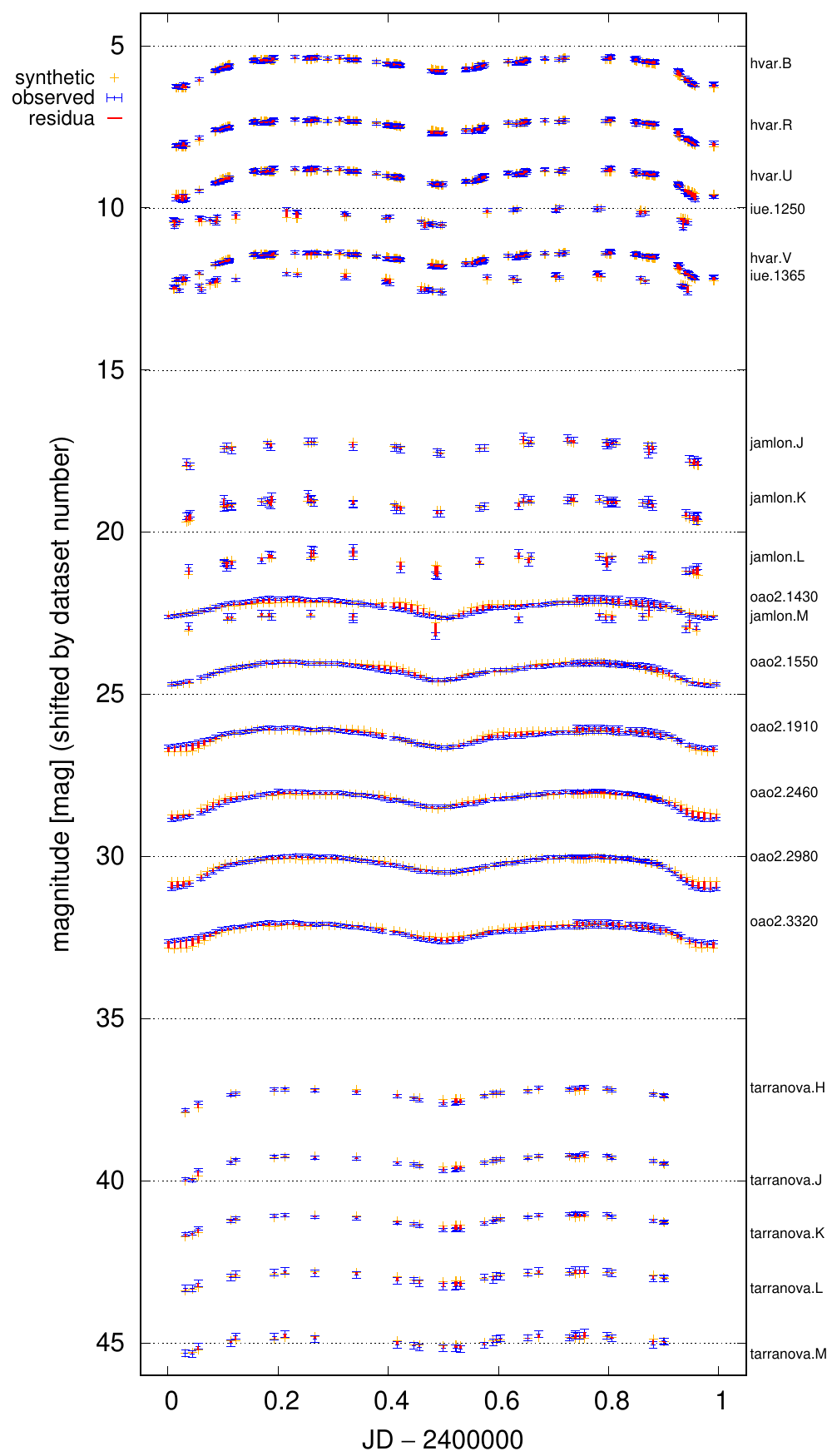}
\caption{
The same as Fig.~\ref{fitting_shell8_20200415__20_chi2_LC_PHASE} for the observation-specific model.
}
\label{fitting_shell9_20200415_SED_chi2_LC_PHASE}
\end{figure}

\begin{figure}
\includegraphics[width=9cm]{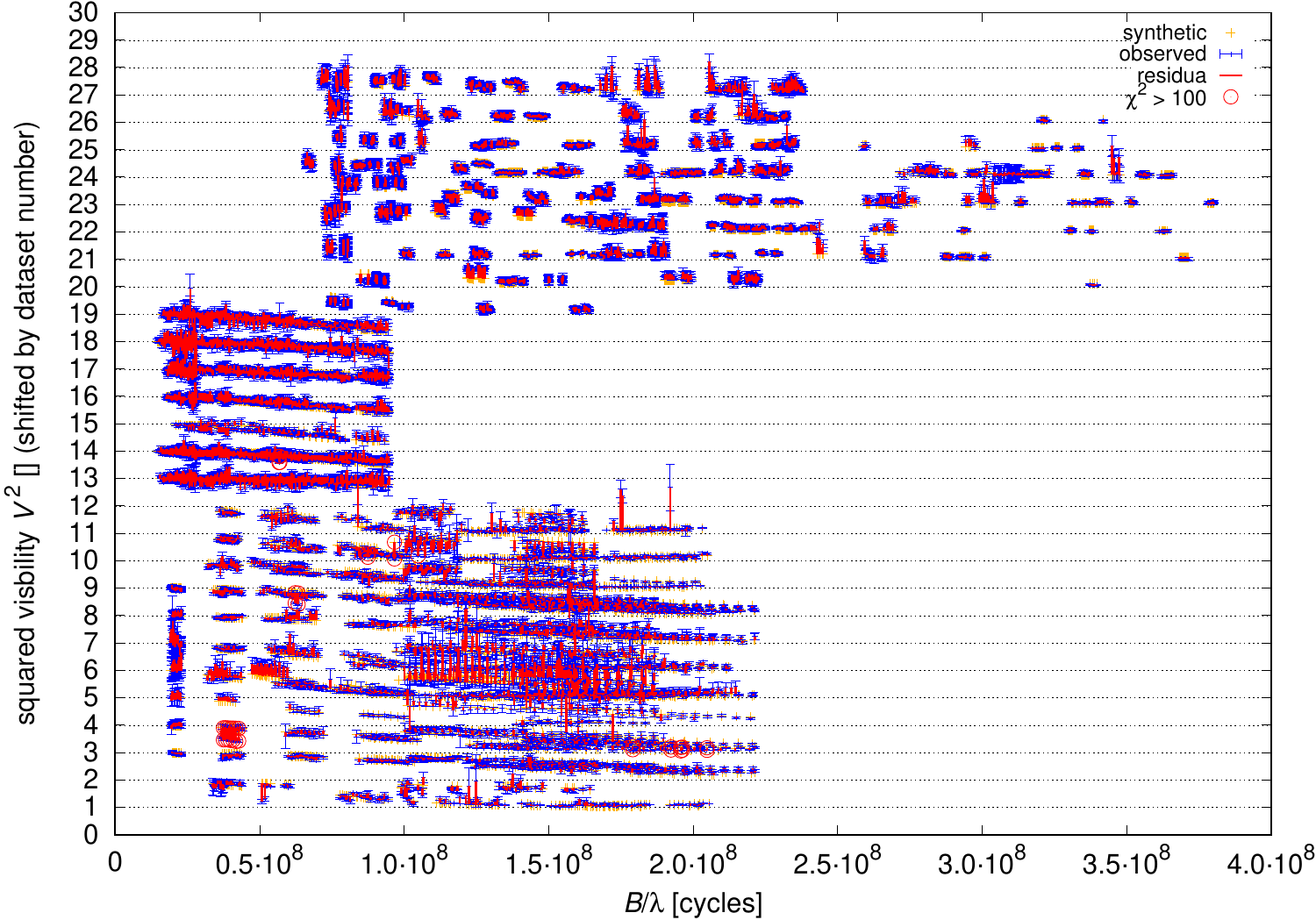}
\caption{
The same as Fig.~\ref{fitting_shell8_20200415__20_chi2_VIS} for the observation-specific model.
}
\label{fitting_shell9_20200415_VIS_chi2_VIS}
\end{figure}

\begin{figure}
\includegraphics[width=9cm]{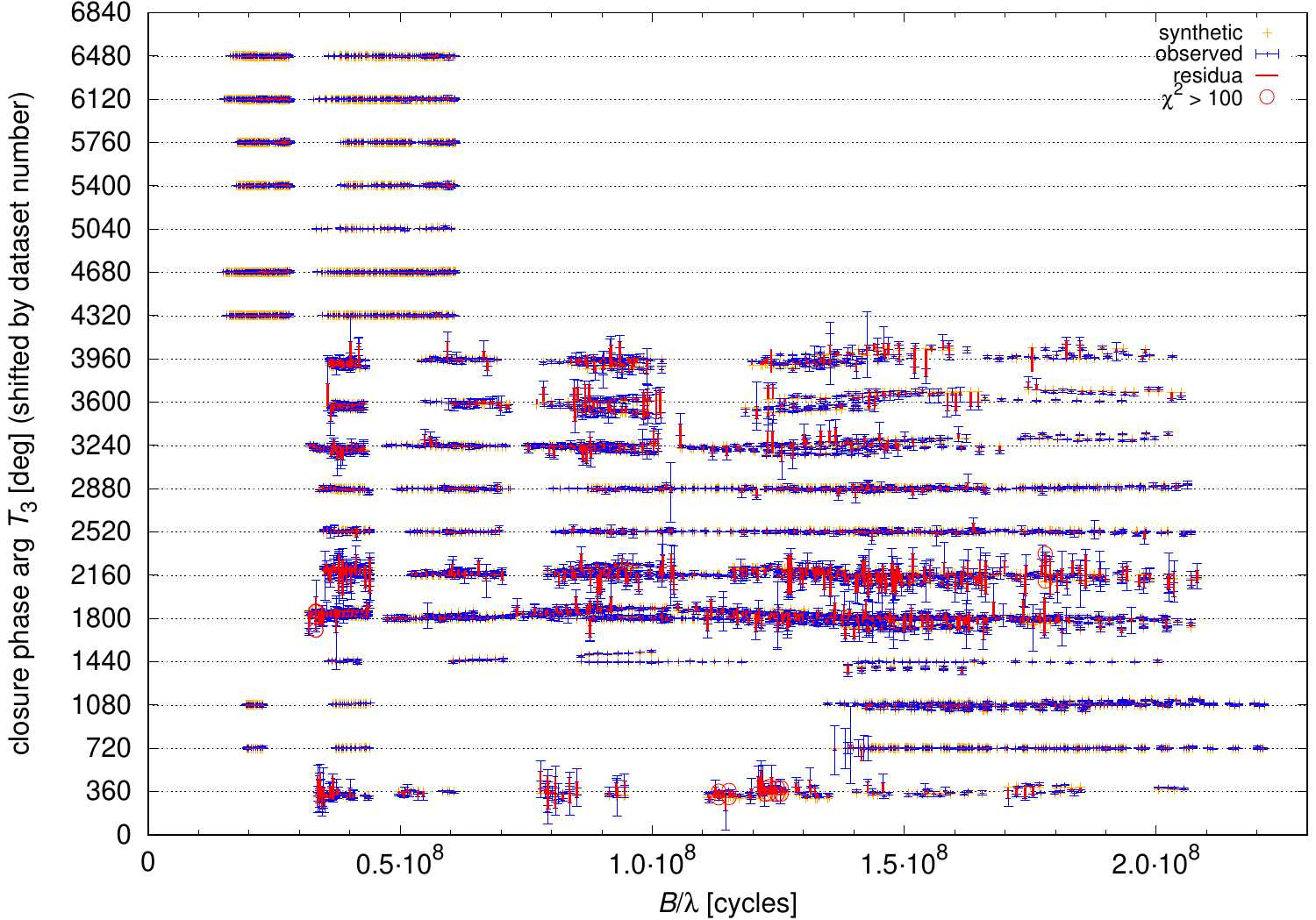}
\caption{
The same as Fig.~\ref{fitting_shell8_20200415__20_chi2_CLO} for the observation-specific model.
}
\label{fitting_shell9_20200415_CLO_chi2_CLO}
\end{figure}

\begin{figure}
\includegraphics[width=9cm]{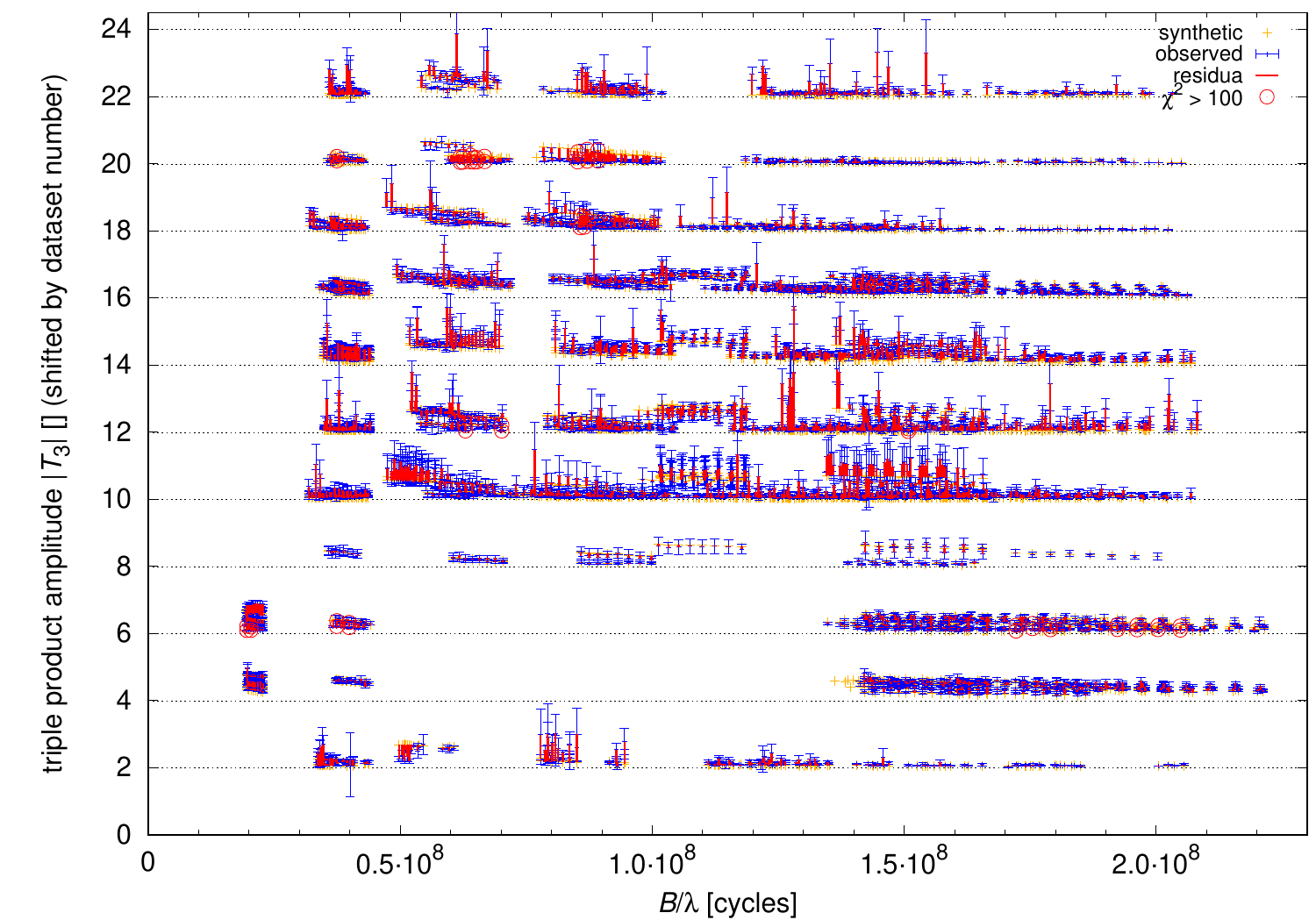}
\caption{
The same as Fig.~\ref{fitting_shell8_20200415__20_chi2_T3} for the observation-specific model.
}
\label{fitting_shell9_20200415_T3_chi2_T3}
\end{figure}

\begin{figure}
\includegraphics[width=9cm]{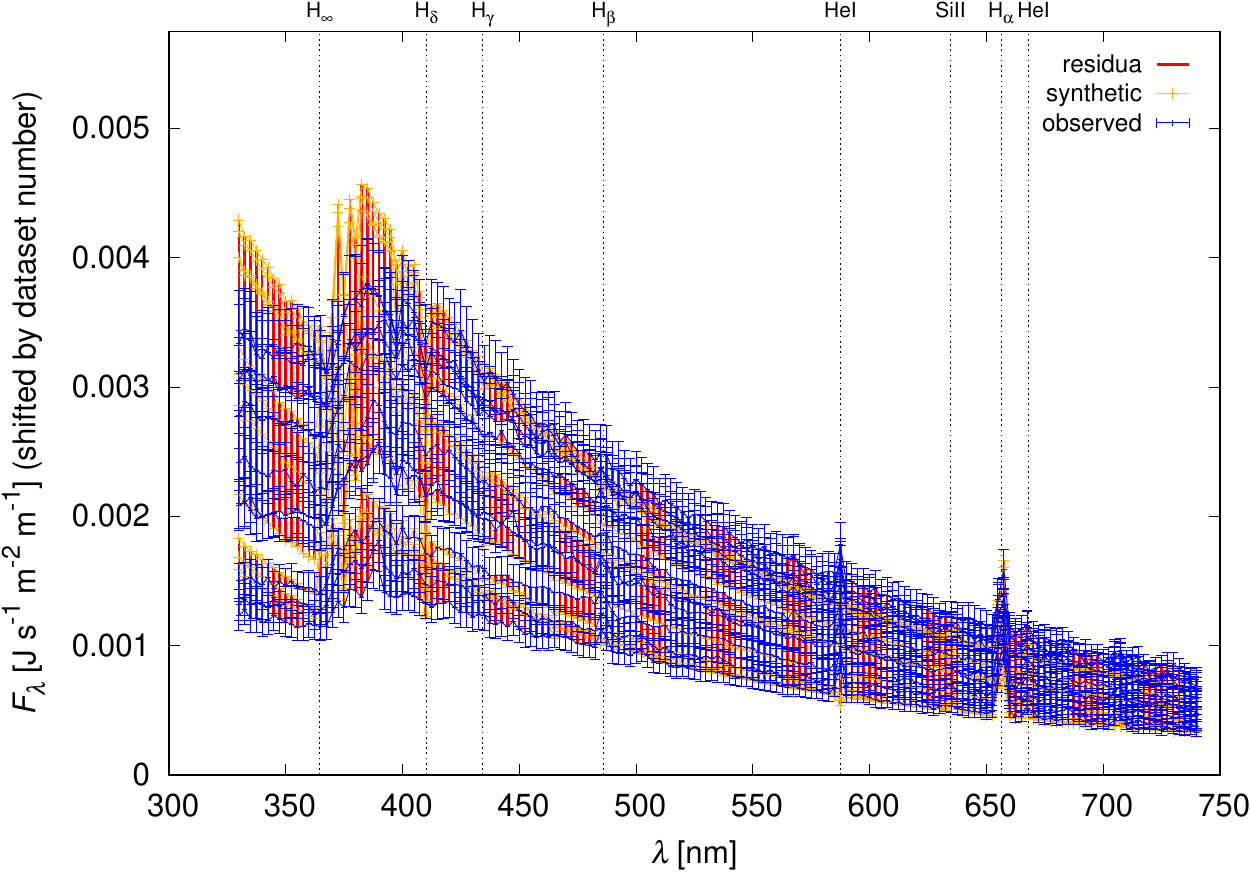}
\caption{
The same as Fig.~\ref{fitting_shell8_20200415__20_chi2_SED} for the observation-specific model.
}
\label{fitting_shell9_20200415_SED_chi2_SED}
\end{figure}

\begin{figure}
\includegraphics[width=9cm]{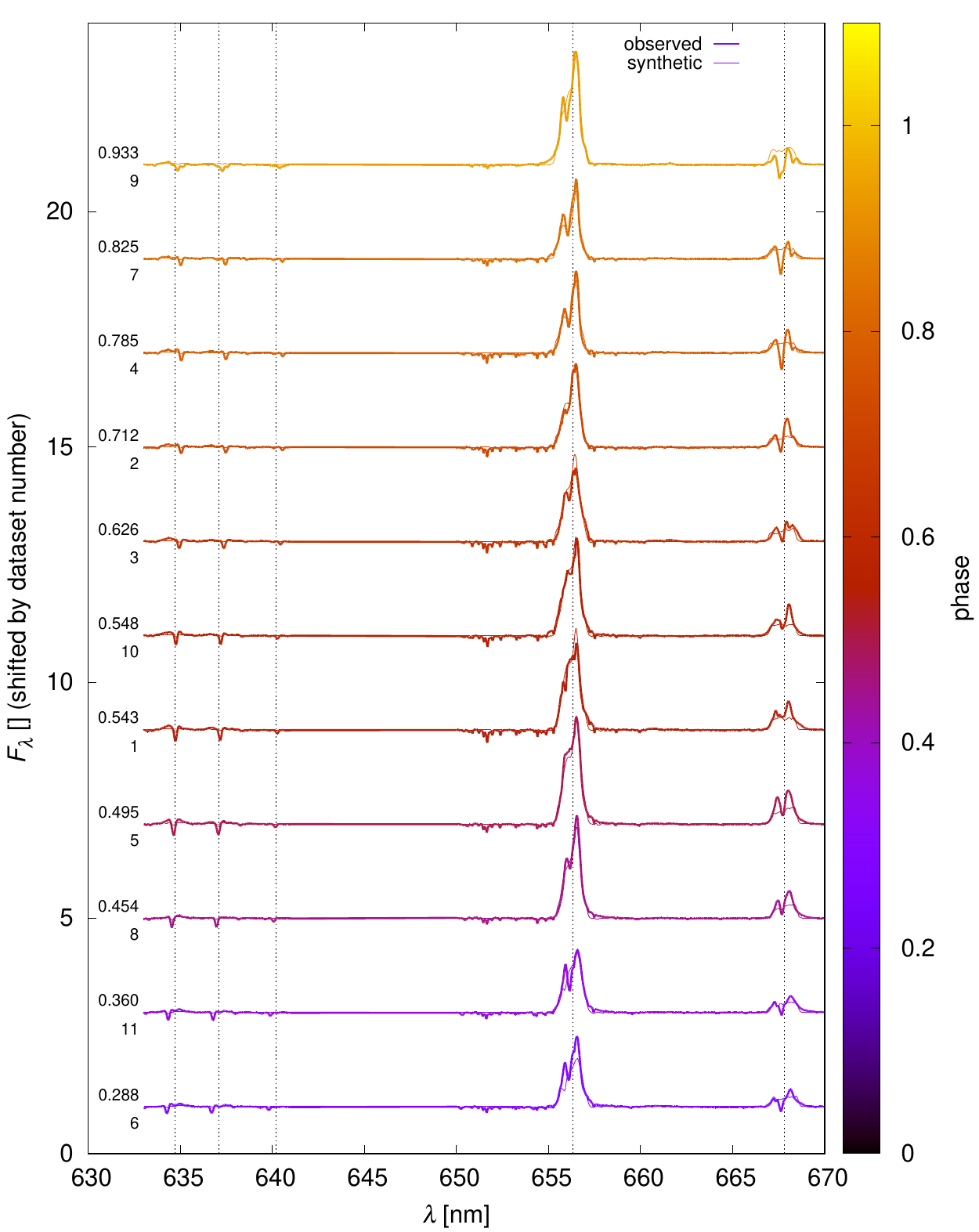}
\caption{
The same as Fig.~\ref{fitting_shell8_20200415__20_chi2_SPE_PHASE} for the observation-specific model.
}
\label{fitting_shell9_20200415_SPE_chi2_SPE_PHASE}
\end{figure}

\begin{figure}
\includegraphics[width=9cm]{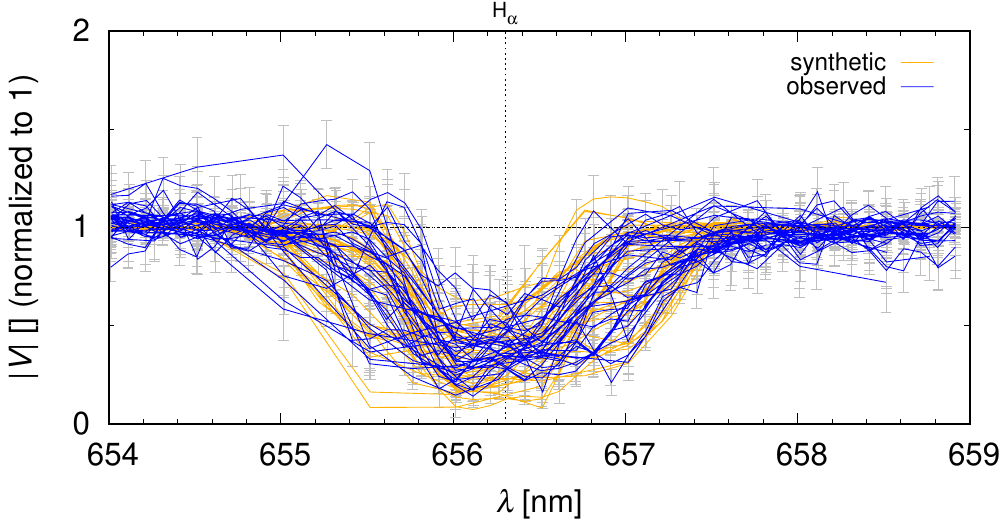}
\caption{
The same as Fig.~\ref{fitting_shell8_20200415__20_chi2_VAMP_null} for the observation-specific model.
}
\label{fitting_shell9_20200415_VAMP_chi2_VAMP_null}
\end{figure}

\begin{figure}
\includegraphics[width=9cm]{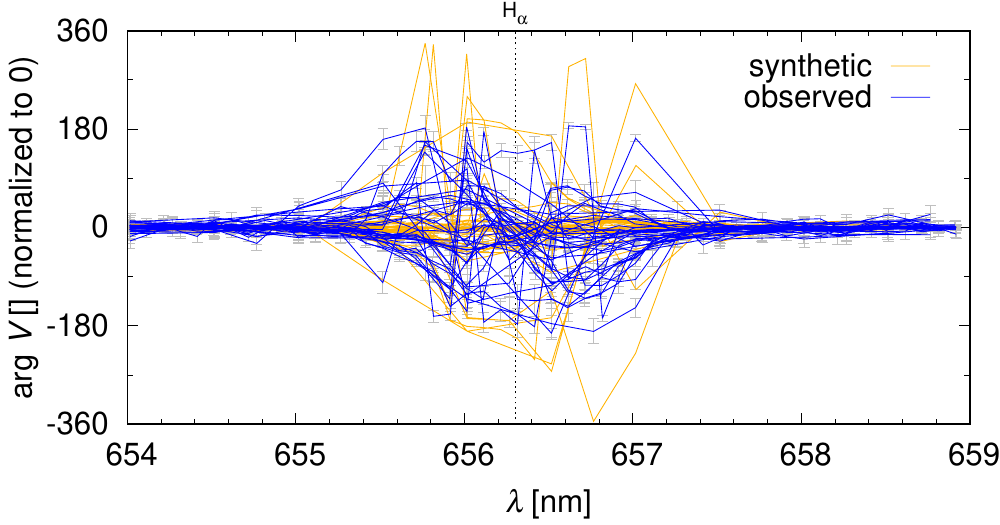}
\caption{
The same as Fig.~\ref{fitting_shell8_20200415__20_chi2_VPHI_null} for the observation-specific model.
}
\label{fitting_shell9_20200415_VPHI_chi2_VPHI_null}
\end{figure}

The observation-specific models converged for a given dataset 
(LC, VIS, CLO, T3, SED, SPE, VAMP, VPHI)
are presented in Figures~\ref{fitting_shell9_20200415_SED_chi2_LC_PHASE}
to~\ref{fitting_shell9_20200415_VPHI_chi2_VPHI_null}.
Generally, the individual $\chi^2$ contributions are smaller
than for the joint model (see Tab.~\ref{tab1}).
We take these results also as a confirmation that our model
converges and is capable to optimally fit these datasets.

\end{document}